%% file: main.tex
\pdfoutput=1
\documentclass[aps,prx,twocolumn,superscriptaddress,floatfix,nofootinbib,longbibliography]{revtex4-2}
\usepackage{graphicx,booktabs,multirow,amsmath,amssymb,xcolor,xspace,microtype}
\usepackage{float}
\usepackage{tikz}
\usetikzlibrary{arrows.meta,decorations.pathreplacing,positioning,calc}
\input{figs/figstyle}
\usepackage[colorlinks=true,allcolors=blue]{hyperref}
\usepackage{orcidlink}
\newcommand{\sys}{Duet\xspace}
\newcommand{\lom}{LoMatching\xspace}
\newcommand{\bplom}{BP+LoMatching\xspace}
\newcommand{\tess}{Tesseract\xspace}

\newcommand{\ie}{\textit{i.e.}}

\begin{document}
\title{Neural Decoding of Logical Clifford Circuits with Transversal Gates}
\author{Yicheng Guang\,\orcidlink{0009-0006-6180-9497}}
\email[Contact author: ]{yicheng.guang@colorado.edu}
\affiliation{Department of Computer Science, University of Colorado Boulder, Boulder, Colorado, USA}
\author{Yuliang Xiong\,\orcidlink{0009-0000-7258-1770}}
\affiliation{Department of Computer Science, University of Colorado Boulder, Boulder, Colorado, USA}
\author{Xun Gao\,\orcidlink{0009-0005-7278-4005}}
\email[Contact author: ]{xun.gao@colorado.edu}
\affiliation{Department of Physics and JILA, University of Colorado Boulder, Boulder, Colorado, USA}
\author{Yueqi Chen\,\orcidlink{0009-0007-4205-3273}}
\email[Contact author: ]{yueqi.chen@colorado.edu}
\affiliation{Department of Computer Science, University of Colorado Boulder, Boulder, Colorado, USA}
\date{\today}

\begin{abstract}
\input{sections/abstract}

\end{abstract}
\maketitle

\input{sections/intro}

\input{sections/background}
\input{sections/design}
\input{sections/eval}
\input{sections/discussion}
\input{sections/related}
\input{sections/conclusion}

\begin{acknowledgments}
We used Anthropic's Claude Opus 5.5 for implementation and for polishing the text, Anthropic's Claude Sonnet 5.5 for implementation, and OpenAI's GPT-6 Astra and Anthropic's Claude Fable 5.1 for review.
We checked every result and take full responsibility. X.G. acknowledges funding from the U.S. Department of Energy, Quantum Systems Accelerator.
\end{acknowledgments}

\section*{Data availability}
The code, the trained networks, the data behind every table and figure, and the raw records of every test are available at \texttt{https://github.com/a85tract/Duet}~\cite{duet_code}.

\appendix
\input{sections/appendix}

\clearpage
\bibliography{refs}
\end{document}

%% file: figs/figstyle.tex
\definecolor{fnet}{HTML}{004488}
\definecolor{fcomp}{HTML}{BB5566}
\definecolor{fzchk}{HTML}{E9C87A}
\definecolor{fxchk}{HTML}{EBEBEB}
\colorlet{fnetbg}{fnet!10}
\colorlet{fcompbg}{fcomp!26}
\colorlet{fline}{black!85}
\tikzset{
  figbase/.style={x=1cm,y=1cm,font=\footnotesize,line width=0.6pt,draw=fline,line join=round,
    >={Stealth[length=4pt,width=3.2pt]}},
  fthin/.style={line width=0.4pt},
  fmid/.style={line width=0.6pt},
  fheavy/.style={line width=1pt},
  fbox/.style={draw=fline,line width=0.6pt,rounded corners=1pt,fill=white,inner sep=2.5pt,minimum height=0.6cm,
    align=center},
  flow/.style={->,draw=fline,line width=0.6pt},
}

%% file: sections/abstract.tex
On neutral-atom and trapped-ion machines, a transversal gate needs only one syndrome round, so a logical gate takes up to $d$ times less time than with lattice surgery.
However, transversal gates make decoding harder, because a transversal CNOT copies errors between two blocks of physical qubits and a decoder must therefore read all blocks of a circuit together.
Algorithmic decoders for such circuits gain accuracy at the cost of speed.
Neural decoders ease this trade-off, but they must infer from data where errors go across each gate and how logical outcomes propagate through the circuit, although both are known exactly.
In this work, we present \sys, a neural decoder for logical Clifford circuits with transversal gates that builds in this known information and learns only which errors occurred.
The ``\sys'' name reflects a duet between the circuit, which supplies what is known, and the network, which learns the rest.
More specifically, \sys leverages three insights.
First, each transversal gate moves an error to a fixed place, so the network exchanges features only between the sites that a gate connects.
Second, a logical outcome propagates through Clifford gates exactly and linearly, so the network predicts only the flips that the errors of each layer cause and combines the layers by a soft XOR.
Third, every circuit is built from the same few gates, so the network takes the gates of each layer as input and one network per code distance decodes circuits of different sizes and depths.
To further reduce failures, we pair \sys with the search decoder \tess, which re-decodes the small fraction of shots on which the network is least confident.
On general circuits with 2 to 4 blocks and up to 16 layers, \sys alone is already more accurate than matching at every distance, and with \tess it fails 44 times less often than LoMatching at distance 9.
On the random circuits of the Lukin group, it fails 0.299 times as often as their neural decoder at distance 7.
The composition stays fast and takes 2.4 to 121.7\,$\mu$s per layer, less than their network takes per gate at every distance.

%% file: sections/intro.tex
\section{Introduction}
\label{sec:intro}

A quantum computer represents each logical qubit in a block of many physical qubits and repeatedly measures parity checks on the block to find errors.
Every round of check measurements is called a syndrome round, and a detection event is a check whose outcome differs from its previous round, so errors appear as detection events around them.
A decoder reads the detection events and decides which logical outcomes the errors flipped.
For example, the surface code of distance $d$ arranges $d\times d$ qubits on a square grid.
With a suitable decoder, such a block can correct any set of up to $\lfloor(d-1)/2\rfloor$ physical errors~\cite{dennis,fowler}.
The decoder must be accurate to avoid logical errors, and fast because the logical qubits keep decohering while they wait.

A quantum circuit computes by applying a sequence of logical gates, such as CNOT, H and S, to the logical qubits.
The implementation of these logical gates differs across hardware platforms.
For example, neutral-atom and trapped-ion machines implement the logical CNOT between two blocks as a transversal CNOT, \ie, one physical CNOT between each pair of matching physical qubits.
These machines can also run the logical H on one block transversally and the logical S fold-transversally~\cite{foldS}.
A transversal CNOT needs only one syndrome round, whereas lattice surgery needs about $d$ rounds~\cite{horsman}, so a logical gate takes up to $d$ times less time~\cite{zhou25,cain24}.

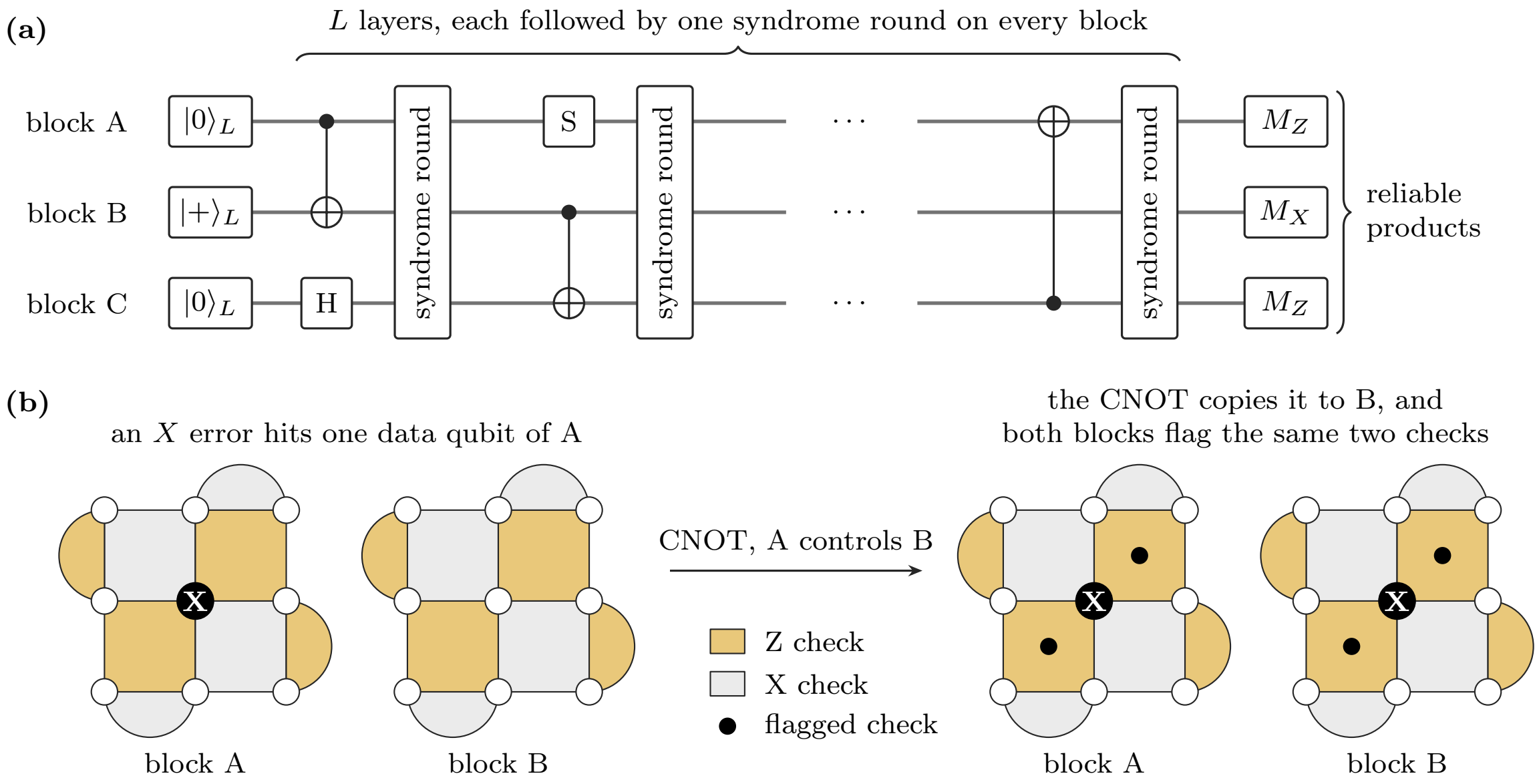
\begin{figure*}[t]
\centering
\input{figs/fig_circuit}

\caption{The circuits and how a CNOT spreads an error. (a) A random logical Clifford circuit on $K=3$ blocks. Each block is prepared in $|0\rangle_L$ or $|+\rangle_L$, each layer holds transversal CNOTs between random pairs of blocks and I, H or S gates on the other blocks, one syndrome round on every block follows each layer, and each block is measured in the $Z$ or $X$ basis. The decoder outputs the reliable logical products of the final measurements. (b) A single $X$ error on one data qubit of block A flips two $Z$ checks of A. A transversal CNOT with A as control copies the error to the same qubit of block B, and then both blocks flag two checks. Drawn for a distance-3 patch with its four weight-4 and four weight-2 checks.}
\label{fig:circuit}
\end{figure*}

However, transversal gates make decoding harder.
A transversal CNOT copies an error on any qubit of one block to the matching qubit of the other block (Figure~\ref{fig:circuit}(b)).
Decoding the blocks one by one discards the correlations that the CNOTs create, and Zhou et al.\ prove that decoding them jointly keeps the circuit fault-tolerant with a constant number of syndrome rounds per gate~\cite{zhou25}.
Therefore, a decoder must read all blocks of a circuit together.
Figure~\ref{fig:circuit}(a) shows the circuits we study, random logical Clifford circuits on $K$ blocks.
Such a circuit runs in layers, and each layer applies gates to all blocks in parallel and is followed by one syndrome round.

The available decoders for circuits with transversal gates fall into two groups.
Algorithmic decoders run a fixed algorithm on the error model of the circuit, gaining accuracy at the cost of speed.
For example, matching decoders, including LoMatching~\cite{lom} and the decoder of Cain et al.~\cite{cain25}, encode how errors cross the gates by fixed rules and take $\mu$s per shot on a CPU core.
\bplom adds belief propagation, which improves on matching but takes 28\,ms to 1.4\,s per shot in our circuits.
The search decoder \tess~\cite{tesseract} and integer programming~\cite{cain24} come close to the most likely error but take up to seconds per shot.

Neural decoders ease this trade-off because they decode a shot without search and run in batches on a GPU~\cite{bonilla,mccd}.
For example, the network of the Lukin group~\cite{bonilla} decodes random logical circuits on four blocks, fails at most 1.65 times as often as the exact most-likely-error decoder up to distance 7, and takes only tens to hundreds of $\mu$s per gate.
However, such a network must also infer from data where errors go across each gate and how logical outcomes propagate through the circuit, although both are known exactly.
Learning them is costly, because the number of gate sequences grows exponentially with the depth and the number of blocks.

In this work, we present \sys, a neural decoder for logical Clifford circuits with transversal gates.
The key idea behind \sys is that, since where errors go across each gate and how logical outcomes propagate through the circuit are both known before any shot is decoded, a network can build in this information and learn only which errors occurred.
The ``\sys'' name reflects a duet between the circuit, which supplies what is known, and the network, which learns the rest.
More specifically, \sys leverages three insights as below.

\vspace{0.05in}
\noindent\textbf{Insight 1: Each transversal gate moves an error to a fixed place.}
A transversal CNOT copies an error to the qubit at the same position of the partner block, or to its rotated partner when an H has rotated one of the two blocks (Figure~\ref{fig:circuit}(b)).
A fold-transversal S copies an X error on a qubit as a Z error onto its mirror across the anti-diagonal of the same block, which links each X check to a fixed Z check of that block.
Therefore, which detection events of different blocks an error can link is fixed by the gates, and the network does not need to discover it from data.
\sys places the detection events of every block on a grid and, at every gate, exchanges features only between the grid sites that the gate connects.

\vspace{0.05in}
\noindent\textbf{Insight 2: A logical outcome propagates through Clifford gates exactly and linearly.}
For the errors of any layer, the gates after that layer determine exactly which errors on which block flip a given logical outcome at the end of the circuit, and propagating the outcome backwards through the gates finds them~\cite{cain25,lom}.
Further, these flips add up modulo 2, so the final flip of the outcome is the parity of the flips that the errors of each layer cause.
Therefore, the network does not need to learn the propagation but only which errors occurred in each layer.
In \sys, the read-out of each layer takes the propagated operator of that layer as input and is trained on an exact label for that layer from the detector error model, and a soft XOR combines the layers~\cite{baireuther,zhang26}.

\vspace{0.05in}
\noindent\textbf{Insight 3: Every circuit is built from the same few gates.}
Although the number of gate sequences grows exponentially with the depth and the number of blocks, every circuit consists of CNOT, H and S gates, and each gate acts on one or two blocks within one layer.
With the gates handled locally (Insight 1) and the propagation computed exactly (Insight 2), the network needs to know only which gate acts on each block at each layer, not the whole sequence.
Therefore, \sys takes the role of every block in the gates of each layer as input, shares all weights across blocks, and ties no parameter to the number of blocks or layers.
In the read-out, the blocks enter only through sums over blocks, and the depth only through the number of terms of the soft XOR.
As a result, one network per code distance decodes circuits with 2 to 4 blocks and up to 16 layers, and with four blocks and up to 20 layers, without retraining.

\vspace{0.05in}
To further reduce failures, we pair \sys with \tess through the confidence-based composition of the CoDecoder work~\cite{codecoder}.
The network decodes every shot, and \tess re-decodes the small fraction of shots on which the network is least confident.

In our experiments, we trained one network per code distance, $d=3$, 5, 7 and 9, under the noise model of the Lukin group, and evaluated it using two benchmarks.
Benchmark I consists of general circuits with 2 to 4 blocks, 1 to 16 layers, CNOT, H and S gates, preparation and measurement in both bases, and three noise strengths.
On these circuits, we ran LoMatching~\cite{lom}, the decoder of Cain et al.~\cite{cain25} and \bplom~\cite{lom} as baselines.
At $p=0.3\,\%$, averaged over block counts and depths, \sys alone fails 0.56 to 0.15 times as often as LoMatching from $d=3$ to 9, and Cain et al.\ performs like LoMatching.
In other words, without any search, \sys is already more accurate than matching at every distance, and its advantage grows with the distance.
Composed with \tess, \sys fails even less often.
At $d=9$, it fails 44 times less often than LoMatching and 7.8 times less often than \bplom.
The composition stays fast because \tess re-decodes only 1 to 2\,\% of the shots.
With batches of 256 shots on one GPU and 32 AMD EPYC cores, it takes 2.4 to 121.7\,$\mu$s per layer from $d=3$ to 9, less than the network of the Lukin group takes per gate at every distance.

Benchmark II consists of random circuits on four blocks with up to 20 layers from the Lukin group, so that we can compare \sys with their neural decoder.
We rebuilt these circuits and took the failure rates of their network from their paper.
On this benchmark, \sys composed with \tess fails 0.866, 0.705 and 0.299 times as often as their network at $d=3$, 5 and 7.
Further, its failure rate is statistically compatible with that of the exact most-likely-error decoder on the shots for which that decoder certified its answer.

In summary, this work makes three major contributions:
\begin{itemize}
\setlength{\itemsep}{2pt}
\item We identify three insights into decoding logical Clifford circuits with transversal gates. They illustrate which parts of the decoding problem are fixed by the circuit and need not be learned.
\item Based on these insights, we design \sys, a neural decoder that builds this known structure into the network and learns only which errors occurred.
\item We compose \sys with \tess and evaluate it on two benchmarks at distances 3 to 9.  The results demonstrate the effectiveness of \sys in both accuracy and speed.
\end{itemize}

We organize the rest of this paper as follows.
Section~\ref{sec:background} defines the circuits, the noise and the decoders.
Section~\ref{sec:design} presents \sys.
Section~\ref{sec:eval} evaluates it.
Section~\ref{sec:discussion} discusses its scope and limitations, Section~\ref{sec:related} reviews related work, and Section~\ref{sec:conclusion} concludes this paper.

%% file: figs/fig_circuit.tex
\begin{tikzpicture}[figbase,
  wire/.style={fheavy,draw=black!55},
  gate/.style={fbox,minimum height=0.5cm,minimum width=0.5cm,inner sep=1.5pt},
  dq/.style={circle,draw=fline,fill=white,fthin,minimum size=2.6mm,inner sep=0pt},
  err/.style={circle,draw=black,fill=black,fthin,minimum size=3.6mm,inner sep=0pt,text=white,font=\footnotesize\bfseries},
  flag/.style={circle,fill=black,minimum size=1.7mm,inner sep=0pt},
  chk/.style={draw=fline,fthin},
  panel/.style={anchor=north west,font=\footnotesize\bfseries,inner sep=0pt}]
\begin{scope}[shift={(0,4.6)}]
  \node[panel] at (0,2.85) {(a)};
  \foreach \y/\b in {1.8/A,0.9/B,0/C}{
    \node[anchor=east] at (1.35,\y) {block \b};
    \draw[wire] (2.05,\y)--(7.75,\y) (9.05,\y)--(12.7,\y);
    \node at (8.4,\y) {$\cdots$};
  }
  \foreach \y/\s in {1.8/0,0.9/+,0/0}{\node[gate,minimum width=0.82cm] at (2.05,\y) {$|\s\rangle_L$};}
  \foreach \y/\m in {1.8/Z,0.9/X,0/Z}{\node[gate,minimum width=0.82cm] at (12.7,\y) {$M_\m$};}
  \foreach \x/\c/\t in {3.2/1.8/0.9, 5.6/0.9/0, 10.4/0/1.8}{
    \draw[fmid] (\x,\c)--(\x,\t);
    \fill[fline] (\x,\c) circle (0.075);
    \draw[fmid,fill=white] (\x,\t) circle (0.15);
    \draw[fmid] (\x-0.15,\t)--(\x+0.15,\t) (\x,\t-0.15)--(\x,\t+0.15);
  }
  \node[gate] at (3.2,0) {H};
  \node[gate] at (5.6,1.8) {S};
  \foreach \x in {4.15,6.55,11.35}{\node[fbox,rotate=90,minimum width=2.5cm,minimum height=0.55cm,inner sep=1pt] at (\x,0.9) {syndrome round};}
  \draw[fmid,decorate,decoration={brace,amplitude=4pt}] (2.9,2.4)--(11.65,2.4)
    node[midway,above=4pt] {$L$ layers, each followed by one syndrome round on every block};
  \draw[fmid,decorate,decoration={brace,amplitude=4pt}] (13.2,2.1)--(13.2,-0.3)
    node[midway,right=5pt,align=left] {reliable\\products};
\end{scope}
\def\patch#1#2#3#4{%
  \begin{scope}[shift={(#1,#2)}]
    \foreach \i in {0,1}{\foreach \j in {0,1}{
      \pgfmathtruncatemacro{\pty}{mod(\i+\j,2)}
      \ifnum\pty=0 \filldraw[chk,fill=fzchk] (0.9*\i,0.9*\j) rectangle +(0.9,0.9);
      \else \filldraw[chk,fill=fxchk] (0.9*\i,0.9*\j) rectangle +(0.9,0.9);\fi
    }}
    \filldraw[chk,fill=fxchk] (0.9,1.8) arc[start angle=180,end angle=0,radius=0.45] -- cycle;
    \filldraw[chk,fill=fxchk] (0,0) arc[start angle=180,end angle=360,radius=0.45] -- cycle;
    \filldraw[chk,fill=fzchk] (0,1.8) arc[start angle=90,end angle=270,radius=0.45] -- cycle;
    \filldraw[chk,fill=fzchk] (1.8,0) arc[start angle=-90,end angle=90,radius=0.45] -- cycle;
    \ifnum#4=1 \node[flag] at (0.45,0.45) {}; \node[flag] at (1.35,1.35) {};\fi
    \foreach \i in {0,1,2}{\foreach \j in {0,1,2}{\node[dq] at (0.9*\i,0.9*\j) {};}}
    \ifnum#3=1 \node[err] at (0.9,0.9) {X};\fi
  \end{scope}}
\node[panel] at (0,3.75) {(b)};
\patch{1.0}{0.75}{1}{0}
\patch{4.0}{0.75}{0}{0}
\patch{9.9}{0.75}{1}{1}
\patch{12.9}{0.75}{1}{1}
\foreach \x/\b in {1.9/A,4.9/B,10.8/A,13.8/B}{\node[anchor=base] at (\x,-0.05) {block \b};}
\node[anchor=south] at (3.4,3.05) {an $X$ error hits one data qubit of A};
\node[anchor=south,align=center] at (12.3,3.05) {the CNOT copies it to B, and\\both blocks flag the same two checks};
\draw[flow] (6.6,1.95)--(9.1,1.95) node[midway,above=1pt] {CNOT, A controls B};
\begin{scope}[shift={(7.0,1.25)}]
  \foreach \y/\c/\t in {0/fzchk/Z check,-0.42/fxchk/X check}{
    \filldraw[chk,fill=\c] (0,\y-0.12) rectangle +(0.34,0.24);
    \node[anchor=west] at (0.42,\y) {\t};
  }
  \node[flag] at (0.17,-0.84) {};
  \node[anchor=west] at (0.42,-0.84) {flagged check};
\end{scope}
\end{tikzpicture}

%% file: sections/background.tex
\section{Circuits, Noise and Decoders}
\label{sec:background}
This section defines the circuits, the noise, the failure rate and the decoders that the later sections use.

\paragraph{Circuits.}
The circuits are those of Fig.~\ref{fig:circuit}(a), with $L$ layers on $K$ rotated surface-code blocks of distance $d$, each with $d^2$ data qubits.
The logical H is a physical H on every data qubit.
It turns the code of the block into its copy rotated by 90 degrees, so every check keeps its place and measures the other Pauli type from then on, and a later CNOT with a block of the other orientation pairs each data qubit with its rotated partner.
The logical S is the fold-transversal S of Chen et al.~\cite{foldS}, which runs inside the syndrome round of its layer and applies CZ gates between the qubits that mirror each other about the anti-diagonal of the patch.
Logical Pauli gates are tracked in software, and $S^\dagger=SZ$ costs one physical S.
The outcomes to decode are the reliable products, the products of final logical measurements whose values are fixed in the noiseless circuit~\cite{cain25}.
We decode an independent set of them, between one and four per circuit.
A shot fails if the decoder gets any of them wrong, and the failure rate is the fraction of failing shots.

\paragraph{Noise.}
We use the noise model of the Lukin group~\cite{bonilla} with one strength $p$.
Every two-qubit gate (the transversal CNOTs, the CNOTs of the syndrome rounds and the CZ gates of the S) is followed by two-qubit depolarizing noise of strength $p$.
Every single-qubit gate is followed by single-qubit depolarizing noise of strength $p/10$, every reset and every measurement fails with probability $p/10$, and idle qubits are free of noise.
Each block starts in its exact code state from a noiseless encoder, so the checks of the first round have fixed values.
We simulate the circuits with Stim~\cite{stim}.

\paragraph{Decoders.}
Matching decoders turn the detection events into a graph and pair the events with minimum total weight~\cite{dennis,pymatching}.
LoMatching~\cite{lom} builds one such graph for each reliable product from the detectors that the product's region covers and decodes the products one after another.
We run LoMatching 0.3.0 on the regions of its own definition, because its circuit reader finds no region for the centre-line logical operators that our H and S gates need.
Cain et al.~\cite{cain25} propagate each product back through the circuit and match on the checks along its path.
We reimplement it because it has no released decoder (Appendix~\ref{app:cain}).
\bplom~\cite{lom} runs min-sum belief propagation with a scaling factor of 0.5 on the whole circuit and, if it has not converged after 100 iterations, matches with the error probabilities of the last iteration.
\tess~\cite{tesseract} searches over sets of errors in a fixed order of the detectors and keeps a beam of the most promising sets.
Its long-beam setting runs 21 trials with 21 detector orders and beams of 0 to 20 and returns the cheapest answer, and we run the trials in parallel, one per CPU core.
The exact most-likely-error decoder, MLE for short, solves an integer program for the most likely set of errors, here with the SCIP solver and a limit of 600\,s per shot.
The optimal decoder instead picks the most likely logical class, which sums the probabilities of all sets of errors in the class, so a decoder can fail less often than the MLE.
The recurrent attention network of Bonilla Ataides et al.~\cite{bonilla} is the reference of Benchmark~II.

%% file: sections/design.tex
\section{Decoder}
\label{sec:design}
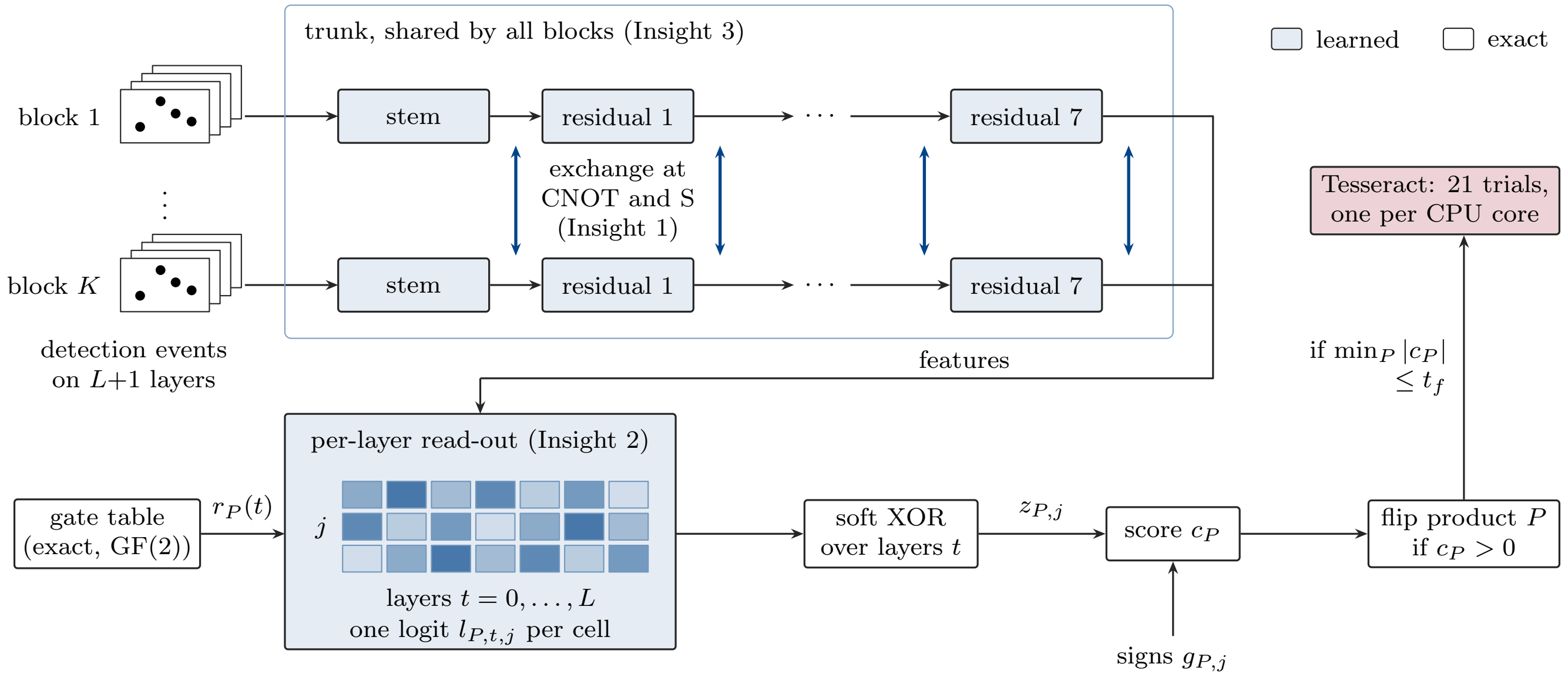
\begin{figure*}[t]
\centering
\input{figs/fig_network}
\caption{The network and the composition. The detection events of all $K$ blocks enter a convolutional trunk with shared weights, and an exchange step (blue arrows) mixes features between the blocks of a CNOT and inside the block of an S at every gate (Insight 1). The gate table gives the parts $r_P(t)$ of every reliable product $P$ on every block and layer, which condition the per-layer logits $l_{P,t,j}$. A soft XOR over the layers combines them into $z_{P,j}$ (Insight 2), and the signs $g_{P,j}$ from the detection events turn these into the score $c_P$ of each product. No weight belongs to one block or one layer (Insight 3). A shot whose smallest $|c_P|$ lies below the threshold of its depth goes to \tess, whose 21 trials run on separate CPU cores.}
\label{fig:network}
\end{figure*}

With the circuits and their reliable products defined, we now present how \sys builds in what the gates fix before any shot is decoded.
We first present the exchange at every gate, which realizes Insight 1 of Sec.~\ref{sec:intro}, the per-layer read-out, which realizes Insight 2, and the gate roles and shared weights, which realize Insight 3.
Then we describe the training, and finally the composition with \tess.
Figure~\ref{fig:network} gives an overview.

\subsection{Insight 1: Exchange at every gate}
\label{sec:network}
Insight 1 says that each transversal gate moves an error to a fixed place.
To build this in, \sys places the detection events on a grid on which every gate connects fixed sites, turns the grid into features with a convolutional trunk, and exchanges features between the connected sites at every gate.

\paragraph{Input grid.}
The grid has one axis for the block, one for the layer $t=0,\dots,L$ and two for the position of the check on the code.
Layer $t<L$ holds the syndrome round after gate layer $t$, and layer $L$ holds the final data measurement.
Each block sits in its own orientation, so the check type of a cell follows the H gates of its block.
Each site carries its detection bit, its check type and flags for the layers where its block is prepared and measured.

\paragraph{Trunk.}
The trunk turns the input grid into features and follows our previous convolutional decoders~\cite{codecoder}.
A stem applies a three-dimensional convolution of size $3\times5\times5$ over layer and position, with separate weights for X and Z checks.
Seven residual blocks follow, each with a bottleneck of width 32 and a $3\times3\times3$ convolution, at width 128.

\paragraph{Exchange between partner sites.}
A transversal CNOT pairs each data qubit of one block with the qubit at the same position of the other block, or with its rotated partner when the two blocks have different orientations.
To build in this pairing, the exchange lets every site read the features of its partner.
After the stem and after every residual block, a site on a layer next to a CNOT receives a linear map of the features that the partner block holds at the partner cell on the layers before and after the gate.
A site next to an S receives the features of its partner in the same block, the check of the other type that the fold links to it.
The map depends on the role of the receiving block (control, target or S), on whether the receiving layer lies before or after the gate, and on the check type.
Hence X-type and Z-type information travels along every gate in both directions, and only between the sites that the gate connects.

\subsection{Insight 2: Per-layer read-out}
\label{sec:readout}
Recall from Insight 2 that a logical outcome propagates through Clifford gates exactly and linearly.
The read-out uses the linearity to predict the answer layer by layer and the exactness to give each layer its propagated operator.
For each reliable product $P$ the code offers $d$ parallel representatives, and representative $j$ has a flip bit $y_{P,j}$.
The flip $y_P$ of the product equals the flip of each representative up to a sign that is a fixed parity of the detection events $D$,
\begin{equation}
y_P=y_{P,j}\oplus g_{P,j}(D).
\end{equation}
The read-out assigns every error to the layer of its first detection event.
Let $u_{P,t,j}$ be the flip of representative $j$ at the end of the circuit that the errors of layer $t$ cause.
Because the flips add up modulo 2, the flip of the representative is the parity of the layer parts,
\begin{equation}
y_{P,j}=\bigoplus_t u_{P,t,j}.
\end{equation}
The head of layer $t$ predicts one logit $l_{P,t,j}$ for $u_{P,t,j}$, and Sec.~\ref{sec:training} gives each $u_{P,t,j}$ an exact label.

Which errors at layer $t$ flip $P$ is known exactly.
Propagating $P$ backwards through the gates over GF(2) gives, for every block, the part of $P$ at layer $t$, an X, Z or Y operator or nothing, and we call this set of parts $r_P(t)$.
Each head receives $r_P(t)$ before it predicts, so a layer's prediction already refers to the final outcome and the network does not propagate soft quantities through the gates.
Each head is a set of line queries in the style of our previous decoder~\cite{codecoder}.
Its query adds one learned vector per part type for every block that carries such a part, and it pools the features of the sites on layer $t$ with attention weights that depend on the site's distance to the nearest line of its block's part.
The pooling uses four attention heads.

A soft XOR over the layers follows the heads.
It treats the layer parts as independent given the detection events, so the logit $z_{P,j}$ of the flip $y_{P,j}$ satisfies
\begin{equation}
\tanh(-z_{P,j}/2)=\prod_t\tanh(-l_{P,t,j}/2).
\end{equation}
We compute the product in log space in 32-bit floats.
The score of product $P$ is $c_P=\sum_j w_j(-1)^{g_{P,j}}z_{P,j}+b$ with learned weights $w_j$ and a bias $b$, and \sys flips $P$ if $c_P>0$.

\subsection{Insight 3: One network for every circuit}
\label{sec:shared}
Insight 3 rests on the fact that every circuit is built from the same few gates.
With the exchange and the read-out in place, the network needs to know only which gate acts on each block at each layer.
Each site therefore also carries the roles of its block in the gates next to its layer (control or target of a CNOT, S, or H).
The roles tell the network the gate sequence, so one network decodes different circuits without retraining.
All surface-code blocks share all trunk weights, and the heads of all layers share their weights.
The number of blocks enters only through sums over blocks, and the depth only through the number of terms of the soft XOR, so no parameter of the read-out is tied to $K$ or $L$.
With the read-out, the networks have 911\,877, 912\,199, 912\,521 and 912\,843 parameters at $d=3$, 5, 7 and 9, whatever the number of blocks and layers.

\subsection{Training}
\label{sec:training}
We obtain the exact per-layer labels from the circuit's detector error model.
Each error mechanism is assigned to the layer of its first detection event, and its effect on every representative is appended to the model as an extra observable of that layer.
One call of the sampler then returns the detection events, the product flips $y_P$, the flips $y_{P,j}$ and the layer parts $u_{P,t,j}$, and these satisfy the two relations above on every shot.
The loss of one shot is
\begin{equation}
\begin{aligned}
\ell=\sum_P\Big[&\mathrm{BCE}(c_P,y_P)+\frac1d\sum_j\mathrm{BCE}(z_{P,j},y_{P,j})\\
&+\frac1{d(L+1)}\sum_{j,t}\mathrm{BCE}(l_{P,t,j},u_{P,t,j})\Big],
\end{aligned}
\end{equation}
where $\mathrm{BCE}(l,y)$ is the binary cross-entropy of the logit $l$ against the bit $y$, and the loss is averaged over the batch.

Every distance uses the same recipe.
It starts from a network that had already trained on random circuits with two to four blocks under the same noise model (Appendix~\ref{app:train}).
The start at $d=3$ has three residual blocks, and we grow it to seven by adding four residual blocks whose exchange and output start at zero, so the grown network computes exactly the outputs of its start on the first training batches.
All four networks then train on $5\times10^7$ samples in batches of 1\,024.
Half of the batches come from the circuit family of Benchmark II (four blocks, 2 to 20 layers), and the other half from random circuits with two to four blocks and 4, 6 or 8 layers.
The noise strength is $p=0.3\,\%$ and rises linearly to $0.6\,\%$ and back over the last 20\,\% of the samples.
We use AdamW with a peak learning rate of $3\times10^{-4}$, a warm-up of 500 steps, a cosine decay to 2\,\% and an exponential moving average of the weights with rate 0.999, and the network is the final averaged weights, without a choice of checkpoint.

\subsection{Composition with \tess}
\label{sec:composition}
The network decodes every shot.
Its confidence on a shot is the smallest $|c_P|$ over the shot's products.
For a routed fraction $f$, each depth and noise strength, with the block counts pooled, has a threshold $t_f$, the largest confidence among the $\lceil fn\rceil$ least confident of its $n$ development shots, and a test shot with confidence at most $t_f$ goes to \tess.
\tess runs its 21 trials on separate cores of a pool of $N$ cores and stops each trial after $cL$ seconds, where $c$ is a cap per layer.
The answer is the cheapest trial that finished within its cap without a low-confidence flag, with the lower order on a tie, and a shot without such a trial keeps the network's answer.

\paragraph{When the composition helps.}
The conditions of CoDecoder~\cite{codecoder} follow from counting.
Let $F_{\mathrm{net}}(S)$ and $F_{\mathrm{T}}(S)$ be the failures of the network and of \tess on a set $S$ of shots, and let $R$ hold the routed shots and $\bar R$ the kept ones.
The composition fails $F_{\mathrm{net}}(\bar R)+F_{\mathrm{T}}(R)$ times.
It therefore fails less often than the network exactly when $F_{\mathrm{T}}(R)<F_{\mathrm{net}}(R)$, that is, when \tess fails less often than the network on the routed shots.
It fails less often than \tess alone exactly when $F_{\mathrm{net}}(\bar R)<F_{\mathrm{T}}(\bar R)$, when the network fails less often on the kept shots.
A good confidence score routes the shots on which the network is likely wrong and keeps those on which it is right.

\paragraph{Operating point.}
The candidates are the routed fractions $f\in\{0,0.5,1,2,5,10,20\,\%\}$, where $f=0$ is the network alone, and the caps per layer $c\in\{0.001,\allowbreak 0.002,\allowbreak 0.005,\allowbreak 0.01,\allowbreak 0.02,\allowbreak 0.05,\allowbreak 0.1\,\mathrm{s},\allowbreak \infty\}$.
For each $f$, benchmark, distance and hardware platform we fix the development-best cap, the $c$ with the lowest failure rate on development shots among those whose time per layer in the simulated paced stream below is less than the Lukin group's time per gate at that distance.
The failure rate of a candidate is the equal-weight geometric mean of its rates at the points of the benchmark, with 0.5 failures in place of a zero count, compared unrounded, and equal means go to the larger cap.
On the Grace platform at $d=3$ in Benchmark~I, every cap at $f=10\,\%$ gives 30\,057 or 30\,058 development failures, and the cap of 0.001\,s has the lowest mean log rate, $-5.4054146$ against $-5.4054106$ without a cap.
Every $f$ with its development-best cap was frozen before the fresh test set of Sec.~\ref{sec:setup} existed, and Appendix~\ref{app:speed} reports all of them on that test.
The operating point is the smallest $f>0$ whose development failures lie within 1\,\% of the fewest development failures of any $f$ on that platform, so it reads development data only.
We fixed this rule after both test sets had shown that the accuracy is flat beyond a few percent of routed shots.
On the fresh test on the Grace platform, $f=20\,\%$ fails at most 2\,\% less often than the operating point, at 1.3 to 4.1 times its time per layer.
On the Spark platform the time budget forces short caps at large $f$, and $f=20\,\%$ fails more often than the operating point at $d=7$ and 9 in Benchmark~I and at $d=5$ and 7 in Benchmark~II.
At $d=7$ in Benchmark~II it fails 3\,605 times against 1\,209, at 159.9 against 193.8 microseconds per layer (Sec.~\ref{sec:rq3}), and that operating point runs above the time budget of 185.2 microseconds per layer (Sec.~\ref{sec:rq4}).
The first test used an earlier rule, the candidate with the fewest development failures under the time budget (Appendix~\ref{app:first}).

\paragraph{GPU engine and paced stream.}
To keep the composition within the time budget, the network runs in an inference engine of fused GPU kernels written in Triton and launched as CUDA graphs.
In the paced stream, the GPU decodes the next batch of 256 shots while the CPU pool runs the trials of the routed shots, from one first-in-first-out queue with each shot's trials ordered from the largest beam down.
The network's next batch starts when the GPU is free and the first trial of the previous batch has started.
The time per layer is the wall time from the first dispatch to the last answer, divided by the number of shots and by $L$, and averaged with equal weight over the points of a benchmark.
We simulate this stream from measured network times and recorded trial times (Sec.~\ref{sec:rq4}).

%% file: figs/fig_network.tex
\begin{tikzpicture}[figbase,
  card/.style={draw=fline,fill=white,fthin},
  ev/.style={circle,fill=black,inner sep=0pt,minimum size=1.1mm},
  learned/.style={fbox,fill=fnetbg},
  exch/.style={<->,fheavy,draw=fnet},
  key/.style={draw=fline,fthin,rounded corners=1pt}]
\def\ya{6.3}\def\yk{4.4}
\foreach \y in {\ya,\yk}{
  \foreach \k in {3,2,1,0}{\draw[card] (0.9+0.12*\k,\y-0.3+0.09*\k) rectangle +(1.0,0.6);}
  \foreach \px/\py in {0.22/0.2,0.62/0.45,0.45/0.68,0.8/0.3}{\node[ev] at (0.9+\px,\y-0.3+\py*0.6+0.06) {};}
  \draw[flow] (2.3,\y)--(3.35,\y);
}
\node[anchor=east] at (0.8,\ya) {block 1};
\node[anchor=east] at (0.8,\yk) {block $K$};
\node at (1.4,5.4) {$\vdots$};
\node[anchor=north,align=center] at (1.05,3.9) {detection events\\on $L{+}1$ layers};
\draw[draw=fnet!50,fthin,rounded corners=2pt] (2.75,3.8) rectangle (12.75,7.55);
\node[anchor=north west] at (2.85,7.5) {trunk, shared by all blocks (Insight 3)};
\foreach \y/\id in {\ya/a,\yk/b}{
  \node[learned,minimum width=1.7cm] (s\id) at (4.2,\y) {stem};
  \node[learned,minimum width=1.7cm] (r1\id) at (6.5,\y) {residual 1};
  \node (r2\id) at (8.8,\y) {$\cdots$};
  \node[learned,minimum width=1.7cm] (r7\id) at (11.1,\y) {residual 7};
  \draw[flow] (s\id)--(r1\id);
  \draw[flow] (r1\id)--(r2\id);
  \draw[flow] (r2\id)--(r7\id);
}
\foreach \x in {5.35,7.65,9.95,12.25}{\draw[exch] (\x,\ya-0.33)--(\x,\yk+0.33);}
\node[align=center] at (6.5,5.35) {exchange at\\CNOT and S\\(Insight 1)};
\draw[fmid] (r7a.east)--(13.2,\ya)--(13.2,3.35)--(4.95,3.35);
\draw[fmid] (r7b.east)--(13.2,\yk);
\draw[flow] (4.95,3.35)--(4.95,2.95);
\node[anchor=south] at (10.4,3.35) {features};
\node[fbox,minimum width=2.0cm] (tab) at (0.75,1.6) {gate table\\(exact, GF(2))};
\draw[learned] (2.75,0.3) rectangle (7.15,2.95);
\node[anchor=north] at (4.95,2.9) {per-layer read-out (Insight 2)};
\foreach \t in {0,...,6}{\foreach \j in {0,1,2}{
  \pgfmathsetmacro{\sh}{18+9*mod(\t*3+\j*5,7)}
  \filldraw[fthin,draw=fnet!55,fill=fnet!\sh] (3.4+0.5*\t,1.17+0.36*\j) rectangle +(0.44,0.3);
}}
\node[anchor=east] at (3.35,1.67) {$j$};
\node[anchor=north] at (5.07,1.1) {layers $t=0,\dots,L$};
\node[anchor=north] at (4.95,0.75) {one logit $l_{P,t,j}$ per cell};
\draw[flow] (tab.east)--(2.75,1.6) node[midway,above=1pt] {$r_P(t)$};
\node[fbox,minimum width=1.95cm] (xor) at (9.575,1.6) {soft XOR\\over layers $t$};
\draw[flow] (7.15,1.6)--(xor.west);
\node[fbox,minimum width=1.5cm] (sc) at (12.75,1.6) {score $c_P$};
\draw[flow] (xor.east)--(sc.west) node[midway,above=1pt] {$z_{P,j}$};
\draw[flow] (12.75,0.45)--(sc.south);
\node[anchor=north] at (12.75,0.45) {signs $g_{P,j}$};
\node[fbox,minimum width=2.15cm] (out) at (16.025,1.6) {flip product $P$\\if $c_P>0$};
\draw[flow] (sc.east)--(out.west);
\node[fbox,fill=fcompbg,minimum width=2.8cm] (tess) at (15.7,5.35) {\tess: 21 trials,\\one per CPU core};
\draw[flow] (out.north)--(out.north|-tess.south) node[midway,left=2pt,align=right] {if $\min_P|c_P|$\\${}\le t_f$};
\node[anchor=east] (k2) at (17.1,7.17) {exact};
\draw[key,fill=white] ($(k2.west)+(-0.38,-0.12)$) rectangle +(0.34,0.24);
\node[anchor=east] (k1) at ($(k2.west)+(-0.75,0)$) {learned};
\draw[key,fill=fnetbg] ($(k1.west)+(-0.38,-0.12)$) rectangle +(0.34,0.24);
\end{tikzpicture}

%% file: sections/eval.tex
\section{Results}
\label{sec:eval}
With the design in place, we test \sys on fresh shots of the two benchmarks of Sec.~\ref{sec:intro}.
We first describe the setup.
Then we present the accuracy on Benchmark~I, the comparison with the Lukin group's network on Benchmark~II, the routing to \tess, the speed, and the contribution of each insight.

\subsection{Setup}
\label{sec:setup}
\paragraph{Benchmark I: general circuits.}
The circuits are those of Sec.~\ref{sec:background} with $K=2$, 3 and 4 blocks.
Each distance has 24 test points: depths $L=1$, 2, 4, 8, 12 and 16 at $p=0.3\,\%$ and $L=8$ at $p=0.1\,\%$ and $0.2\,\%$, for each $K$.
Every point holds 64 random circuits, and each circuit draws the probability that a pair of blocks runs a CNOT in a layer from $\{0,1/3,2/3,1\}$.
A circuit has 100\,000 shots at $L=1$, falling to 2\,800 at $L=16$, and 16\,000 at the two lower noise strengths, so each distance has 43\,833\,600 test shots.
$L=12$ and 16 are untrained within this family, whose last training stage used $L=4$, 6 and 8, while the family of Benchmark~II, which also enters training, covers $L=2$ to 20 with four blocks only.

\paragraph{Benchmark II: the Lukin group's circuits.}
We rebuild the benchmark of Bonilla Ataides et al.~\cite{bonilla} from their Methods and their released compiler (Appendix~\ref{app:circuits}).
Four blocks start in $|+\rangle_L$.
A circuit of depth $D=1,\dots,10$ repeats $D$ times a layer of single-qubit gates drawn from $\{I,X,Y,Z,H\}$, with physical Pauli gates, and a layer of two CNOTs on a random pairing of the blocks, so it has $L=2D$ layers with one syndrome round each.
A shot fails when the decoder gets the parity of one target product wrong, a stabilizer of the final logical state.
Each depth has 20 fixed circuits, and fresh noise samples give 299\,400, 1\,464\,000 and 16\,456\,000 test shots at $d=3$, 5 and 7 at $p=0.3\,\%$.

\paragraph{Why the two benchmarks differ.}
Benchmark~II rebuilds the circuits of the Lukin group, four blocks whose shot fails when the parity of one target product is wrong, so that we can compare with their published rates.
Benchmark~I uses general settings, with two to four blocks, S gates, preparation and measurement in both bases, a failure whenever any reliable product of the circuit is wrong, and three noise strengths.
Both families enter the training of every network, and the same network per distance decodes both.

\paragraph{Test protocol.}
All test shots are fresh, drawn from seeds that no training or development set used.
We drew two test sets of the same design and size.
The first scored the operating points of an earlier rule, frozen before it was read (Appendix~\ref{app:first}).
The second, the fresh test set, scores every routed fraction with its development-best cap, all frozen before it existed (Sec.~\ref{sec:design}).
It draws new circuits and noise samples for Benchmark~I and new noise samples on the same circuits for Benchmark~II.
The results in the main text are those of the fresh test at the operating points, and Appendix~\ref{app:speed} gives every other row.
LoMatching and Cain et al.\ decode every test shot of Benchmark~I, and \bplom the first 1\,000 shots of every circuit, 1\,536\,000 shots per distance, on which alone we compare against it.
In Benchmark~II the exact MLE decodes a fixed prefix of the shots of every circuit, 280\,000 shots at $d=3$, all shots at $d=5$ and 1\,611\,720 shots at depths 1 to 8 at $d=7$.
We leave out the 3 prefix shots at each of $d=5$ and 7 for which the solver did not certify its answer.

\paragraph{Metrics.}
In Benchmark~I we compare two decoders by the ratio of their failures on the same shots.
At each point the 95\,\% interval comes from 10\,000 bootstrap draws that redraw the 64 circuits with their paired counts, and we pool points by the geometric mean of their ratios, over the 18 points at $p=0.3\,\%$ and over the six points at the lower noise strengths (Appendix~\ref{app:stats}).
Benchmark~II uses two measures.
Measure (a) divides our failures on every test shot by the failures that the Lukin group's network would make on the same shots, computed from the failure rates per depth in their published figure.
Measure (b) divides our failure rate relative to the exact MLE on the certified prefix shots, with every depth weighted equally, by the same quantity for their network and their MLE, so it compares each network with an exact MLE on its own circuits, and a difference between our rebuild and theirs cancels when it changes a network and the MLE by the same factor.
Our exact MLE fails 1.01, 0.83 and 0.58 times as often as their published MLE at $d=3$, 5 and 7 (Appendix~\ref{app:circuits}), so the two measures differ most at $d=7$.
Below 1, both measures favour \sys, and the intervals include the error bars of their figure.

\paragraph{Hardware platforms.}
We give speed on two platforms.
The Grace platform runs the network on an NVIDIA RTX PRO 6000 Blackwell Server Edition GPU and the trials on $N=64$ cores of an NVIDIA Grace CPU.
The Spark platform is an NVIDIA DGX Spark, whose GB10 chip holds a GPU and $N=20$ CPU cores.
We measure the network on each GPU at batch 256 and the trials on AMD EPYC cores, whose times we convert to each platform's CPU (Appendix~\ref{app:speed}).
Each platform has its own operating points, which route the same shots except at $d=3$ in Benchmark~I and $d=7$ in Benchmark~II.

\begin{table*}[t]
\caption{Benchmark I on the fresh test set at the operating points: failures of the network alone and of the composition with \tess, divided by those of three decoders on the same shots, pooled over $K=2$, 3 and 4 and the depths 1, 2, 4, 8, 12 and 16 at $p=0.3\,\%$ (18 points) and over the depth-8 points at $p=0.1$ and $0.2\,\%$ (6 points), with 95\,\% intervals below. The column $f$ gives the routed fraction. The composition is the Grace platform's. At $d=3$ the Spark platform's operating point routes 2\,\% and has its own rows, and at $d=5$, 7 and 9 both platforms share the operating point and its values.}
\label{tab:mr1}
\centering\footnotesize\setlength{\tabcolsep}{3.4pt}
\input{tables/tab_mr1}
\end{table*}

\subsection{Accuracy on Benchmark I}
\label{sec:rq1}
\begin{figure}[t]
\centering
\includegraphics[width=\columnwidth]{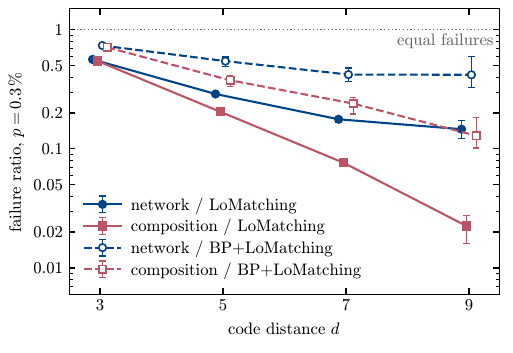}
\caption{Benchmark I at $p=0.3\,\%$ on the fresh test set at the operating points: failures of the network alone and of the composition divided by those of LoMatching and of \bplom on the same shots, pooled over $K=2$, 3, 4 and six depths, with 95\,\% intervals. One network per distance.}
\label{fig:mr1}
\end{figure}

We compare the network alone and the composition with LoMatching, Cain et al.\ and \bplom on the fresh test shots of Benchmark~I.
Table~\ref{tab:mr1} lists the pooled ratios, and Fig.~\ref{fig:mr1} plots them against the distance.
At $p=0.3\,\%$ the network alone fails 0.563, 0.289, 0.177 and 0.146 times as often as LoMatching at $d=3$, 5, 7 and 9, and the composition 0.547, 0.205, 0.0765 and 0.0225 times as often.
Every interval of Table~\ref{tab:mr1} lies below 1.
Cain et al.\ fails 0.4 to 0.9\,\% more often than LoMatching in total, so the ratios against it are slightly lower.
Against \bplom on its 1\,152\,000 matched shots per distance at $p=0.3\,\%$, the network alone fails 0.737, 0.546, 0.421 and 0.419 times as often, and the composition 0.711, 0.376, 0.240 and 0.129 times as often.

\begin{table*}[t]
\caption{Benchmark II on the fresh test set at the operating points, at $p=0.3\,\%$. Measure (a) is our failures on every test shot over those expected from their network's published rates on the same shots, and measure (b) divides our failure rate relative to the exact MLE by theirs, with equal-depth sums of rates on the certified prefix shots. Both favour \sys below 1. The last two columns divide equal-weight sums of per-depth failure rates on all test shots, ours by those of LoMatching and of the network alone. Grace and Spark name the platform of a composition, given with its routed fraction, and both platforms share one composition at $d=3$ and 5. 95\,\% intervals below the values.}
\label{tab:mr2}
\centering\footnotesize\setlength{\tabcolsep}{4pt}
\input{tables/tab_mr2}
\end{table*}

Against LoMatching and Cain et al., the pooled gains at the lower noise strengths are larger.
There the composition fails 0.405, 0.115, 0.0206 and 0.0128 times as often as LoMatching at $d=3$, 5, 7 and 9.
At $d=9$ the counts are small, 5 failures of the composition against 903 of LoMatching, and the first test gives 0.0456 there (Appendix~\ref{app:first}).
The ratios differ little between block counts.
At $d=7$, for example, the composition fails 0.0788, 0.0829 and 0.0685 times as often as LoMatching at $K=2$, 3 and 4.
The two systems scale differently with the distance.
The network alone gains less from $d=7$ to 9, from 0.177 to 0.146, than from $d=5$ to 7, from 0.289 to 0.177.
The composition keeps gaining, and its ratio to the network falls from 0.971 at $d=3$ to 0.154 at $d=9$.
Section~\ref{sec:rq3} explains this with the routed shots.
At $p=0.3\,\%$, \sys is thus more accurate than all three practical decoders at every distance, and \tess on its least confident shots multiplies the gain at the larger distances.
The first test scored the operating points of an earlier rule, and at $p=0.3\,\%$ the fresh ratios for these points lie inside its 95\,\% intervals, except one that misses by 0.0001 (Appendix~\ref{app:first}).

\subsection{Comparison with the Lukin group's network on Benchmark II}
\label{sec:rq2}
We compare the network alone and the composition with the Lukin group's network on Benchmark~II (Table~\ref{tab:mr2}).
At $d=3$ the network alone fails 0.875 [0.804, 0.952] times as often as their network by measure (a), with a sensitivity interval of [0.790, 0.969] when the variation between circuits is added, and 0.871 [0.816, 0.929] times by measure (b).
The composition routes 0.5\,\% of the shots with a cap of 0.002\,s per layer on both platforms and reaches 0.866 [0.796, 0.943] and 0.868 [0.817, 0.923], and it fails 1.030 [0.984, 1.077] times as often as the exact MLE on the prefix shots.
With so few routed shots it fails 0.992 [0.981, 1.003] times as often as the network alone.
At $d=5$ both intervals of the network alone include 1, with 0.956 [0.838, 1.091] by (a) and 1.051 [0.920, 1.201] by (b).
The composition routes 5\,\% of the shots with a cap of 0.02\,s per layer and reaches 0.705 [0.617, 0.807] and 0.751 [0.672, 0.838].
On the certified prefix shots it fails 1\,520 times, against 1\,517 times for the exact MLE and 2\,060 times for the network alone.
At $d=7$ both intervals of the network alone include 1, with 0.743 [0.482, 1.146] by (a), an interval that the error bars of the published rates widen, and 1.459 [0.854, 2.495] by (b).
On the certified prefix shots the network alone fails 208 times against 92 for the exact MLE, counted without the equal weighting of depths that (b) uses, and on the first test its (b) lay above 1, at 1.772 [1.026, 3.063] (Appendix~\ref{app:first}).
The composition removes this gap.
On the Grace platform it routes 10\,\% of the shots without a cap and reaches 0.299 [0.193, 0.464] by (a) and 0.600 [0.408, 0.881] by (b), and on the Spark platform it routes 2\,\% without a cap and reaches 0.305 [0.197, 0.473] and 0.607 [0.406, 0.906].
On the certified prefix shots it fails 92 and 93 times, against 92 for the exact MLE.
In this setting the network alone fails 0.560, 0.259 and 0.170 times as often as LoMatching at $d=3$, 5 and 7, and the composition 0.556, 0.185 and 0.063 times.
For the composition, then, the intervals of both measures, and the sensitivity intervals of (a), lie below 1 at every distance, and its failure rate is statistically compatible with the exact MLE on the certified prefixes.
For the network alone, both measures lie below 1 at $d=3$, and both include 1 at $d=5$ and 7.

\subsection{Routing to \tess}
\label{sec:rq3}
We apply the conditions of Sec.~\ref{sec:design} to the kept and the routed shots (Table~\ref{tab:routing}).
In Benchmark~I the operating points route $f=1$, 2, 2 and 2\,\% of the development shots at $d=3$, 5, 7 and 9 on the Grace platform and 2\,\% at every distance on the Spark platform, and the thresholds route 1.02, 2.03, 2.02 and 2.03\,\% of the test shots at $p=0.3\,\%$ on the Grace platform.

\begin{table*}[t]
\caption{Where the failures go at the operating points on the fresh test set. Benchmark I counts its 18 points at $p=0.3\,\%$, and Benchmark II every test shot. The network's failures split into those on the kept and on the routed shots, with the routed share of all its failures. On the routed shots the composition takes the answer of \tess, which corrects some of the network's failures and introduces new errors, so its failures there equal the network's minus the corrections plus the introduced errors. Ratios divide the composition's failures by the network's, on the routed shots and on all shots. Both marks a composition that is the same on both platforms.}
\label{tab:routing}
\centering\footnotesize\setlength{\tabcolsep}{3.2pt}
\input{tables/tab_routing}
\end{table*}

\paragraph{When the composition beats the network.}
The \mbox{composition} changes answers only on the routed shots.
There \tess corrects $C$ of the network's failures and introduces $I$ new errors, so $F_{\mathrm{T}}(R)=F_{\mathrm{net}}(R)-C+I$, and the first condition holds exactly when $C>I$.
The gain then depends on two things: the share of the network's failures that the confidence score sends to \tess, and how often \tess fails on the routed shots compared with the network.
In Benchmark~I at $p=0.3\,\%$ the routed 1 to 2\,\% of the shots hold 54.0, 91.7, 96.6 and 98.7\,\% of the network's failures at $d=3$, 5, 7 and 9 on the Grace platform.
The share approaches its ceiling from $d=5$ on, and what grows with the distance is the advantage of \tess on the routed shots, where the composition fails 0.946, 0.674, 0.383 and 0.137 times as often as the network.
At $d=3$ \tess introduces almost as many errors as it corrects, 15\,132 against 18\,198, so the composition gains only 3\,\% overall.
At $d=9$ it corrects 600 failures and introduces 31, and the composition fails 99 times against 668 for the network.
The remaining failures are mostly routed shots on which \tess fails too, 947 of 1\,033 at $d=7$ and 90 of 99 at $d=9$.
Benchmark~II shows the same pattern, with ratios on the routed shots of 0.938, 0.714 and 0.395 at $d=3$, 5 and 7 on the Grace platform, although at $d=3$ the routed 0.48\,\% of the shots hold only 15.9\,\% of the network's failures.
The first condition holds in every row of Table~\ref{tab:routing}.

\paragraph{When the composition beats \tess alone.}
The second condition asks the network to fail less often than \tess on the kept shots.
In Benchmark~II the exact MLE decodes the prefix shots, and long-beam \tess decides as the exact MLE on all but one of 960\,000 shots of these circuits (Appendix~\ref{app:circuits}).
There the failure rate of the composition is statistically compatible with the exact MLE on the certified prefixes, with ratios of 1.030, 1.008 and 0.992 at $d=3$, 5 and 7.

\paragraph{How much to route.}
Every routed fraction with its development-best cap is a frozen system, and Appendix~\ref{app:speed} scores each one on the fresh test (Tables~\ref{tab:rt1} and~\ref{tab:rt2}).
Most of the gain comes from the least confident few percent of the shots.
In Benchmark~I on the Grace platform the composition fails 0.980, 0.757, 0.458 and 0.171 times as often as the network at $f=0.5\,\%$, and from 2 to 20\,\% the ratio changes by less than 0.01 at every distance, while the time per layer grows from 1.2 to 1.8, 6.8 to 24.3, 27.1 to 96.4 and 66.4 to 270.0 microseconds.
At $d=9$ the failures saturate at 115, 104 and 102 for $f=0.5$, 2 and 20\,\%, at 26.8, 66.4 and 270.0 microseconds per layer.
On the Grace platform $f=20\,\%$ fails at most 2\,\% less often than the operating point on both benchmarks, at 1.3 to 4.1 times its time per layer.
On the Spark platform the time budget forces short caps at large $f$, and $f=20\,\%$ fails more often than the operating point at $d=7$ and 9 in Benchmark~I and at $d=5$ and 7 in Benchmark~II.
The trade-off is not monotone.
On the Spark platform at $d=7$ in Benchmark~II, a cap of 0.001\,s per layer at $f=20\,\%$ makes the composition fail 1.173 [1.112, 1.237] times as often as the network alone, 3\,605 times against 1\,209 at the operating point, in less time, 159.9 against 193.8 microseconds per layer.
The first test's rule took the fewest development failures under the time budget, which on Benchmark~II put three operating points near their budgets, where the curve is already flat, and on the fresh test these three run above the budget (Appendix~\ref{app:first}).

\begin{table*}[t]
\caption{Time per layer in microseconds at batch 256. The Lukin group's network ran on an H100 GPU. The network alone was measured on the GPU of each platform, averaged over the points of each benchmark. The composition gives the paced stream at the operating points, simulated on the fresh test shots from measured network and trial times with converted CPU times, on the Grace platform (RTX PRO 6000 and 64 Grace cores) and the Spark platform (DGX Spark, 20 cores), with the routed fraction and the cap per layer of its operating point below. Appendix~\ref{app:speed} gives the live stream.}
\label{tab:speed}
\centering\footnotesize\setlength{\tabcolsep}{3.6pt}
\input{tables/tab_speed}
\end{table*}

\subsection{Speed}
\label{sec:rq4}
\begin{figure}[t]
\centering
\includegraphics[width=\columnwidth]{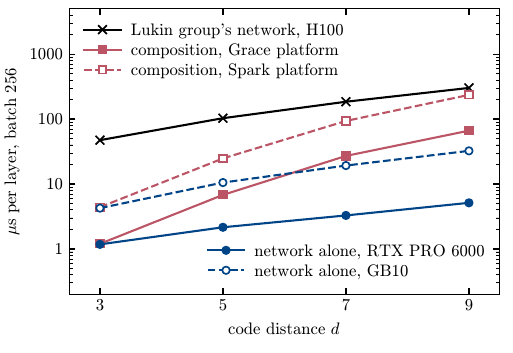}
\caption{Time per layer at batch 256 against the distance. The Lukin group's network ran on an H100 GPU. The composition is Benchmark I's operating point in the paced stream, simulated on the fresh test shots from measured network and trial times, and the network alone was measured on each platform's GPU.}
\label{fig:speed}
\end{figure}
We compare the time per layer of \sys at batch 256 with the Lukin group's time per gate, one layer with one syndrome round, at the same batch size, both amortized over the batch (Table~\ref{tab:speed} and Fig.~\ref{fig:speed}).
The network alone takes 1.18, 2.16, 3.28 and 5.14 microseconds per layer at $d=3$, 5, 7 and 9 on the RTX PRO 6000 and 4.26, 10.5, 19.3 and 32.4 on the GB10, which is 9.3 to 59 times less than their 47.4, 103.3, 185.2 and 302.2 microseconds on the H100.
The composition adds the trials of the routed shots.
We simulate its paced stream from the network times measured on each GPU and the trial times recorded on the fresh test shots, converted to each platform's CPU by measured factors, where $d=3$ borrows the factor of $d=5$ and $d=9$ that of $d=7$ (Appendix~\ref{app:speed}).
At the operating points it takes 1.2, 6.8, 27.1 and 66.4 microseconds per layer on the Grace platform and 4.4, 24.8, 93.7 and 236.1 on the Spark platform in Benchmark~I, which is 4.5 to 40 and 1.3 to 11 times less than their time per gate.
In Benchmark~II the operating points take 1.2, 20.8 and 152.7 microseconds per layer on the Grace platform and 5.1, 72.6 and 193.8 on the Spark platform at $d=3$, 5 and 7.
The Spark platform's point at $d=7$ runs above their 185.2 microseconds on the fresh test shots, after 176.5 on the development shots that fixed it.
We also ran the paced stream live on development shots, for every routed fraction at its development-best cap, on one RTX PRO 6000 with 32 cores of an AMD EPYC 9555, the most CPU cores per GPU that the cluster grants, with the Grace platform's caps, and on a DGX Spark with its 20 cores with the Spark platform's caps (Appendix~\ref{app:speed}).
With the live trial times converted to the platform's CPU through paired trials, the live stream on the EPYC cores took 0.87 to 0.99 times the time that the simulation from the recorded trial times predicts for the same batches at $N=32$, with a median of 0.94 over the 42 configurations that route shots, so for these cores the simulation is faithful and slightly conservative.
These are not the 64 Grace cores of the Grace platform.
On the DGX Spark the live stream agrees with the simulation only approximately, at 0.86 to 1.14 times the simulated time with a median of 0.99, and above it in 14 of the 42 configurations.
Measured live, the operating points of Benchmark~I take 2.4, 13.4, 52.8 and 121.7 microseconds per layer on the 32 EPYC cores and 7.1, 25.3, 103.8 and 257.4 on the DGX Spark, less than the Lukin group's time per gate at every distance.
Those of Benchmark~II take 1.7, 23.6 and 218.2 on the EPYC cores and 5.3, 49.9 and 199.7 on the DGX Spark at $d=3$, 5 and 7, above their time per gate at $d=7$ on both.
Matching decoders are faster still: on one busy core of an AMD EPYC 7713, LoMatching takes 1.4 microseconds per shot at $d=9$, $K=4$ and $L=1$ and 28.9 at $L=16$, while it fails 13 and 44 times as often as the composition at $d=7$ and 9 at $p=0.3\,\%$.
On the fresh test shots, the composition of Benchmark~I therefore decodes a layer in less time than the Lukin group's network at every distance on both platforms, and so does its live stream on both hosts.

\subsection{Contribution of each insight}
\label{sec:rq5}
To test each insight, we remove or replace the mechanism that realizes it, and we also shrink the trunk.
We train these variants of the full network at $d=5$, each from scratch with $10^8$ samples, one seed and one recipe for all variants (Appendix~\ref{app:train}), and read them on the development shots of both benchmarks (Table~\ref{tab:abl}).
Each entry is the equal-weight geometric mean of paired per-point failure ratios of a variant against the full network on the same shots.
The intervals come from the development shots and do not include the variation between training seeds, so every conclusion of this section holds for $d=5$, one seed and development data.

\begin{table*}[t]
\caption{Variants of the network at $d=5$, each trained from scratch with $10^8$ samples. Each entry is the equal-weight geometric mean of paired per-point failure ratios of the variant against the full network on the same development shots, with the 95\,\% interval in brackets. Benchmark I: its circuit family per $K$ and pooled over $K=2$, 3, 4 at $p=0.3\,\%$, and the depth-8 points at $p=0.1$ and $0.2\,\%$. Benchmark II: the Lukin group's circuits, depths 1 to 10. The two exchange variants test Insight 1, the pooled read-out Insight 2 and the specialists Insight 3. The last row is the main network of this paper.}
\label{tab:abl}
\centering\footnotesize\setlength{\tabcolsep}{3pt}
\input{tables/tab_abl}
\end{table*}

\paragraph{Insight 1: exchange between blocks.}
Without the exchange, every exchange step reads the site's own block, with the same parameters, and the read-out reads each block alone and joins the blocks only by the parity rule.
This variant fails 2.00 [1.91, 2.08] times as often as the full network on Benchmark~I and 3.26 [3.14, 3.37] times on Benchmark~II.
A late-fusion variant keeps the blocks apart in the trunk and lets the read-out attend to all blocks.
It fails 1.60 [1.54, 1.65] and 2.05 [2.01, 2.10] times as often, so the read-out recovers part of what the trunk lacks, and the exchange at every gate supplies the rest.

\paragraph{Insight 2: per-layer read-out.}
A read-out with one query per product that pools the features of all sites and layers, without representatives, per-layer logits and the soft XOR, and trained on the final flips alone, fails 28.8 [24.4, 31.6] times as often on Benchmark~I and 12.1 [10.4, 13.4] times on Benchmark~II.
It still sees which blocks carry a part of the product at each layer.
The variant changes the read-out and the loss together, so its ratio measures that combined change and singles out no part of it.
On the circuits of Benchmark~II the ratio is 4.35 at $D=1$ and 39.9 at $D=10$, with nonmonotonic intermediate values.

\paragraph{Insight 3: one network for all block counts.}
A specialist trained only on $K=2$, 3 or 4 fails 1.046 [0.992, 1.106], 0.977 [0.936, 1.021] and 0.976 [0.942, 1.010] times as often as the shared network on its own block count, and the $K=4$ specialist fails 1.001 [0.983, 1.019] times as often on the circuits of Benchmark~II.
At the lower noise strengths the $K=4$ specialist fails more often, 1.315 [1.055, 1.826] times.
At $p=0.3\,\%$ all three intervals include 1, so a specialist per block count brings no measurable gain, and the specialists' training mixes differ from that of the shared network (Appendix~\ref{app:train}).

\paragraph{Trunk size and training history.}
A trunk with three residual blocks instead of seven has 523\,719 parameters and fails 1.23 [1.20, 1.26] and 1.24 [1.22, 1.27] times as often.
The main network of this paper, which continued its start for $5\times10^7$ samples, fails 0.943 [0.921, 0.967] and 1.038 [1.020, 1.058] times as often as the full variant trained from scratch for $10^8$ samples.

Overall, at $d=5$, with one seed and on development shots, replacing the per-layer heads and the auxiliary losses together (Insight 2) produces the largest observed penalty and removing the exchange (Insight 1) the second largest, the larger trunk adds a smaller gain, and specialists per block count bring no measurable gain over the one shared network (Insight 3).

%% file: tables/tab_mr1.tex
\begin{tabular}{@{}cccccccccc@{}}
\toprule
& & & \multicolumn{2}{c}{against LoMatching} & \multicolumn{2}{c}{against Cain et al.} & \multicolumn{2}{c}{against \bplom} & composition \\
\cmidrule(lr){4-5}\cmidrule(lr){6-7}\cmidrule(lr){8-9}
$d$ & $f$ & $p$ (\%) & network & composition & network & composition & network & composition & / network \\
\midrule
3 & 1\,\% & 0.3 & 0.5634 & 0.5471 & 0.5603 & 0.5442 & 0.7371 & 0.7114 & 0.9711 \\
 &  &  & {\tiny[0.5578, 0.5696]} & {\tiny[0.5412, 0.5536]} & {\tiny[0.5547, 0.5668]} & {\tiny[0.5382, 0.5509]} & {\tiny[0.7170, 0.7570]} & {\tiny[0.6914, 0.7311]} & {\tiny[0.9675, 0.9748]} \\[2pt]
 &  & 0.1, 0.2 & 0.4294 & 0.4049 & 0.4286 & 0.4041 & 0.6823 & 0.6514 & 0.9430 \\
 &  &  & {\tiny[0.4175, 0.4425]} & {\tiny[0.3927, 0.4182]} & {\tiny[0.4166, 0.4415]} & {\tiny[0.3920, 0.4173]} & {\tiny[0.6427, 0.7237]} & {\tiny[0.6148, 0.6886]} & {\tiny[0.9311, 0.9548]} \\
 & Spark: 2\,\% & 0.3 &  & 0.5429 &  & 0.5400 &  & 0.7059 & 0.9636 \\
 &  &  &  & {\tiny[0.5370, 0.5495]} &  & {\tiny[0.5340, 0.5467]} &  & {\tiny[0.6860, 0.7255]} & {\tiny[0.9597, 0.9677]} \\[2pt]
 &  & 0.1, 0.2 &  & 0.4049 &  & 0.4041 &  & 0.6536 & 0.9429 \\
 &  &  &  & {\tiny[0.3928, 0.4181]} &  & {\tiny[0.3920, 0.4172]} &  & {\tiny[0.6167, 0.6913]} & {\tiny[0.9310, 0.9547]} \\
\midrule
5 & 2\,\% & 0.3 & 0.2889 & 0.2045 & 0.2840 & 0.2011 & 0.5456 & 0.3763 & 0.7081 \\
 &  &  & {\tiny[0.2826, 0.2954]} & {\tiny[0.1996, 0.2099]} & {\tiny[0.2774, 0.2909]} & {\tiny[0.1960, 0.2065]} & {\tiny[0.4911, 0.5912]} & {\tiny[0.3342, 0.4082]} & {\tiny[0.6953, 0.7207]} \\[2pt]
 &  & 0.1, 0.2 & 0.1681 & 0.1154 & 0.1675 & 0.1149 & 0.3678 & 0.2547 & 0.6862 \\
 &  &  & {\tiny[0.1528, 0.1831]} & {\tiny[0.1036, 0.1274]} & {\tiny[0.1523, 0.1823]} & {\tiny[0.1031, 0.1268]} & {\tiny[0.2400, 0.4709]} & {\tiny[0.1603, 0.3378]} & {\tiny[0.6333, 0.7402]} \\
\midrule
7 & 2\,\% & 0.3 & 0.1772 & 0.0765 & 0.1717 & 0.0741 & 0.4207 & 0.2400 & 0.4317 \\
 &  &  & {\tiny[0.1677, 0.1868]} & {\tiny[0.0706, 0.0816]} & {\tiny[0.1620, 0.1817]} & {\tiny[0.0684, 0.0793]} & {\tiny[0.3659, 0.4771]} & {\tiny[0.1959, 0.2699]} & {\tiny[0.4025, 0.4576]} \\[2pt]
 &  & 0.1, 0.2 & 0.0707 & 0.0206 & 0.0703 & 0.0205 & 0.2670 & 0.2379 & 0.2720 \\
 &  &  & {\tiny[0.0429, 0.1003]} & {\tiny[0.0157, 0.0283]} & {\tiny[0.0426, 0.0998]} & {\tiny[0.0156, 0.0281]} & {\tiny[0.1746, 0.4943]} & {\tiny[0.1427, 0.4673]} & {\tiny[0.2009, 0.4428]} \\
\midrule
9 & 2\,\% & 0.3 & 0.1462 & 0.0225 & 0.1397 & 0.0215 & 0.4188 & 0.1288 & 0.1541 \\
 &  &  & {\tiny[0.1215, 0.1740]} & {\tiny[0.0161, 0.0276]} & {\tiny[0.1164, 0.1650]} & {\tiny[0.0154, 0.0263]} & {\tiny[0.3289, 0.5922]} & {\tiny[0.1015, 0.1835]} & {\tiny[0.1123, 0.1836]} \\[2pt]
 &  & 0.1, 0.2 & 0.0359 & 0.0128 & 0.0359 & 0.0128 & 0.2149 & 0.1684 & 0.3403 \\
 &  &  & {\tiny[0.0217, 0.1273]} & {\tiny[0.0081, 0.0457]} & {\tiny[0.0217, 0.1271]} & {\tiny[0.0080, 0.0455]} & {\tiny[0.1381, 0.9925]} & {\tiny[0.1167, 0.6934]} & {\tiny[0.2174, 0.5952]} \\
\bottomrule
\end{tabular}

%% file: tables/tab_mr2.tex
\begin{tabular}{@{}clccccccc@{}}
\toprule
& & \multicolumn{2}{c}{measure (a)} & \multicolumn{3}{c}{measure (b)} & & \\
\cmidrule(lr){3-4}\cmidrule(lr){5-7}
$d$ & system & failures / expected & ratio & ours / MLE & theirs / MLE & ratio & vs LoMatching & vs network \\
\midrule
3 & network alone & 2846 / 3254.0 & 0.875 & 1.032 & 1.185 & 0.871 & 0.560 & --- \\
 &  &  & {\tiny[0.804, 0.952]} & {\tiny[0.982, 1.085]} &  & {\tiny[0.816, 0.929]} & {\tiny[0.541, 0.581]} &  \\
 & composition, both, 0.5\,\% & 2818 / 3254.0 & 0.866 & 1.030 & 1.185 & 0.868 & 0.556 & 0.992 \\
 &  &  & {\tiny[0.796, 0.943]} & {\tiny[0.984, 1.077]} &  & {\tiny[0.817, 0.923]} & {\tiny[0.536, 0.576]} & {\tiny[0.981, 1.003]} \\
\midrule
5 & network alone & 2060 / 2155.1 & 0.956 & 1.411 & 1.342 & 1.051 & 0.259 & --- \\
 &  &  & {\tiny[0.838, 1.091]} & {\tiny[1.308, 1.522]} &  & {\tiny[0.920, 1.201]} & {\tiny[0.246, 0.272]} &  \\
 & composition, both, 5\,\% & 1520 / 2155.1 & 0.705 & 1.008 & 1.342 & 0.751 & 0.185 & 0.714 \\
 &  &  & {\tiny[0.617, 0.807]} & {\tiny[0.991, 1.025]} &  & {\tiny[0.672, 0.838]} & {\tiny[0.174, 0.196]} & {\tiny[0.679, 0.751]} \\
\midrule
7 & network alone & 2946 / 3962.4 & 0.743 & 2.414 & 1.654 & 1.459 & 0.170 & --- \\
 &  &  & {\tiny[0.482, 1.146]} & {\tiny[1.658, 3.515]} &  & {\tiny[0.854, 2.495]} & {\tiny[0.162, 0.178]} &  \\
 & composition, Grace, 10\,\% & 1186 / 3962.4 & 0.299 & 0.992 & 1.654 & 0.600 & 0.063 & 0.372 \\
 &  &  & {\tiny[0.193, 0.464]} & {\tiny[0.954, 1.033]} &  & {\tiny[0.408, 0.881]} & {\tiny[0.059, 0.068]} & {\tiny[0.347, 0.399]} \\
 & composition, Spark, 2\,\% & 1209 / 3962.4 & 0.305 & 1.004 & 1.654 & 0.607 & 0.065 & 0.385 \\
 &  &  & {\tiny[0.197, 0.473]} & {\tiny[0.891, 1.131]} &  & {\tiny[0.406, 0.906]} & {\tiny[0.061, 0.070]} & {\tiny[0.360, 0.412]} \\
\bottomrule
\end{tabular}

%% file: tables/tab_routing.tex
\begin{tabular}{@{}cclrrrrrrr@{}}
\toprule
& & & & \multicolumn{2}{c}{network failures} & composition on & & & failures, network \\
\cmidrule(lr){5-6}
benchmark & $d$ & platform, $f$, cap & routed & kept & routed (share) & routed (ratio) & corrected & introduced & $\to$ composition (ratio) \\
\midrule
I & 3 & Grace, 1\,\%, 0.001\,s & 1.02\,\% & 48\,121 & 56\,517 (54.01\,\%) & 53\,451 (0.946) & 18\,198 & 15\,132 & 104\,638 $\to$ 101\,572 (0.971) \\
I & 3 & Spark, 2\,\%, none & 2.00\,\% & 33\,740 & 70\,898 (67.76\,\%) & 67\,169 (0.947) & 21\,516 & 17\,787 & 104\,638 $\to$ 100\,909 (0.964) \\
I & 5 & both, 2\,\%, 0.05\,s & 2.03\,\% & 1254 & 13\,911 (91.73\,\%) & 9375 (0.674) & 7167 & 2631 & 15\,165 $\to$ 10\,629 (0.701) \\
I & 7 & both, 2\,\%, none & 2.02\,\% & 86 & 2474 (96.64\,\%) & 947 (0.383) & 1858 & 331 & 2560 $\to$ 1033 (0.404) \\
I & 9 & both, 2\,\%, none & 2.03\,\% & 9 & 659 (98.65\,\%) & 90 (0.137) & 600 & 31 & 668 $\to$ 99 (0.148) \\
\midrule
II & 3 & both, 0.5\,\%, 0.002\,s & 0.48\,\% & 2393 & 453 (15.92\,\%) & 425 (0.938) & 193 & 165 & 2846 $\to$ 2818 (0.990) \\
II & 5 & both, 5\,\%, 0.02\,s & 4.92\,\% & 171 & 1889 (91.70\,\%) & 1349 (0.714) & 927 & 387 & 2060 $\to$ 1520 (0.738) \\
II & 7 & Grace, 10\,\%, none & 9.47\,\% & 39 & 2907 (98.68\,\%) & 1147 (0.395) & 2133 & 373 & 2946 $\to$ 1186 (0.403) \\
II & 7 & Spark, 2\,\%, none & 1.86\,\% & 161 & 2785 (94.53\,\%) & 1048 (0.376) & 2093 & 356 & 2946 $\to$ 1209 (0.410) \\
\bottomrule
\end{tabular}

%% file: tables/tab_speed.tex
\begin{tabular}{@{}cccccccccc@{}}
\toprule
& their network & \multicolumn{4}{c}{network alone} & \multicolumn{2}{c}{composition, Benchmark I} & \multicolumn{2}{c}{composition, Benchmark II} \\
\cmidrule(lr){3-6}\cmidrule(lr){7-8}\cmidrule(lr){9-10}
& & \multicolumn{2}{c}{Benchmark I} & \multicolumn{2}{c}{Benchmark II} & & & & \\
$d$ & H100 & RTX PRO 6000 & GB10 & RTX PRO 6000 & GB10 & Grace platform & Spark platform & Grace platform & Spark platform \\
\midrule
3 & 47.4 & 1.18 & 4.26 & 1.18 & 5.01 & 1.2 & 4.4 & 1.2 & 5.1 \\
 &  &  &  &  &  & {\tiny 1\,\%, 0.001\,s} & {\tiny 2\,\%, none} & {\tiny 0.5\,\%, 0.002\,s} & {\tiny 0.5\,\%, 0.002\,s} \\
5 & 103.3 & 2.16 & 10.54 & 2.37 & 12.99 & 6.8 & 24.8 & 20.8 & 72.6 \\
 &  &  &  &  &  & {\tiny 2\,\%, 0.05\,s} & {\tiny 2\,\%, 0.05\,s} & {\tiny 5\,\%, 0.02\,s} & {\tiny 5\,\%, 0.02\,s} \\
7 & 185.2 & 3.28 & 19.26 & 4.08 & 24.38 & 27.1 & 93.7 & 152.7 & 193.8 \\
 &  &  &  &  &  & {\tiny 2\,\%, none} & {\tiny 2\,\%, none} & {\tiny 10\,\%, none} & {\tiny 2\,\%, none} \\
9 & 302.2 & 5.14 & 32.44 & --- & --- & 66.4 & 236.1 & --- & --- \\
 &  &  &  &  &  & {\tiny 2\,\%, none} & {\tiny 2\,\%, none} &  &  \\
\bottomrule
\end{tabular}

%% file: tables/tab_abl.tex
\begin{tabular}{@{}lrcccccc@{}}
\toprule
& & \multicolumn{5}{c}{Benchmark I} & Benchmark II \\
\cmidrule(lr){3-7}
variant & parameters & $K=2$ & $K=3$ & $K=4$ & pooled & $p=0.1, 0.2\,\%$ & pooled \\
\midrule
no exchange between blocks & 912\,199 & 2.088 & 1.813 & 2.111 & 1.999 & 2.755 & 3.260 \\
 &  & {\tiny[1.924, 2.269]} & {\tiny[1.695, 1.931]} & {\tiny[1.967, 2.254]} & {\tiny[1.912, 2.081]} & {\tiny[2.334, 3.382]} & {\tiny[3.135, 3.373]} \\
exchange only at the read-out & 912\,199 & 1.647 & 1.475 & 1.672 & 1.596 & 1.919 & 2.054 \\
 &  & {\tiny[1.540, 1.769]} & {\tiny[1.393, 1.557]} & {\tiny[1.573, 1.774]} & {\tiny[1.539, 1.654]} & {\tiny[1.666, 2.313]} & {\tiny[2.007, 2.102]} \\
pooled read-out & 911\,415 & 19.818 & 47.082 & 25.486 & 28.757 & 47.795 & 12.138 \\
 &  & {\tiny[15.031, 24.241]} & {\tiny[35.176, 56.258]} & {\tiny[20.565, 29.970]} & {\tiny[24.360, 31.611]} & {\tiny[32.334, 64.044]} & {\tiny[10.364, 13.385]} \\
small trunk (3 blocks) & 523\,719 & 1.264 & 1.202 & 1.218 & 1.228 & 1.348 & 1.242 \\
 &  & {\tiny[1.204, 1.333]} & {\tiny[1.154, 1.253]} & {\tiny[1.175, 1.261]} & {\tiny[1.198, 1.259]} & {\tiny[1.199, 1.565]} & {\tiny[1.219, 1.266]} \\
specialist, $K=2$ only & 912\,199 & 1.046 & --- & --- & --- & 1.056 & --- \\
 &  & {\tiny[0.992, 1.106]} &  &  &  & {\tiny[0.908, 1.245]} &  \\
specialist, $K=3$ only & 912\,199 & --- & 0.977 & --- & --- & 1.113 & --- \\
 &  &  & {\tiny[0.936, 1.021]} &  &  & {\tiny[0.904, 1.445]} &  \\
specialist, $K=4$ only & 912\,199 & --- & --- & 0.976 & --- & 1.315 & 1.001 \\
 &  &  &  & {\tiny[0.942, 1.010]} &  & {\tiny[1.055, 1.826]} & {\tiny[0.983, 1.019]} \\
\midrule
main network (fine-tuned) & 912\,199 & 0.974 & 0.910 & 0.947 & 0.943 & 1.013 & 1.038 \\
 &  & {\tiny[0.929, 1.024]} & {\tiny[0.875, 0.948]} & {\tiny[0.913, 0.985]} & {\tiny[0.921, 0.967]} & {\tiny[0.888, 1.181]} & {\tiny[1.020, 1.058]} \\
\bottomrule
\end{tabular}

%% file: sections/discussion.tex
\section{Discussion}
\label{sec:discussion}
\paragraph{Scope.}
The three insights hold for every logical Clifford circuit built from transversal CNOT, H and S gates.
\sys decodes such circuits on two to four surface-code blocks under one noise model, at distances 3 to 9 and depths up to 16 layers on Benchmark~I and 20 on Benchmark~II.
Non-Clifford gates, lattice surgery and other noise models need new training.
The network decodes a whole circuit at once, so circuits of thousands of layers need windowed decoding, which is future work.

\vspace{0pt plus 3\baselineskip}\penalty-200\vspace{0pt plus -3\baselineskip}%
\paragraph{The network alone against the Lukin group's network.}
The gains on Benchmark~II belong to the composition.
At $d=5$ and 7 the intervals of the network alone against the Lukin group's network include 1 on the fresh test, and on the first test its measure (b) at $d=7$ lay above 1 (Appendix~\ref{app:first}).
On Benchmark~I, from $d=7$ to 9 the network alone's ratio to LoMatching falls from 0.177 to 0.146 and the composition's from 0.0765 to 0.0225.
At $d=7$ the composition fails 92 times on the certified prefix shots at the Grace platform's operating point and 93 times at the Spark platform's, against 92 for the exact MLE, so with \tess on 2 to 10\,\% of the shots its failure rate is statistically compatible with the exact MLE.

\paragraph{Throughput and delay.}
The Lukin group's time per gate and our time per layer are both amortized over batches of 256 shots, so both measure throughput.
A routed shot waits for its trials, and at the operating points of Benchmark~I the 99th percentile of the batch delay in the simulated paced stream grows from 19 milliseconds at $d=3$ to 9.4 seconds at $d=9$ on the Grace platform.
A gate that waits for a decoded outcome needs a bound on this delay, which the composition does not give.

\paragraph{Hardware conversion.}
The trials ran on AMD EPYC cores.
For the Grace platform we convert their times through a measured ratio to an AMD Ryzen 9 9950X and the Grace-to-Ryzen ratio of PyMatching from our earlier work~\cite{codecoder}, so \tess itself did not run on a Grace CPU, and for the Spark platform we scale the Grace times by a factor of 1.124 measured on the DGX Spark at $d=7$.
The live stream on an RTX PRO 6000 with 32 EPYC cores tests the stream model and the recorded trial times, and the stream from its trial times took a median 0.94 of the simulated time, but no Grace CPU ran \tess, so the Grace times rest on the conversion.
The DGX Spark also ran the stream live, so the Spark platform is measured directly.
There the stream from the live trial times took a median 0.99 of the simulated time, with gaps of up to 14\,\% in either direction, and a calibration of \tess on its 20 cores puts the trials at 0.84 of the time that the factor of 1.124 assumes.

\paragraph{Training history.}
The four networks share one architecture and one final training recipe, following different pretraining histories (Appendix~\ref{app:train}).
Pooled over block counts, each network fails 0.872 to 0.995 times as often as its start on the development shots of Benchmark~I.
At $K=2$ the upper ends of the intervals reach 1.059, 1.204 and 1.139 at $d=5$, 7 and 9, and 1.057 at $K=3$ and $d=7$, so a rise of the two-block failures of up to 20\,\% at $d=7$ after the training on the four-block circuits of Benchmark~II lies within the interval.

\paragraph{Choice of the routed fraction.}
The first test used the candidate with the fewest development failures under the time budget.
On Benchmark~II that rule put three operating points near their budgets, where the accuracy is already flat, and these three run above the budget on the fresh test (Appendix~\ref{app:first}).
We fixed the present rule after both tests had shown the flat curve, so the test results shaped its form.
It reads development data only, every routed fraction was frozen before the fresh test, and Tables~\ref{tab:rt1} and~\ref{tab:rt2} report all of them, so a reader can apply another rule to the same rows.

%% file: sections/related.tex
\section{Related Work}
\label{sec:related}
The works most relevant to ours include decoders for transversal circuits, neural decoders for logical circuits, read-out rules that compose predictions by parity, and the routing of uncertain shots to a stronger decoder.

\paragraph{Decoders for transversal circuits.}
Cain et al.\ decode interacting blocks jointly on a circuit hypergraph with integer programming and union-find~\cite{cain24}, and Zhou et al.\ prove that this joint decoding gives fault tolerance with a constant number of syndrome rounds per gate~\cite{zhou25}.
Sahay et al.\ use one block's inferred errors to update the partner block, and Wan et al.\ repeat such passes over the blocks~\cite{sahay,wan}.
Cain et al.\ later propagate reliable logical products back through the circuit and match on the relevant checks~\cite{cain25}, and LoMatching builds a matching graph for an observable through the transversal Clifford gates, with belief propagation as an optional first stage~\cite{lom}.
\tess searches over sets of errors for the most likely one~\cite{tesseract}.
Chen et al.\ realize the logical S of the rotated surface code with a fold-transversal gate inside a syndrome round~\cite{foldS}, which our circuits use.
These decoders carry the correlations between blocks by hand-built rules or by search.
\sys builds the same knowledge into a network instead.
It exchanges features at the sites that each gate connects (Insight 1), takes the exact propagation of the reliable products of~\cite{cain25,lom} as input to its read-out (Insight 2), and uses \tess on its least confident shots.

\paragraph{Neural decoders for logical circuits.}
The recurrent attention network of Bonilla Ataides et al.\ decodes random logical circuits on four blocks up to distance 7, and the authors time it up to distance 9~\cite{bonilla}.
Benchmark~II compares \sys with their network on their circuits.
Zhou et al.\ compose gate-specific recurrent modules and decode logical circuits at distances 3 and 5 with higher accuracy and lower speed than LoMatching~\cite{mccd}.
Lange et al.\ decode a distance-3 lattice-surgery merge on hardware~\cite{lange}.
AlphaQubit and AlphaQubit 2 decode memory experiments, the latter in under a microsecond per round up to distance 11~\cite{alphaqubit,alphaqubit2}.
These decoders learn from data how information crosses gates or time, and \sys differs in the read-out (Insight 2).
Its per-layer logits refer to a fixed final product through its exact parts on every block, and their parity composition makes the number of read-out terms the only dependence on the depth.

\paragraph{Read-out by parity.}
Baireuther et al.\ combine two network branches by the parity formula that our read-out uses to combine the layers~\cite{baireuther}, and Zhang et al.\ train overlapping windows to predict their contributions to a global logical flip and compose them by parity~\cite{zhang26}.
Chamberland et al.\ track Pauli corrections through Clifford gates~\cite{chamberland}.
We borrow these parity and propagation rules and apply them to per-layer contributions of a convolutional decoder for Clifford circuits on several blocks.

\paragraph{Routing to a stronger decoder.}
Switching from a fast decoder to a stronger one when the confidence is low is an established idea~\cite{toshio,chongder}, and Shutty et al.\ pass the shots on which a small ensemble of matching decoders disagrees to a larger ensemble~\cite{shutty}.
Our earlier CoDecoder work states when such a composition beats each of its parts~\cite{codecoder}.
We apply these conditions to a neural decoder and a search decoder on transversal circuits and measure the composition in a stream of batches against a fixed time budget per gate.

%% file: sections/conclusion.tex
\section{Conclusion}
\label{sec:conclusion}
Transversal gates promise a fast logical clock, but they need a decoder that reads all blocks of a circuit together, accurately and in little time per gate.
Where errors go across each gate and how logical outcomes propagate through the circuit are both known before any shot is decoded, so a neural decoder does not need to learn them.
In this work, we present \sys, which builds them in through an exchange at every gate, a per-layer read-out on the exact propagation, and weights shared across blocks and layers.
We show that it decodes the random logical Clifford circuits of Benchmark~I on two to four blocks more accurately than LoMatching, Cain et al.\ and \bplom at distances 3 to 9 at $p=0.3\,\%$, with one network per distance.
With \tess on its least confident shots, it fails 0.547, 0.205, 0.0765 and 0.0225 times as often as LoMatching on the fresh test set at the operating points, because the routed 1 to 2\,\% of the shots hold 54 to 99\,\% of the network's failures and \tess corrects more of them than it introduces.
On the routed shots the composition fails 0.946, 0.674, 0.383 and 0.137 times as often as the network, a relative reduction that grows with the distance.
An earlier test set, scored at the operating points of an earlier rule, gives consistent ratios at $p=0.3\,\%$.
On Benchmark~II, the Lukin group's circuits at distances 3, 5 and 7, the composition fails less often than their network both against their published rates and relative to the exact MLE, and its failure rate is statistically compatible with the exact MLE on the certified prefixes.
On this benchmark its time per layer in a paced stream simulated from measured components stays within their time per gate at every distance on the Grace platform, and on the Spark platform it exceeds it at $d=7$, at 193.8 against 185.2 microseconds per layer, as do the live streams on both hosts at $d=7$.
A network that builds in what the gates fix and learns only which errors occurred, paired with a search on the shots it doubts, is therefore a practical route to decoding transversal logical circuits.

%% file: sections/appendix.tex
\section{Circuits, noise and test sets}
\label{app:circuits}
This appendix gives the circuit details that Secs.~\ref{sec:background} and~\ref{sec:setup} summarize.
The code is the rotated surface code of odd distance $d$ with data qubits at odd coordinates and weight-2 checks on the boundaries.
A syndrome round resets the check qubits, applies a Hadamard to the X-type ones, runs four layers of CNOTs in orders that keep the circuit distance at $d$, applies a Hadamard and measures.
An exact search with the SCIP solver, one integer program per logical observable, proves circuit distance $d$ for every observable it checked: 39 observables of two-block circuits with H and S gates at each of $d=3$ and 5, and 206 observables of three-block circuits at $d=3$ and 5 and four-block circuits at $d=3$.
Every detector compares a check result with the value that earlier results and the gates since then require.

\paragraph{Gates.}
The logical H applies a physical H to every data qubit in one step before the round, and the block's code becomes the code rotated by 90 degrees in place.
Every check qubit keeps its cell and its CNOT order and measures the other Pauli type from then on, which keeps the hook errors perpendicular to the logical operators.
A CNOT between blocks of different orientations pairs the control's data qubit at position $x$ with the target's data qubit at the rotated position, and we take the move that brings them together as free of noise, as the Lukin group does.
The logical S follows Chen et al.~\cite{foldS}.
After the second CNOT step of the round, the block's data and check qubits hold an unrotated surface code, and CZ gates act on every pair of qubits mirrored about the anti-diagonal, with S on the data qubits of the fold line and $S^\dagger$ on its check qubits, before the round goes on.
The S costs one extra step with $(d^2-d)/2+(d-1)(d-2)/2$ CZ gates and no extra round.
With the fold about the anti-diagonal, circuits with S, H and rotated CNOTs keep the full circuit distance.
Every block reads its logical operators on its centre row and column, which no gate moves.

\paragraph{Benchmark I: general circuits.}
A circuit of Benchmark~I draws the probability $q$ that a pair of blocks runs a CNOT from $\{0,1/3,2/3,1\}$, 16 circuits of each per point.
In each layer the blocks are shuffled into consecutive pairs, each pair runs a CNOT with probability $q$, and each block outside a CNOT gets I, H or S, so at $K=2$ a layer is either a CNOT or single-qubit gates, and at $K=3$ one block always stays outside the pairs.
The preparation and measurement bases are drawn per block from $Z$ and $X$, and a circuit without a reliable product is drawn again.
Logical Paulis and the $Z$ of $S^\dagger=SZ$ are tracked in software, so they change neither the detectors nor the error model.
The 64 circuits of a point are the same at every distance and noise strength, and development and test circuits differ.

\paragraph{Benchmark II: the Lukin group's circuits.}
We rebuild the benchmark of Bonilla Ataides et al.\ from their Methods, their Fig.~1a and their released compiler.
Four blocks start in $|+\rangle_L$ from a noiseless encoder.
A depth unit applies a single-qubit Clifford drawn from $\{I,X,Y,Z,H\}$ to every block, with the Paulis as physical gates, and then CNOTs on a random perfect matching of the four blocks with random orientations.
Their Methods place one syndrome round after the CNOTs, while their compiler gives every layer of an algorithm its own round, so a single-qubit layer and a CNOT layer receive two.
We follow the compiler, and 230 of our circuits have the same detector error model as their released simulator gives them.
Under exact MLE our circuits fail 1.01 [0.93, 1.09], 0.83 [0.72, 0.97] and 0.58 [0.28, 1.19] times as often as their published MLE at $d=3$, 5 and 7, with intervals that account for their error bars.
With one round per depth unit, long-beam \tess fails 0.48 to 0.70 times as often as their MLE across the depths at $d=3$.
Each variant we tested, among them other target rules, other bases for the blocks outside the target, no round after the last layer and noisy identity gates, either leaves $d=5$ almost unchanged or moves $d=3$ out of agreement, and their release does not contain the evaluation behind their published figure.
On our circuits, long-beam \tess decides as the exact MLE on all but one of 960\,000 shots at $d=3$, 5 and 7.
Their time per gate is the time of one step of their decoder, which takes one step per layer of their compiler, each layer with one syndrome round.
The logical H is a physical H followed by a noiseless rotation of the patch.
The target is a stabilizer of the final logical state with factors I, X or Z, drawn uniformly, and blocks outside its support are measured in $Z$.
The 20 circuits per depth are the same at every distance.
The test shots of Benchmark~II are fresh noise samples of these circuits.

\paragraph{Noise.}
Two-qubit depolarizing noise of strength $p$ follows every two-qubit gate.
Single-qubit depolarizing noise of strength $p/10$ follows every single-qubit gate, resets and measurements fail with probability $p/10$, and idle qubits have no noise.
In the circuits of Benchmark~II the identity is free of noise.

\section{Networks and training}
\label{app:train}
Table~\ref{tab:starts} gives the history of each start.
An earlier network for two-block CNOT circuits initialized a network for Clifford circuits, which trained on two-block circuits under the Lukin noise model with a learning rate of $10^{-3}$ and then on two to four blocks with a learning rate of $3\times10^{-4}$.
The start at $d=3$ has three residual blocks.
We grow it to seven by adding four residual blocks whose exchange and output start at zero, so the parameters rise from 523\,397 to 911\,877.
On the first 16 training batches, the grown network and its start give equal outputs in 32-bit floats and in bfloat16, and the first monitor row of each network equals its start's last row.

\begin{table}[t]
\caption{The start of each network and the training that preceded it, in samples, first on two blocks with a learning rate of $10^{-3}$ and then on two to four blocks with $3\times10^{-4}$. All of it used the Lukin noise model at $p=0.3\,\%$.}
\label{tab:starts}
\centering\footnotesize\setlength{\tabcolsep}{4pt}
\begin{tabular}{@{}crrr@{}}
\toprule
$d$ & two blocks & two to four blocks & residual blocks \\
\midrule
3 & $1\times10^8$ & $2.5\times10^7$ & 3, grown to 7 \\
5 & $2\times10^8$ & $5\times10^7$ & 7 \\
7 & $3\times10^8$ & $7.5\times10^7$ & 7 \\
9 & $2\times10^8$ & $5\times10^7$ & 7 \\
\bottomrule
\end{tabular}
\end{table}

\paragraph{The common recipe.}
Every network then trains on $5\times10^7$ samples, 48\,828 steps of 1\,024 samples, with four chunks of 256 shots of one block count and depth per step, and with seeds 9103, 9105, 9107 and 9109 at $d=3$, 5, 7 and 9.
Half of the steps, spread evenly, draw the family of Benchmark~II with $D=1,\dots,10$ in turn.
The other half draw the start's mix: block counts from $\{2,2,3,4\}$, depths 4, 6 and 8, a quarter of CNOT-only circuits, and for $K>2$ a tenth of circuits that entangle the blocks and then idle.
AdamW uses betas $(0.9,0.99)$, weight decay 0.01, gradient clipping at 1 and bfloat16 mixed precision.
Training ran at 8\,180, 3\,531, 1\,902 and 1\,161 samples per second on an RTX PRO 6000 GPU at $d=3$, 5, 7 and 9.

\paragraph{Comparison with the start.}
On the development shots of Benchmark~I, each network fails 0.995, 0.990, 0.993 and 0.872 times as often as its start at $d=3$, 5, 7 and 9, pooled over block counts, with 95\,\% upper bounds of 0.9995, 1.004, 1.039 and 0.955.
Per block count the upper bound exceeds 1.05 at $K=2$ for $d=5$ (1.059), $d=7$ (1.204) and $d=9$ (1.139) and at $K=3$ for $d=7$ (1.057).
On the development shots of Benchmark~II the networks fail 0.932, 0.911 and 0.753 times as often as their starts at $d=3$, 5 and 7.

\paragraph{Variants at $d=5$.}
Every variant trains from scratch on $10^8$ samples with the mix of the main networks and seed 9205, using AdamW with a learning rate of $10^{-3}$, a warm-up of 1\,000 steps and a cosine decay to 2\,\%.
The variant without exchange keeps the exchange modules but reads the site's own block, so it has the same parameters as the full network.
The specialists for $K=2$ and 3 train on their block count only, and the specialist for $K=4$ also on the family of Benchmark~II, which has four blocks.

\section{Statistics}
\label{app:stats}
\paragraph{Benchmark I.}
Every circuit of a point has the same number of shots, so the failure rate of a point equals the mean of its circuits' rates.
The ratio of two decoders at a point divides their failures on the same shots.
Its 95\,\% interval comes from 10\,000 bootstrap draws that redraw the 64 circuits with replacement, each with its observed paired counts, with a seed fixed by the point.
When either decoder has no failure at a point, 0.5 is added to both counts in the estimate and in every draw with a zero count, and the point is flagged.
A point where both decoders fail on exactly the same shots gives 1 [1, 1], which we count as inconclusive.
A pooled ratio is the geometric mean of the per-point ratios, with its interval from each draw's mean log ratio, and flagged points stay in every pooled row.

\paragraph{Benchmark II.}
Measure (a) is the ratio $A=\sum_D F_D\big/\allowbreak\sum_D n_D p_D$, with our failures $F_D$ and shots $n_D$ at depth $D$ and the published failure rate $p_D$ of their network.
Their text calls these rates normalized per round, but every plotted value of their network is $k/25\,600$ for an integer $k$, with an error bar of one binomial standard deviation, so we read $p_D$ as a failure fraction per shot.
The variance of $\log A$ adds a part from 10\,000 bootstrap draws with seed $186014+d$, each of which redraws the 20 circuits of every depth and then each drawn circuit's failures binomially, and a part from the published error bars.
Measure (b) is
\begin{equation}
B=\frac{\sum_D x_D/\sum_D m_D}{\sum_D p_D/\sum_D q_D},
\end{equation}
with our failure rate $x_D$ and the exact MLE's $m_D$ on the same prefix shots and the published rates $p_D$ and $q_D$ of their network and their MLE, over depths 1 to 10 at $d=3$ and 5 and depths 1 to 8 at $d=7$, where the exact MLE was not run at depths 9 and 10.
The variance of $\log B$ adds a part from 10\,000 bootstrap draws with seed $186114+d$, each of which redraws the circuits of every depth and then each drawn circuit's paired shots in four categories (both fail, only ours fails, only the exact MLE fails, neither fails), and a part from the published error bars.
Table~\ref{tab:app_mr2} lists these inputs per depth for the first test, and the rates $x_D$ and $m_D$ are the failures on the certified prefix shots of depth $D$ divided by those shots.

\begin{table*}[tp]
\caption{The first test, Benchmark~II per depth: the inputs of measures (a) and (b). Every test shot enters (a), with the published failure rate $p_D$ of their network and its error bar, and the line ``all'' gives the sums, with $E=\sum_D n_D p_D$. The certified prefix enters (b), with the uncertified shots that we leave out in parentheses, our failures and those of the exact MLE on these shots, and the published rate $q_D$ of their MLE. At $d=7$ the exact MLE ran on depths 1 to 8. $^{\ast}$Their figure gives this value without an error bar, and the bar shown is the binomial one at their 17\,777 shots of depth 1. At $d=3$ one composition serves both platforms.}
\label{tab:app_mr2}
\centering\scriptsize\setlength{\tabcolsep}{3pt}
\input{tables/tab_app_mr2_depth}
\end{table*}
The four prefix shots at $d=5$ and three at $d=7$ without a certificate of optimality are left out of both $x_D$ and $m_D$.
An interval below 1 means that \sys fails less often, an interval above 1 that it fails more often, and an interval that contains 1 is inconclusive.

\section{The first test}
\label{app:first}
The first test set has the design and size of the fresh one, with its own circuits in Benchmark~I and its own noise samples on both benchmarks.
It scored the operating points of an earlier rule, which took, per benchmark, distance and platform, the candidate with the fewest development failures under the time budget.
These points were frozen distance by distance before that distance's test was read.
They route $f=10$, 10, 5 and 2\,\% at $d=3$, 5, 7 and 9 in Benchmark~I on both platforms, and 10, 10 and 20\,\% on the Grace platform and 10, 10 and 2\,\% on the Spark platform at $d=3$, 5 and 7 in Benchmark~II (Table~\ref{tab:choices}).
Tables~\ref{tab:first1} and~\ref{tab:first2} give its results in the form of Tables~\ref{tab:mr1} and~\ref{tab:mr2}.

\paragraph{Consistency with the fresh test.}
The fresh test scores these points too, as rows of Tables~\ref{tab:rt1} and~\ref{tab:rt2}.
In Benchmark~I at $p=0.3\,\%$ the network alone fails 0.5634, 0.2889, 0.1772 and 0.1462 times as often as LoMatching on the fresh test at $d=3$, 5, 7 and 9, against 0.5645, 0.2911, 0.1808 and 0.1694 on the first, and the composition at these points 0.5406, 0.2028, 0.0762 and 0.0225 times as often, against 0.5390, 0.2027, 0.0731 and 0.0248.
Its ratio to the network is 0.9595, 0.7021, 0.4301 and 0.1541, against 0.9550, 0.6963, 0.4042 and 0.1458.
In Benchmark~II the composition reaches a measure (a) of 0.848, 0.705 and 0.299 at $d=3$, 5 and 7 on the Grace platform, against 0.874, 0.696 and 0.300, and a measure (b) of 0.849, 0.747 and 0.600, against 0.847, 0.745 and 0.628.
Every one of these fresh values, and every fresh ratio of these points against Cain et al.\ and \bplom\ at $p=0.3\,\%$, lies inside the 95\,\% interval of the first test, except the composition's ratio to the network at $d=3$, 0.9595 against an upper end of 0.9594.
At the lower noise strengths, six points per distance with few failures, 15 of the 28 fresh ratios lie outside the first test's intervals.
Every pair of intervals overlaps, in both noise groups.
At $d=7$ the network alone's measure (b) lay above 1 on the first test, at 1.772 [1.026, 3.063], and is inconclusive on the fresh test, at 1.459 [0.854, 2.495].
At $d=9$ the composition's ratio to LoMatching is smaller at the lower noise strengths than at $p=0.3\,\%$ on the fresh test, 0.0128 against 0.0225, and larger on the first, 0.0456 against 0.0248.

\paragraph{Time budget.}
On Benchmark~II the earlier rule put three of these points near their time budgets, where the accuracy curve is already flat.
On development shots the Spark platform's point at $d=5$ took 93.9 microseconds per layer against a budget of 103.3, and the points at $d=7$ took 183.5 on the Grace platform and 176.5 on the Spark platform against 185.2 (Table~\ref{tab:choices}).
On the fresh test all three run above the budget, at 115.8, 198.0 and 193.8 microseconds per layer.
Frozen rows of the same distance and platform with smaller $f$ match their accuracy under the budget.
At $d=5$ on the Spark platform, $f=5\,\%$ fails 1\,520 times at 72.6 microseconds per layer, against 1\,519 at 115.8 for $f=10\,\%$.
At $d=7$ on the Grace platform, $f=10\,\%$ fails 1\,186 times at 152.7 and $f=5\,\%$ 1\,195 times at 101.1, against 1\,186 at 198.0 for $f=20\,\%$, and on the Spark platform $f=1\,\%$ fails 1\,231 times at 116.0, against 1\,209 at 193.8 for $f=2\,\%$.
Table~\ref{tab:sweep} gives Benchmark~II on the Grace platform at other routed fractions on the first test, rows picked from the development columns and scored from the trials of the chosen runs, a first view of the trade-off that Tables~\ref{tab:rt1} and~\ref{tab:rt2} give in full.

\paragraph{Failure counts and per-point results.}
For the first test, Table~\ref{tab:counts} gives the failure counts and Table~\ref{tab:perk} the ratios per block count.
Against \bplom on its matched shots, some points have zero failures of one decoder, and the 0.5 rule applies.
At $d=7$ these are 3 of the 18 points at $p=0.3\,\%$ for both the network and the composition, and 2 and 3 of the six points at the lower noise strengths.
At $d=9$ they are 7 and 12 of the 18 points and 3 and 4 of the six, and for each system 5 of the 18 points and 3 of the six have no shot that separates the decoders.
Against LoMatching and Cain et al., the composition has zero failures at 1 point at $d=7$ and at 6 points at $d=9$, and the network at 3 points at $d=9$.
Tables~\ref{tab:app_pt_counts_a} and~\ref{tab:app_pt_counts_b} list the failures at every point of Benchmark~I, with the routed shots and the failures on the matched shots of \bplom, the first 1\,000 shots of each circuit and 64\,000 per point.
Tables~\ref{tab:app_pt_ratios_a} and~\ref{tab:app_pt_ratios_b} give the paired ratios with their intervals.
The composition is the Grace platform's, and the Spark platform's differs from it by at most 2 failures at any point.

\begin{table*}[t]
\caption{The first test, Benchmark~I: failures of the network alone and of the composition with \tess at the first test's operating points, divided by those of three decoders on the same shots, pooled over $K=2$, 3 and 4 and the depths 1, 2, 4, 8, 12 and 16 at $p=0.3\,\%$ (18 points) and over the depth-8 points at $p=0.1$ and $0.2\,\%$ (6 points), with 95\,\% intervals below. The column $f$ gives the routed fraction. The composition is the Grace platform's, and the Spark platform's differs by at most 0.0007 in any entry.}
\label{tab:first1}
\centering\footnotesize\setlength{\tabcolsep}{3.4pt}
\input{tables/tab_first_mr1}
\end{table*}

\begin{table*}[t]
\caption{The first test, Benchmark~II, at $p=0.3\,\%$, as Table~\ref{tab:mr2} at the first test's operating points. Grace and Spark name the platform of a composition, and both platforms share one composition at $d=3$. 95\,\% intervals below the values.}
\label{tab:first2}
\centering\footnotesize\setlength{\tabcolsep}{4pt}
\input{tables/tab_first_mr2}
\end{table*}

\begin{table*}[t]
\caption{The first test: Benchmark~II on the Grace platform at several routed fractions $f$ and caps per layer. The development columns give the failures on the development shots and the time per layer in the paced stream, simulated from measured network and trial times. Rows marked $\ast$ are the Grace platform's operating points of the first test, and the row marked $\dagger$ is the Spark platform's at $d=7$, timed here on the Grace platform. At $d=7$ the 5 and 10\,\% rows carry a cap of 0.1\,s per layer, in place of the uncapped rows first named for this table. The other rows were picked from the development columns and scored from the trials of the chosen runs. Intervals at 95\,\%.}
\label{tab:sweep}
\centering\footnotesize\setlength{\tabcolsep}{4pt}
\input{tables/tab_sweep}
\end{table*}

\begin{table*}[t]
\caption{The first test: failures on the test shots of Benchmark~I. \bplom decodes the first 1\,000 shots of every circuit.}
\label{tab:counts}
\centering\footnotesize\setlength{\tabcolsep}{3.4pt}
\input{tables/tab_app_counts}
\end{table*}

\begin{table*}[t]
\caption{The first test: pooled ratios of Benchmark~I at $p=0.3\,\%$ per block count, over the six depths, with 95\,\% intervals. The composition is the Grace platform's.}
\label{tab:perk}
\centering\footnotesize\setlength{\tabcolsep}{4pt}
\input{tables/tab_app_k}
\end{table*}

\begin{table*}[tp]
\caption{The first test, Benchmark~I per point at $d=3$ and 5: failures on the test shots. Routed counts the shots that the thresholds send to \tess. The last three columns count failures on the matched shots of \bplom. The composition is the Grace platform's.}
\label{tab:app_pt_counts_a}
\centering\scriptsize\setlength{\tabcolsep}{4pt}
\input{tables/tab_app_pt_counts_a}
\end{table*}

\begin{table*}[tp]
\caption{The first test, Benchmark~I per point at $d=7$ and 9: failures on the test shots, as in Table~\ref{tab:app_pt_counts_a}. One composition serves both platforms.}
\label{tab:app_pt_counts_b}
\centering\scriptsize\setlength{\tabcolsep}{4pt}
\input{tables/tab_app_pt_counts_b}
\end{table*}

\begin{table*}[tp]
\caption{The first test, Benchmark~I per point at $d=3$ and 5: paired failure ratios on the same shots, with 95\,\% intervals from 10\,000 draws of the 64 circuits. Ratios against \bplom use its matched shots. The composition is the Grace platform's.}
\label{tab:app_pt_ratios_a}
\centering\scriptsize\setlength{\tabcolsep}{3pt}
\input{tables/tab_app_pt_ratios_a}
\end{table*}

\begin{table*}[tp]
\caption{The first test, Benchmark~I per point at $d=7$ and 9: paired failure ratios as in Table~\ref{tab:app_pt_ratios_a}. $^{\mathrm{z}}$The 0.5 rule for a zero count applies. $^{\mathrm{d}}$No discordant shot, so the interval is degenerate and the point is inconclusive.}
\label{tab:app_pt_ratios_b}
\centering\scriptsize\setlength{\tabcolsep}{3pt}
\input{tables/tab_app_pt_ratios_b}
\end{table*}

\section{Operating points and speed}
\label{app:speed}
Table~\ref{tab:value} lists the operating points with their development failures and the fewest development failures of any $f$ on that platform, and their failures and time on the fresh test beside those of $f=20\,\%$.
Table~\ref{tab:choices} lists the first test's operating points with their development failures and their time in the paced stream simulated on development shots.
The geometric mean over the points of the composition's failure rate decides between candidates.
The simulated paced stream runs in circuit-major batches, and the network's next batch starts when the GPU is free and the first trial of the previous batch has started.
A trial past its cap holds its core for the cap and is dropped.

\paragraph{Network times.}
The network times come from the inference engine at batch 256 on a whole NVIDIA RTX PRO 6000 Blackwell Server Edition (driver 595.91.07, CUDA 13.2) and on the GB10 of a DGX Spark (driver 580.126.09, CUDA 13.0), for every distance, block count and depth of the development shots.

\paragraph{Trial times.}
The long beam is \tess, version 0.1.1.dev20260822020007, with a beam of 20 and beam climbing over 21 detector orders from a breadth-first search with seed 2384753, a priority-queue limit of $10^6$ and no revisits of detectors, which we run as 21 trials, trial $j$ with order $j$ and beam $j$ for $j=0,\dots,20$.
Each trial was timed alone on one core of an AMD EPYC node with every core busy.
A calibration on common shots converts each CPU model to a reference model, and a second factor converts the reference to a Grace core: 0.577412 from the EPYC 7713P at $d=5$ and 0.884923 from the EPYC 9534 at $d=7$.
These factors come from a measured ratio between each EPYC and an AMD Ryzen 9 9950X and the ratio of PyMatching's times on Grace and on that Ryzen.
No Grace factor exists at $d=3$ and 9, so $d=3$ uses that of $d=5$ and $d=9$ that of $d=7$.
The Spark platform multiplies the Grace factors by 1.124, a factor measured on the DGX Spark at $d=7$.
The fresh test simulates the same stream with the trial times recorded on its own shots, converted in the same way, and every routed shot up to $f=20\,\%$ ran its trials without a cap, so each cap acts exactly on the recorded trial times.

\paragraph{Every routed fraction on the fresh test.}
Every routed fraction at its development-best cap on each platform was frozen before the fresh test existed, and Tables~\ref{tab:rt1} and~\ref{tab:rt2} give all of them.
In Benchmark I on the Grace platform, routing the least confident 0.5\,\% of the shots brings the composition to 0.980, 0.757, 0.458 and 0.171 times the network's failures at $d=3$, 5, 7 and 9, and from 2 to 20\,\% the ratio changes by less than 0.01 at every distance.
At $d=3$ and $p=0.3\,\%$ the shots routed beyond 5\,\% bring slightly more introduced errors than corrections, 459 against 431.
At $d=9$ \tess already corrects 584 of the network's 668 failures at $f=0.5\,\%$, so routing 20\,\% adds only 19 corrections and 6 introduced errors.
In Benchmark~II measure (a) falls from 0.875, 0.956 and 0.743 for the network alone at $d=3$, 5 and 7 to 0.852, 0.716 and 0.305 at 2\,\% and 0.849, 0.705 and 0.299 at 20\,\%.
On the Spark platform the time budget forces a cap on the trials beyond 5\,\% at $d=7$ and beyond 2\,\% at $d=9$ in Benchmark~I, and at 20\,\% the composition fails 1\,237 and 310 times, against 1\,080 and 102 on the Grace platform.
In Benchmark~II at $d=7$ its cap tightens to 0.001\,s at 20\,\%, where the composition fails 3\,605 times against 2\,946 for the network alone, and the live check below finds 28 failures against 11 simulated for this configuration.

\paragraph{Live check.}
A live stream ran every routed fraction at its development-best cap on development shots, 49 configurations per host: 55\,296 shots over the 24 points of Benchmark~I and 22\,528 to 25\,600 over the 10 depths of Benchmark~II per configuration, in batches of 256, with the frozen development thresholds, the engine on the GPU and each trial stopped at its cap in that CPU's time.
The first host is a whole NVIDIA RTX PRO 6000 Blackwell Server Edition with 32 cores of an AMD EPYC 9555, the most CPU cores per GPU that the cluster grants, with the Grace platform's caps at $N=32$.
The second is a DGX Spark, whose GB10 holds the GPU and 10 Cortex-X925 and 10 Cortex-A725 cores, with the Spark platform's caps at $N=20$.
A calibration of \tess on these 20 cores against the reference EPYC models puts its trials at 0.84 of the time that the Spark platform assumes.
Table~\ref{tab:live} gives both hosts at the operating points of both tests, with each time averaged with equal weight over the points.
On both hosts the simulation routed one borderline shot in Benchmark~I at $d=3$ and $f=0.5\,\%$ that the live stream did not route, and live that shot kept the network's answer, which is correct, so this configuration fails 353 times live against 354 simulated.
Two more, Benchmark~II at $d=3$ with $f=2$ and 20\,\%, routed one shot fewer live without changing an answer, and the failures agree in every other configuration on the EPYC cores.
On the DGX Spark six configurations at $d=7$ have more failures live, given here live against simulated.
Five have one more: in Benchmark~I the one with $f=20\,\%$, 8 against 7, and in Benchmark~II those with $f=1$, 2, 5 and 10\,\%, 5 against 4, 5 against 4, 4 against 3 and 5 against 4, the 2\,\% row being the operating point.
The configuration of Benchmark~II with $f=20\,\%$ and a cap of 0.001\,s per layer has 28 against 11 at 118.8 microseconds per layer, as 21 answers turned into failures and 4 into correct ones, so its simulated accuracy does not hold at its measured speed.
We compare the time with two simulations of the same batches and $N$, both with the GPU times measured in the run.
The replay of the live trial times gives the measured wall time within 0.8\,\% at the median and 22\,\% at most on the EPYC cores and within 0.4\,\% and 11\,\% on the DGX Spark, the largest gaps at $d=3$, where per-batch overheads that the simulator leaves out weigh most at a few microseconds per layer.
Against the simulation from the recorded trial times, the live trial times, converted to the platform by the median ratio of paired uncapped trials, give 0.866 to 0.991 times the simulated time on the EPYC cores, below 1 in all 42 configurations that route shots with a median of 0.94, so there the converted simulation is slightly conservative.
On the DGX Spark they give 0.863 to 1.140 times, with a median of 0.99 and above 1 in 14 of the 42 configurations, among them the first test's operating points of Benchmark~I at $d=7$ and 9 (1.140 and 1.068) and of Benchmark~II at $d=3$ and 7 (1.097 and 1.086), so there the agreement is approximate, with gaps of either sign.
An earlier check on 24 cores of the same EPYC model ran the first test's Grace operating points of $d=5$ and 7 at 0.955 to 0.959 times the simulated time and changed no answer.
None of these checks runs on a Grace CPU.

\begin{table*}[t]
\caption{The operating points: for each benchmark, distance and platform, the smallest $f$ whose development failures lie within 1\,\% of the fewest development failures of any $f$ (best row, with its $f$), at its development-best cap, with its failures and time per layer on the fresh test beside those of $f=20\,\%$ at its own development-best cap. Times in the simulated paced stream at batch 256.}
\label{tab:value}
\centering\footnotesize\setlength{\tabcolsep}{3.5pt}
\input{tables/tab_app_value}
\end{table*}

\begin{table*}[t]
\caption{Every routed fraction on the fresh test set, Benchmark~I, with new circuits and noise samples. Each $f$ at the cap per layer in seconds (cap) that was best on the development shots for that $f$ on each platform, all frozen before this test, where $f=0$ is the network alone. Failures count all 24 points, 43\,833\,600 test shots of a distance, while the ratios are geometric means over the 18 points at $p=0.3\,\%$ of the paired per-point failure ratios, with 95\,\% intervals, on the Grace platform. Times ($\mu$s) are in microseconds per layer in the paced stream at batch 256, simulated from the measured network times and the trial times recorded on these test shots. A circle marks the operating point of each platform and a star that of the first test.}
\label{tab:rt1}
\centering\footnotesize\setlength{\tabcolsep}{3.2pt}
\input{tables/tab_rt_mr1}
\end{table*}

\begin{table*}[t]
\caption{Every routed fraction on the fresh test set, Benchmark~II, on fresh noise samples of its circuits, as Table~\ref{tab:rt1}. Failures count the target failures on all test shots, measure (a) uses every test shot, and the ratios divide equal-weight sums of per-depth failure rates on all test shots, on the Grace platform. The dagger marks the configuration whose live check on development shots failed 28 times against 11 in the simulation.}
\label{tab:rt2}
\centering\footnotesize\setlength{\tabcolsep}{3.2pt}
\input{tables/tab_rt_mr2}
\end{table*}

\begin{table*}[t]
\caption{Live paced stream on development shots at the operating points of the main text (main) and of the first test, on each host with that platform's caps. Time per layer in microseconds, averaged with equal weight over the points. Live / sim divides the time of the stream from the live trial times, converted to the platform's CPU through paired uncapped trials, by that from the recorded trial times for the same batches and $N$. Failures live / simulated.}
\label{tab:live}
\centering\footnotesize\setlength{\tabcolsep}{3.5pt}
\input{tables/tab_app_live}
\end{table*}

\begin{table*}[t]
\caption{The first test's operating points. Development failures of the composition and of the network alone, development shots and routed shots, the time per layer in the paced stream simulated on development shots at batch 256 with the 99th percentile of the batch delay, and the number of the 56 candidates below the time budget.}
\label{tab:choices}
\centering\footnotesize\setlength{\tabcolsep}{3.5pt}
\input{tables/tab_app_choices}
\end{table*}

\section{The Cain et al.\ reimplementation}
\label{app:cain}
The reimplementation treats every reliable product as one logical Pauli product and decodes it with minimum-weight matching on the checks along the product's back-propagation path.
The paths computed from the circuit equal the detector tags of the authors' circuits for all 10 products of their depth-14 and depth-18 circuits at distances 5, 9 and 13.
Over 44 points of their published error rates per layer in their own circuits and noise model, with depths 14 and 18 and distances 5 to 17, our rate is 0.78 to 1.03 times theirs, with a median of 0.94.
We decompose two-qubit errors next to a transversal CNOT into edges, which their matching argument does not cover.
On our circuits, the reimplementation and LoMatching use the same matching subgraphs detector for detector, and their corrections differ only on a few shots in $10^4$ on circuits with S.

%% file: tables/tab_app_mr2_depth.tex
\begin{tabular}{@{}ccrlrrrrrrrrl@{}}
\toprule
& & \multicolumn{5}{c}{every test shot, measure (a)} & \multicolumn{6}{c}{certified prefix, measure (b)} \\
\cmidrule(lr){3-7}\cmidrule(lr){8-13}
& & & & & \multicolumn{2}{c}{composition} & & & \multicolumn{2}{c}{composition} & exact & \\
\cmidrule(lr){6-7}\cmidrule(lr){10-11}
$d$ & $D$ & shots & their $p_D$ & network & Grace & Spark & shots & network & Grace & Spark & MLE & their $q_D$ \\
\midrule
3 & 1 & 164\,000 & 0.003281 {\tiny$\pm$0.00036} & 373 & 387 & 387 & 164\,000 & 373 & 387 & 387 & 387 & 0.002533 {\tiny$\pm$0.00038} \\
 & 2 & 24\,000 & 0.005508 {\tiny$\pm$0.00046} & 162 & 146 & 146 & 24\,000 & 162 & 146 & 146 & 145 & 0.00446 {\tiny$\pm$0.0005} \\
 & 3 & 26\,000 & 0.008907 {\tiny$\pm$0.00059} & 247 & 242 & 242 & 26\,000 & 247 & 242 & 242 & 241 & 0.008501 {\tiny$\pm$0.00069} \\
 & 4 & 22\,000 & 0.01125 {\tiny$\pm$0.00066} & 313 & 315 & 315 & 22\,000 & 313 & 315 & 315 & 315 & 0.01204 {\tiny$\pm$0.00082} \\
 & 5 & 13\,400 & 0.01777 {\tiny$\pm$0.00082} & 247 & 235 & 235 & 13\,400 & 247 & 235 & 235 & 235 & 0.01526 {\tiny$\pm$0.00093} \\
 & 6 & 10\,000 & 0.02653 {\tiny$\pm$0.001} & 218 & 213 & 213 & 9400 & 204 & 199 & 199 & 200 & 0.02282 {\tiny$\pm$0.0011} \\
 & 7 & 10\,000 & 0.03039 {\tiny$\pm$0.0011} & 287 & 279 & 279 & 4400 & 118 & 111 & 111 & 113 & 0.02534 {\tiny$\pm$0.0012} \\
 & 8 & 10\,000 & 0.0341 {\tiny$\pm$0.0011} & 341 & 317 & 317 & 7600 & 251 & 237 & 237 & 234 & 0.02626 {\tiny$\pm$0.0012} \\
 & 9 & 10\,000 & 0.04864 {\tiny$\pm$0.0013} & 398 & 360 & 360 & 4200 & 175 & 160 & 160 & 158 & 0.04045 {\tiny$\pm$0.0015} \\
 & 10 & 10\,000 & 0.04699 {\tiny$\pm$0.0013} & 375 & 349 & 349 & 5000 & 184 & 170 & 170 & 168 & 0.03919 {\tiny$\pm$0.0015} \\
 & all & 299\,400 & $E=3254.0$ & 2961 & 2843 & 2843 & 280\,000 & 2274 & 2202 & 2202 & 2196 &  \\
\midrule
5 & 1 & 660\,000 & 0.0003516 {\tiny$\pm$0.00012} & 160 & 137 & 137 & 660\,000 & 160 & 137 & 137 & 137 & 0.0003377 {\tiny$\pm$0.00014} \\
 & 2 & 260\,000 & 0.0005469 {\tiny$\pm$0.00015} & 253 & 190 & 190 & 260\,000 & 253 & 190 & 190 & 191 & 0.0007905 {\tiny$\pm$0.00021} \\
 & 3 & 152\,000 & 0.001367 {\tiny$\pm$0.00023} & 208 & 149 & 149 & 152\,000 & 208 & 149 & 149 & 148 & 0.001133 {\tiny$\pm$0.00025} \\
 & 4 & 94\,000 & 0.001523 {\tiny$\pm$0.00024} & 226 & 168 & 169 & 94\,000 & 226 & 168 & 169 & 168 & 0.001476 {\tiny$\pm$0.00029} \\
 & 5 & 92\,000 & 0.002734 {\tiny$\pm$0.00033} & 217 & 159 & 159 & 92\,000 & 217 & 159 & 159 & 158 & 0.002344 {\tiny$\pm$0.00037} \\
 & 6 & 50\,000 & 0.004297 {\tiny$\pm$0.00041} & 179 & 132 & 132 & 49\,999 (1) & 179 & 132 & 132 & 131 & 0.002702 {\tiny$\pm$0.00039} \\
 & 7 & 60\,000 & 0.005274 {\tiny$\pm$0.00045} & 260 & 176 & 176 & 59\,999 (1) & 260 & 176 & 176 & 178 & 0.003926 {\tiny$\pm$0.00048} \\
 & 8 & 42\,000 & 0.005039 {\tiny$\pm$0.00044} & 216 & 148 & 148 & 41\,999 (1) & 216 & 148 & 148 & 147 & 0.003188 {\tiny$\pm$0.00043} \\
 & 9 & 22\,000 & 0.008203 {\tiny$\pm$0.00056} & 147 & 102 & 102 & 22\,000 & 147 & 102 & 102 & 102 & 0.006004 {\tiny$\pm$0.00059} \\
 & 10 & 32\,000 & 0.007969 {\tiny$\pm$0.00056} & 204 & 139 & 139 & 31\,999 (1) & 203 & 139 & 139 & 140 & 0.005887 {\tiny$\pm$0.00059} \\
 & all & 1\,464\,000 & $E=2155.1$ & 2070 & 1500 & 1501 & 1\,463\,996 (4) & 2069 & 1500 & 1501 & 1500 &  \\
\midrule
7 & 1 & 10\,600\,000 & 0.0001563 {\tiny$\pm$0.000078} & 309 & 164 & 165 & 979\,280 & 25 & 21 & 21 & 21 & 0.00005629$^{\ast}$ {\tiny$\pm$0.000056} \\
 & 2 & 1\,720\,000 & 0.00007813 {\tiny$\pm$0.000055} & 224 & 101 & 102 & 262\,279 (1) & 33 & 20 & 20 & 19 & 0.0001129 {\tiny$\pm$0.00008} \\
 & 3 & 1\,100\,000 & 0.0001953 {\tiny$\pm$0.000087} & 273 & 133 & 131 & 135\,879 (1) & 30 & 22 & 21 & 22 & 0.0001134 {\tiny$\pm$0.00008} \\
 & 4 & 920\,000 & 0.0002344 {\tiny$\pm$0.000096} & 396 & 144 & 144 & 79\,560 & 26 & 11 & 11 & 11 & 0.0001704 {\tiny$\pm$0.000098} \\
 & 5 & 740\,000 & 0.0003516 {\tiny$\pm$0.00012} & 396 & 168 & 175 & 55\,380 & 34 & 12 & 12 & 12 & 0.0002859 {\tiny$\pm$0.00013} \\
 & 6 & 420\,000 & 0.0007422 {\tiny$\pm$0.00017} & 228 & 88 & 87 & 48\,340 & 18 & 8 & 7 & 8 & 0.0004598 {\tiny$\pm$0.00016} \\
 & 7 & 260\,000 & 0.0008984 {\tiny$\pm$0.00019} & 224 & 91 & 96 & 29\,479 (1) & 27 & 9 & 10 & 9 & 0.0002887 {\tiny$\pm$0.00013} \\
 & 8 & 300\,000 & 0.0008594 {\tiny$\pm$0.00018} & 315 & 108 & 115 & 21\,520 & 24 & 5 & 5 & 4 & 0.0006376 {\tiny$\pm$0.00019} \\
 & 9 & 156\,000 & 0.001758 {\tiny$\pm$0.00026} & 210 & 79 & 84 & --- & --- & --- & --- & --- & --- \\
 & 10 & 240\,000 & 0.00168 {\tiny$\pm$0.00026} & 335 & 112 & 119 & --- & --- & --- & --- & --- & --- \\
 & all & 16\,456\,000 & $E=3962.4$ & 2910 & 1188 & 1218 & 1\,611\,717 (3) & 217 & 108 & 107 & 106 &  \\
\bottomrule
\end{tabular}

%% file: tables/tab_first_mr1.tex
\begin{tabular}{@{}cccccccccc@{}}
\toprule
& & & \multicolumn{2}{c}{against LoMatching} & \multicolumn{2}{c}{against Cain et al.} & \multicolumn{2}{c}{against \bplom} & composition \\
\cmidrule(lr){4-5}\cmidrule(lr){6-7}\cmidrule(lr){8-9}
$d$ & $f$ & $p$ (\%) & network & composition & network & composition & network & composition & / network \\
\midrule
3 & 10\,\% & 0.3 & 0.5645 & 0.5390 & 0.5605 & 0.5352 & 0.7387 & 0.7120 & 0.9550 \\
 &  &  & {\tiny[0.5585, 0.5710]} & {\tiny[0.5328, 0.5459]} & {\tiny[0.5544, 0.5671]} & {\tiny[0.5290, 0.5423]} & {\tiny[0.7174, 0.7595]} & {\tiny[0.6921, 0.7326]} & {\tiny[0.9505, 0.9594]} \\[2pt]
 &  & 0.1, 0.2 & 0.4148 & 0.3909 & 0.4141 & 0.3903 & 0.6866 & 0.6469 & 0.9424 \\
 &  &  & {\tiny[0.4032, 0.4274]} & {\tiny[0.3788, 0.4042]} & {\tiny[0.4025, 0.4267]} & {\tiny[0.3782, 0.4036]} & {\tiny[0.6436, 0.7294]} & {\tiny[0.6101, 0.6845]} & {\tiny[0.9310, 0.9538]} \\
\midrule
5 & 10\,\% & 0.3 & 0.2911 & 0.2027 & 0.2861 & 0.1992 & 0.5170 & 0.3831 & 0.6963 \\
 &  &  & {\tiny[0.2851, 0.2974]} & {\tiny[0.1980, 0.2076]} & {\tiny[0.2797, 0.2928]} & {\tiny[0.1944, 0.2043]} & {\tiny[0.4453, 0.5717]} & {\tiny[0.3384, 0.4192]} & {\tiny[0.6844, 0.7085]} \\[2pt]
 &  & 0.1, 0.2 & 0.1916 & 0.1161 & 0.1909 & 0.1157 & 0.3951 & 0.2208 & 0.6058 \\
 &  &  & {\tiny[0.1775, 0.2054]} & {\tiny[0.1049, 0.1269]} & {\tiny[0.1768, 0.2046]} & {\tiny[0.1046, 0.1264]} & {\tiny[0.2607, 0.5061]} & {\tiny[0.1391, 0.3042]} & {\tiny[0.5564, 0.6552]} \\
\midrule
7 & 5\,\% & 0.3 & 0.1808 & 0.0731 & 0.1759 & 0.0711 & 0.4778 & 0.2061 & 0.4042 \\
 &  &  & {\tiny[0.1700, 0.1903]} & {\tiny[0.0671, 0.0782]} & {\tiny[0.1653, 0.1855]} & {\tiny[0.0652, 0.0762]} & {\tiny[0.3946, 0.5680]} & {\tiny[0.1591, 0.2493]} & {\tiny[0.3755, 0.4302]} \\[2pt]
 &  & 0.1, 0.2 & 0.1099 & 0.0257 & 0.1077 & 0.0252 & 0.4800 & 0.2249 & 0.2309 \\
 &  &  & {\tiny[0.0719, 0.1426]} & {\tiny[0.0187, 0.0360]} & {\tiny[0.0703, 0.1396]} & {\tiny[0.0183, 0.0353]} & {\tiny[0.2993, 0.8327]} & {\tiny[0.1476, 0.3904]} & {\tiny[0.1666, 0.3394]} \\
\midrule
9 & 2\,\% & 0.3 & 0.1694 & 0.0248 & 0.1626 & 0.0238 & 0.4913 & 0.1389 & 0.1458 \\
 &  &  & {\tiny[0.1447, 0.1922]} & {\tiny[0.0185, 0.0300]} & {\tiny[0.1377, 0.1853]} & {\tiny[0.0177, 0.0286]} & {\tiny[0.4045, 0.6681]} & {\tiny[0.1124, 0.1970]} & {\tiny[0.1101, 0.1772]} \\[2pt]
 &  & 0.1, 0.2 & 0.1223 & 0.0456 & 0.1154 & 0.0430 & 0.7418 & 0.5810 & 0.3705 \\
 &  &  & {\tiny[0.0899, 0.2084]} & {\tiny[0.0291, 0.0793]} & {\tiny[0.0862, 0.1864]} & {\tiny[0.0277, 0.0711]} & {\tiny[0.4926, 1.3480]} & {\tiny[0.3865, 1.1225]} & {\tiny[0.2592, 0.4536]} \\
\bottomrule
\end{tabular}

%% file: tables/tab_first_mr2.tex
\begin{tabular}{@{}clccccccc@{}}
\toprule
& & \multicolumn{2}{c}{measure (a)} & \multicolumn{3}{c}{measure (b)} & & \\
\cmidrule(lr){3-4}\cmidrule(lr){5-7}
$d$ & system & failures / expected & ratio & ours / MLE & theirs / MLE & ratio & vs LoMatching & vs network \\
\midrule
3 & network alone & 2961 / 3254.0 & 0.910 & 1.064 & 1.185 & 0.898 & 0.578 & --- \\
 &  &  & {\tiny[0.838, 0.988]} & {\tiny[1.011, 1.119]} &  & {\tiny[0.841, 0.958]} & {\tiny[0.558, 0.599]} &  \\
 & composition, both & 2843 / 3254.0 & 0.874 & 1.004 & 1.185 & 0.847 & 0.546 & 0.945 \\
 &  &  & {\tiny[0.806, 0.948]} & {\tiny[0.988, 1.020]} &  & {\tiny[0.810, 0.885]} & {\tiny[0.527, 0.566]} & {\tiny[0.920, 0.971]} \\
\midrule
5 & network alone & 2070 / 2155.1 & 0.961 & 1.421 & 1.342 & 1.059 & 0.263 & --- \\
 &  &  & {\tiny[0.843, 1.095]} & {\tiny[1.322, 1.528]} &  & {\tiny[0.928, 1.207]} & {\tiny[0.250, 0.277]} &  \\
 & composition, Grace & 1500 / 2155.1 & 0.696 & 1.000 & 1.342 & 0.745 & 0.185 & 0.703 \\
 &  &  & {\tiny[0.607, 0.798]} & {\tiny[0.990, 1.010]} &  & {\tiny[0.667, 0.831]} & {\tiny[0.174, 0.197]} & {\tiny[0.666, 0.741]} \\
 & composition, Spark & 1501 / 2155.1 & 0.696 & 1.000 & 1.342 & 0.745 & 0.185 & 0.703 \\
 &  &  & {\tiny[0.608, 0.798]} & {\tiny[0.990, 1.011]} &  & {\tiny[0.667, 0.832]} & {\tiny[0.174, 0.197]} & {\tiny[0.666, 0.742]} \\
\midrule
7 & network alone & 2910 / 3962.4 & 0.734 & 2.932 & 1.654 & 1.772 & 0.169 & --- \\
 &  &  & {\tiny[0.476, 1.133]} & {\tiny[1.983, 4.335]} &  & {\tiny[1.026, 3.063]} & {\tiny[0.161, 0.177]} &  \\
 & composition, Grace & 1188 / 3962.4 & 0.300 & 1.040 & 1.654 & 0.628 & 0.064 & 0.376 \\
 &  &  & {\tiny[0.194, 0.464]} & {\tiny[0.944, 1.145]} &  & {\tiny[0.423, 0.932]} & {\tiny[0.059, 0.068]} & {\tiny[0.351, 0.403]} \\
 & composition, Spark & 1218 / 3962.4 & 0.307 & 1.044 & 1.654 & 0.631 & 0.066 & 0.393 \\
 &  &  & {\tiny[0.199, 0.476]} & {\tiny[0.919, 1.187]} &  & {\tiny[0.422, 0.945]} & {\tiny[0.062, 0.071]} & {\tiny[0.368, 0.419]} \\
\bottomrule
\end{tabular}

%% file: tables/tab_sweep.tex
\begin{tabular}{@{}cccrrcccc@{}}
\toprule
& & & \multicolumn{2}{c}{development} & \multicolumn{4}{c}{test} \\
\cmidrule(lr){4-5}\cmidrule(lr){6-9}
$d$ & $f$ & cap per layer & failures & $\mu$s per layer & (a) & (b) ours / MLE & (b) & vs network \\
\midrule
3 & 1\,\% & none & 6336 & 1.4 & 0.896 [0.827, 0.972] & 1.044 [0.998, 1.093] & 0.881 [0.828, 0.937] & 0.979 [0.965, 0.993] \\
 & 2\,\% & none & 6325 & 1.9 & 0.887 [0.819, 0.962] & 1.033 [0.994, 1.074] & 0.871 [0.823, 0.922] & 0.966 [0.948, 0.984] \\
 & 5\,\% & none & 6314 & 2.1 & 0.879 [0.810, 0.953] & 1.012 [0.987, 1.037] & 0.854 [0.813, 0.896] & 0.953 [0.930, 0.976] \\
 & 10\,\%$^{\ast}$ & none & 6300 & 2.6 & 0.874 [0.806, 0.948] & 1.004 [0.988, 1.020] & 0.847 [0.810, 0.885] & 0.945 [0.920, 0.971] \\
\midrule
5 & 2\,\% & 0.002\,s & 4588 & 4.4 & 0.712 [0.622, 0.815] & 1.030 [0.998, 1.063] & 0.767 [0.685, 0.860] & 0.724 [0.689, 0.761] \\
 & 2\,\% & 0.02\,s & 4552 & 8.3 & 0.706 [0.617, 0.809] & 1.024 [0.993, 1.056] & 0.762 [0.681, 0.854] & 0.719 [0.685, 0.756] \\
 & 5\,\% & 0.02\,s & 4490 & 16.8 & 0.698 [0.609, 0.800] & 1.004 [0.991, 1.018] & 0.748 [0.670, 0.835] & 0.706 [0.669, 0.744] \\
 & 10\,\%$^{\ast}$ & 0.02\,s & 4484 & 26.6 & 0.696 [0.607, 0.798] & 1.000 [0.990, 1.010] & 0.745 [0.667, 0.831] & 0.703 [0.666, 0.741] \\
\midrule
7 & 2\,\%$^{\dagger}$ & none & 372 & 50.7 & 0.307 [0.199, 0.476] & 1.044 [0.919, 1.187] & 0.631 [0.422, 0.945] & 0.393 [0.368, 0.419] \\
 & 2\,\% & 0.1\,s & 372 & 41.4 & 0.308 [0.199, 0.476] & 1.044 [0.919, 1.187] & 0.631 [0.422, 0.945] & 0.394 [0.369, 0.420] \\
 & 5\,\% & 0.1\,s & 364 & 78.7 & 0.301 [0.194, 0.466] & 1.023 [0.918, 1.140] & 0.619 [0.416, 0.921] & 0.378 [0.354, 0.405] \\
 & 10\,\% & 0.1\,s & 359 & 122.0 & 0.300 [0.194, 0.464] & 1.040 [0.944, 1.145] & 0.628 [0.423, 0.932] & 0.375 [0.351, 0.402] \\
 & 20\,\%$^{\ast}$ & 0.1\,s & 357 & 183.5 & 0.300 [0.194, 0.464] & 1.040 [0.944, 1.145] & 0.628 [0.423, 0.932] & 0.376 [0.351, 0.403] \\
\bottomrule
\end{tabular}

%% file: tables/tab_app_counts.tex
\begin{tabular}{@{}ccrrrrrrr@{}}
\toprule
$d$ & $p$ (\%) & shots & network & composition, Grace & composition, Spark & LoMatching & Cain et al. & \bplom (matched shots) \\
\midrule
3 & 0.3 & 37\,689\,600 & 106\,557 & 101\,977 & 101\,974 & 191\,663 & 192\,696 & 17\,393 of 1\,152\,000 \\
 & 0.1, 0.2 & 6\,144\,000 & 20\,283 & 19\,088 & 19\,088 & 46\,198 & 46\,267 & 1722 of 384\,000 \\
5 & 0.3 & 37\,689\,600 & 15\,445 & 10\,644 & 10\,644 & 58\,239 & 58\,836 & 4244 of 1\,152\,000 \\
 & 0.1, 0.2 & 6\,144\,000 & 1939 & 1220 & 1220 & 9548 & 9585 & 260 of 384\,000 \\
7 & 0.3 & 37\,689\,600 & 2609 & 975 & 975 & 17\,393 & 17\,645 & 1018 of 1\,152\,000 \\
 & 0.1, 0.2 & 6\,144\,000 & 231 & 74 & 74 & 1965 & 1984 & 38 of 384\,000 \\
9 & 0.3 & 37\,689\,600 & 714 & 110 & 110 & 6171 & 6344 & 435 of 1\,152\,000 \\
 & 0.1, 0.2 & 6\,144\,000 & 43 & 8 & 8 & 488 & 515 & 9 of 384\,000 \\
\bottomrule
\end{tabular}

%% file: tables/tab_app_k.tex
\begin{tabular}{@{}clccc@{}}
\toprule
$d$ & ratio & $K=2$ & $K=3$ & $K=4$ \\
\midrule
3 & network / LoMatching & 0.5588 [0.5478, 0.5707] & 0.5790 [0.5675, 0.5906] & 0.5558 [0.5468, 0.5654] \\
 & composition / LoMatching & 0.5351 [0.5236, 0.5475] & 0.5502 [0.5380, 0.5624] & 0.5320 [0.5222, 0.5422] \\
 & network / BP+LoMatching & 0.7614 [0.7213, 0.8007] & 0.7768 [0.7387, 0.8139] & 0.6815 [0.6491, 0.7125] \\
 & composition / BP+LoMatching & 0.7207 [0.6804, 0.7628] & 0.7518 [0.7159, 0.7889] & 0.6663 [0.6378, 0.6960] \\
 & composition / network & 0.9576 [0.9495, 0.9660] & 0.9501 [0.9427, 0.9578] & 0.9571 [0.9499, 0.9646] \\
\midrule
5 & network / LoMatching & 0.2913 [0.2805, 0.3023] & 0.3121 [0.3015, 0.3230] & 0.2715 [0.2613, 0.2823] \\
 & composition / LoMatching & 0.1997 [0.1918, 0.2081] & 0.2167 [0.2075, 0.2260] & 0.1926 [0.1850, 0.2006] \\
 & network / BP+LoMatching & 0.4992 [0.3925, 0.5757] & 0.4973 [0.3584, 0.5961] & 0.5566 [0.4693, 0.6921] \\
 & composition / BP+LoMatching & 0.3858 [0.3277, 0.4424] & 0.3848 [0.2993, 0.4586] & 0.3788 [0.3078, 0.4552] \\
 & composition / network & 0.6855 [0.6641, 0.7082] & 0.6943 [0.6738, 0.7153] & 0.7093 [0.6890, 0.7302] \\
\midrule
7 & network / LoMatching & 0.1769 [0.1571, 0.1967] & 0.2045 [0.1851, 0.2225] & 0.1633 [0.1488, 0.1780] \\
 & composition / LoMatching & 0.0761 [0.0651, 0.0867] & 0.0749 [0.0646, 0.0847] & 0.0685 [0.0604, 0.0764] \\
 & network / BP+LoMatching & 0.4526 [0.3191, 0.6299] & 0.4912 [0.3599, 0.6884] & 0.4908 [0.3509, 0.5958] \\
 & composition / BP+LoMatching & 0.2524 [0.1657, 0.3689] & 0.1817 [0.1145, 0.2659] & 0.1908 [0.1271, 0.2512] \\
 & composition / network & 0.4301 [0.3774, 0.4844] & 0.3660 [0.3200, 0.4131] & 0.4193 [0.3766, 0.4621] \\
\midrule
9 & network / LoMatching & 0.1604 [0.1228, 0.2026] & 0.1995 [0.1521, 0.2384] & 0.1520 [0.1145, 0.1932] \\
 & composition / LoMatching & 0.0184 [0.0111, 0.0269] & 0.0298 [0.0181, 0.0441] & 0.0278 [0.0177, 0.0373] \\
 & network / BP+LoMatching & 0.3858 [0.2556, 0.6385] & 0.5268 [0.3883, 0.8152] & 0.5835 [0.3840, 0.9706] \\
 & composition / BP+LoMatching & 0.0761 [0.0577, 0.1336] & 0.2014 [0.1266, 0.3354] & 0.1747 [0.1087, 0.3188] \\
 & composition / network & 0.1140 [0.0708, 0.1668] & 0.1487 [0.0930, 0.2245] & 0.1828 [0.1193, 0.2476] \\
\bottomrule
\end{tabular}

%% file: tables/tab_app_pt_counts_a.tex
\begin{tabular}{@{}ccccrrrrrrrrr@{}}
\toprule
& & & & & & & & & & \multicolumn{3}{c}{matched shots of \bplom} \\
\cmidrule(lr){11-13}
$d$ & $K$ & $L$ & $p$ (\%) & shots & routed & network & composition & LoMatching & Cain et al. & \bplom & network & composition \\
\midrule
3 & 2 & 1 & 0.3 & 6\,400\,000 & 543\,457 & 3907 & 3986 & 5050 & 5267 & 45 & 38 & 38 \\
 & 2 & 2 & 0.3 & 3\,840\,000 & 323\,932 & 6773 & 6562 & 12\,068 & 12\,169 & 140 & 119 & 111 \\
 & 2 & 4 & 0.3 & 1\,344\,000 & 111\,915 & 5314 & 5222 & 9949 & 9992 & 337 & 245 & 245 \\
 & 2 & 8 & 0.3 & 512\,000 & 43\,612 & 5054 & 4630 & 10\,170 & 10\,188 & 840 & 632 & 547 \\
 & 2 & 12 & 0.3 & 288\,000 & 20\,731 & 4169 & 3888 & 7998 & 8017 & 1302 & 907 & 876 \\
 & 2 & 16 & 0.3 & 179\,200 & 13\,639 & 3806 & 3537 & 7508 & 7509 & 1921 & 1368 & 1262 \\
 & 3 & 1 & 0.3 & 6\,400\,000 & 657\,961 & 4720 & 4687 & 5951 & 6119 & 56 & 47 & 53 \\
 & 3 & 2 & 0.3 & 3\,840\,000 & 351\,035 & 7038 & 6926 & 11\,228 & 11\,239 & 146 & 118 & 113 \\
 & 3 & 4 & 0.3 & 1\,344\,000 & 145\,831 & 6870 & 6556 & 12\,884 & 12\,952 & 423 & 324 & 313 \\
 & 3 & 8 & 0.3 & 512\,000 & 55\,911 & 6283 & 5958 & 12\,050 & 12\,075 & 1038 & 800 & 733 \\
 & 3 & 12 & 0.3 & 288\,000 & 28\,290 & 5277 & 4781 & 10\,117 & 10\,134 & 1547 & 1183 & 1104 \\
 & 3 & 16 & 0.3 & 179\,200 & 17\,285 & 4575 & 4201 & 8751 & 8758 & 2224 & 1596 & 1470 \\
 & 4 & 1 & 0.3 & 6\,400\,000 & 787\,096 & 5682 & 5730 & 7412 & 7529 & 77 & 52 & 57 \\
 & 4 & 2 & 0.3 & 3\,840\,000 & 481\,032 & 9966 & 9854 & 17\,016 & 17\,119 & 229 & 151 & 154 \\
 & 4 & 4 & 0.3 & 1\,344\,000 & 158\,199 & 7754 & 7466 & 14\,885 & 14\,973 & 512 & 353 & 346 \\
 & 4 & 8 & 0.3 & 512\,000 & 63\,485 & 6759 & 6383 & 13\,513 & 13\,524 & 1236 & 812 & 772 \\
 & 4 & 12 & 0.3 & 288\,000 & 33\,916 & 6147 & 5708 & 12\,275 & 12\,284 & 2048 & 1427 & 1329 \\
 & 4 & 16 & 0.3 & 179\,200 & 24\,341 & 6463 & 5902 & 12\,838 & 12\,848 & 3272 & 2333 & 2099 \\
 & 2 & 8 & 0.1 & 1\,024\,000 & 89\,089 & 1170 & 1123 & 3271 & 3277 & 102 & 66 & 63 \\
 & 2 & 8 & 0.2 & 1\,024\,000 & 87\,371 & 4581 & 4251 & 10\,336 & 10\,349 & 389 & 263 & 248 \\
 & 3 & 8 & 0.1 & 1\,024\,000 & 110\,354 & 1414 & 1347 & 3791 & 3800 & 132 & 88 & 86 \\
 & 3 & 8 & 0.2 & 1\,024\,000 & 110\,750 & 5464 & 5220 & 11\,945 & 11\,967 & 460 & 330 & 308 \\
 & 4 & 8 & 0.1 & 1\,024\,000 & 117\,370 & 1522 & 1404 & 3795 & 3800 & 131 & 96 & 86 \\
 & 4 & 8 & 0.2 & 1\,024\,000 & 123\,727 & 6132 & 5743 & 13\,060 & 13\,074 & 508 & 347 & 330 \\
\midrule
5 & 2 & 1 & 0.3 & 6\,400\,000 & 577\,092 & 367 & 301 & 647 & 709 & 6 & 3 & 6 \\
 & 2 & 2 & 0.3 & 3\,840\,000 & 307\,839 & 920 & 701 & 2835 & 2891 & 38 & 19 & 17 \\
 & 2 & 4 & 0.3 & 1\,344\,000 & 107\,476 & 772 & 552 & 2886 & 2960 & 71 & 38 & 23 \\
 & 2 & 8 & 0.3 & 512\,000 & 45\,717 & 817 & 492 & 3724 & 3744 & 218 & 103 & 54 \\
 & 2 & 12 & 0.3 & 288\,000 & 19\,068 & 643 & 415 & 2766 & 2785 & 289 & 137 & 88 \\
 & 2 & 16 & 0.3 & 179\,200 & 12\,720 & 681 & 407 & 2798 & 2800 & 484 & 250 & 146 \\
 & 3 & 1 & 0.3 & 6\,400\,000 & 681\,820 & 423 & 393 & 753 & 811 & 8 & 3 & 5 \\
 & 3 & 2 & 0.3 & 3\,840\,000 & 327\,829 & 892 & 689 & 2437 & 2455 & 27 & 12 & 9 \\
 & 3 & 4 & 0.3 & 1\,344\,000 & 147\,645 & 1030 & 706 & 3776 & 3816 & 96 & 56 & 38 \\
 & 3 & 8 & 0.3 & 512\,000 & 59\,539 & 1066 & 667 & 4239 & 4252 & 253 & 137 & 92 \\
 & 3 & 12 & 0.3 & 288\,000 & 28\,658 & 979 & 559 & 3664 & 3676 & 383 & 209 & 129 \\
 & 3 & 16 & 0.3 & 179\,200 & 17\,716 & 813 & 518 & 3317 & 3320 & 579 & 305 & 186 \\
 & 4 & 1 & 0.3 & 6\,400\,000 & 875\,649 & 517 & 467 & 961 & 986 & 6 & 7 & 4 \\
 & 4 & 2 & 0.3 & 3\,840\,000 & 455\,998 & 1210 & 988 & 3716 & 3773 & 39 & 25 & 19 \\
 & 4 & 4 & 0.3 & 1\,344\,000 & 171\,442 & 1102 & 795 & 4880 & 4973 & 108 & 44 & 36 \\
 & 4 & 8 & 0.3 & 512\,000 & 63\,630 & 1074 & 694 & 4938 & 4955 & 310 & 144 & 101 \\
 & 4 & 12 & 0.3 & 288\,000 & 32\,804 & 1010 & 647 & 4908 & 4915 & 481 & 207 & 146 \\
 & 4 & 16 & 0.3 & 179\,200 & 24\,219 & 1129 & 653 & 4994 & 5015 & 848 & 414 & 234 \\
 & 2 & 8 & 0.1 & 1\,024\,000 & 91\,754 & 65 & 42 & 358 & 360 & 12 & 2 & 2 \\
 & 2 & 8 & 0.2 & 1\,024\,000 & 90\,550 & 480 & 279 & 2393 & 2402 & 64 & 35 & 18 \\
 & 3 & 8 & 0.1 & 1\,024\,000 & 117\,669 & 69 & 37 & 405 & 403 & 9 & 5 & 1 \\
 & 3 & 8 & 0.2 & 1\,024\,000 & 120\,120 & 636 & 407 & 2753 & 2768 & 77 & 27 & 21 \\
 & 4 & 8 & 0.1 & 1\,024\,000 & 126\,663 & 86 & 49 & 475 & 479 & 11 & 5 & 3 \\
 & 4 & 8 & 0.2 & 1\,024\,000 & 130\,082 & 603 & 406 & 3164 & 3173 & 87 & 41 & 26 \\
\bottomrule
\end{tabular}

%% file: tables/tab_app_pt_counts_b.tex
\begin{tabular}{@{}ccccrrrrrrrrr@{}}
\toprule
& & & & & & & & & & \multicolumn{3}{c}{matched shots of \bplom} \\
\cmidrule(lr){11-13}
$d$ & $K$ & $L$ & $p$ (\%) & shots & routed & network & composition & LoMatching & Cain et al. & \bplom & network & composition \\
\midrule
7 & 2 & 1 & 0.3 & 6\,400\,000 & 268\,454 & 29 & 28 & 78 & 91 & 0 & 0 & 0 \\
 & 2 & 2 & 0.3 & 3\,840\,000 & 150\,278 & 115 & 56 & 598 & 631 & 3 & 1 & 1 \\
 & 2 & 4 & 0.3 & 1\,344\,000 & 49\,915 & 119 & 52 & 701 & 724 & 9 & 4 & 2 \\
 & 2 & 8 & 0.3 & 512\,000 & 23\,147 & 178 & 49 & 1237 & 1265 & 50 & 22 & 7 \\
 & 2 & 12 & 0.3 & 288\,000 & 7782 & 146 & 47 & 986 & 1007 & 81 & 28 & 14 \\
 & 2 & 16 & 0.3 & 179\,200 & 5626 & 115 & 40 & 970 & 971 & 97 & 37 & 14 \\
 & 3 & 1 & 0.3 & 6\,400\,000 & 329\,646 & 35 & 19 & 84 & 90 & 1 & 0 & 0 \\
 & 3 & 2 & 0.3 & 3\,840\,000 & 156\,684 & 116 & 57 & 446 & 457 & 8 & 5 & 2 \\
 & 3 & 4 & 0.3 & 1\,344\,000 & 77\,527 & 192 & 72 & 989 & 1010 & 13 & 9 & 3 \\
 & 3 & 8 & 0.3 & 512\,000 & 30\,638 & 205 & 60 & 1300 & 1305 & 50 & 32 & 9 \\
 & 3 & 12 & 0.3 & 288\,000 & 13\,136 & 187 & 51 & 1205 & 1208 & 95 & 39 & 7 \\
 & 3 & 16 & 0.3 & 179\,200 & 8873 & 176 & 53 & 1238 & 1235 & 170 & 63 & 24 \\
 & 4 & 1 & 0.3 & 6\,400\,000 & 385\,251 & 41 & 29 & 95 & 96 & 0 & 0 & 0 \\
 & 4 & 2 & 0.3 & 3\,840\,000 & 214\,412 & 166 & 89 & 706 & 735 & 8 & 4 & 4 \\
 & 4 & 4 & 0.3 & 1\,344\,000 & 87\,003 & 174 & 63 & 1399 & 1446 & 15 & 10 & 1 \\
 & 4 & 8 & 0.3 & 512\,000 & 34\,945 & 199 & 81 & 1671 & 1676 & 69 & 26 & 12 \\
 & 4 & 12 & 0.3 & 288\,000 & 16\,660 & 186 & 64 & 1723 & 1731 & 111 & 35 & 11 \\
 & 4 & 16 & 0.3 & 179\,200 & 13\,364 & 230 & 65 & 1967 & 1967 & 238 & 84 & 20 \\
 & 2 & 8 & 0.1 & 1\,024\,000 & 43\,043 & 5 & 0 & 49 & 54 & 1 & 0 & 0 \\
 & 2 & 8 & 0.2 & 1\,024\,000 & 44\,490 & 68 & 19 & 565 & 576 & 13 & 9 & 1 \\
 & 3 & 8 & 0.1 & 1\,024\,000 & 61\,191 & 7 & 1 & 43 & 43 & 0 & 0 & 0 \\
 & 3 & 8 & 0.2 & 1\,024\,000 & 62\,861 & 68 & 22 & 576 & 578 & 11 & 1 & 2 \\
 & 4 & 8 & 0.1 & 1\,024\,000 & 76\,806 & 3 & 1 & 47 & 47 & 1 & 1 & 0 \\
 & 4 & 8 & 0.2 & 1\,024\,000 & 75\,338 & 80 & 31 & 685 & 686 & 12 & 7 & 1 \\
\midrule
9 & 2 & 1 & 0.3 & 6\,400\,000 & 119\,957 & 4 & 0 & 5 & 5 & 0 & 0 & 0 \\
 & 2 & 2 & 0.3 & 3\,840\,000 & 55\,721 & 24 & 9 & 106 & 114 & 1 & 0 & 0 \\
 & 2 & 4 & 0.3 & 1\,344\,000 & 18\,770 & 21 & 2 & 191 & 200 & 3 & 2 & 0 \\
 & 2 & 8 & 0.3 & 512\,000 & 9931 & 41 & 6 & 444 & 487 & 23 & 4 & 0 \\
 & 2 & 12 & 0.3 & 288\,000 & 2550 & 37 & 6 & 397 & 467 & 23 & 6 & 0 \\
 & 2 & 16 & 0.3 & 179\,200 & 1905 & 43 & 1 & 432 & 432 & 55 & 18 & 0 \\
 & 3 & 1 & 0.3 & 6\,400\,000 & 137\,973 & 3 & 0 & 4 & 5 & 0 & 0 & 0 \\
 & 3 & 2 & 0.3 & 3\,840\,000 & 52\,659 & 16 & 3 & 83 & 85 & 0 & 0 & 0 \\
 & 3 & 4 & 0.3 & 1\,344\,000 & 34\,321 & 49 & 5 & 263 & 267 & 1 & 0 & 0 \\
 & 3 & 8 & 0.3 & 512\,000 & 12\,994 & 72 & 14 & 412 & 415 & 14 & 12 & 3 \\
 & 3 & 12 & 0.3 & 288\,000 & 4822 & 60 & 12 & 502 & 500 & 31 & 10 & 2 \\
 & 3 & 16 & 0.3 & 179\,200 & 3277 & 59 & 6 & 527 & 527 & 69 & 16 & 1 \\
 & 4 & 1 & 0.3 & 6\,400\,000 & 159\,723 & 6 & 2 & 9 & 9 & 0 & 0 & 0 \\
 & 4 & 2 & 0.3 & 3\,840\,000 & 75\,312 & 33 & 8 & 149 & 155 & 0 & 0 & 0 \\
 & 4 & 4 & 0.3 & 1\,344\,000 & 35\,004 & 57 & 8 & 406 & 426 & 3 & 3 & 0 \\
 & 4 & 8 & 0.3 & 512\,000 & 14\,516 & 40 & 6 & 489 & 492 & 7 & 5 & 1 \\
 & 4 & 12 & 0.3 & 288\,000 & 6662 & 66 & 10 & 755 & 759 & 59 & 17 & 3 \\
 & 4 & 16 & 0.3 & 179\,200 & 5786 & 83 & 12 & 997 & 999 & 146 & 28 & 4 \\
 & 2 & 8 & 0.1 & 1\,024\,000 & 16\,866 & 0 & 0 & 17 & 22 & 0 & 0 & 0 \\
 & 2 & 8 & 0.2 & 1\,024\,000 & 18\,442 & 15 & 0 & 217 & 239 & 6 & 2 & 0 \\
 & 3 & 8 & 0.1 & 1\,024\,000 & 23\,988 & 0 & 0 & 0 & 0 & 0 & 0 & 0 \\
 & 3 & 8 & 0.2 & 1\,024\,000 & 25\,526 & 11 & 3 & 122 & 122 & 1 & 1 & 1 \\
 & 4 & 8 & 0.1 & 1\,024\,000 & 32\,898 & 0 & 0 & 3 & 3 & 0 & 0 & 0 \\
 & 4 & 8 & 0.2 & 1\,024\,000 & 30\,836 & 17 & 5 & 129 & 129 & 2 & 1 & 1 \\
\bottomrule
\end{tabular}

%% file: tables/tab_app_pt_ratios_a.tex
\begin{tabular}{@{}cccclllll@{}}
\toprule
& & & & \multicolumn{2}{c}{against LoMatching} & \multicolumn{2}{c}{against \bplom} & composition \\
\cmidrule(lr){5-6}\cmidrule(lr){7-8}
$d$ & $K$ & $L$ & $p$ (\%) & network & composition & network & composition & / network \\
\midrule
3 & 2 & 1 & 0.3 & 0.7737 {\tiny[0.7568, 0.7928]} & 0.7893 {\tiny[0.7705, 0.8098]} & 0.8444 {\tiny[0.6458, 1.0638]} & 0.8444 {\tiny[0.6383, 1.1000]} & 1.0202 {\tiny[1.0010, 1.0402]} \\
 & 2 & 2 & 0.3 & 0.5612 {\tiny[0.5148, 0.6118]} & 0.5438 {\tiny[0.4936, 0.5986]} & 0.8500 {\tiny[0.7556, 0.9626]} & 0.7929 {\tiny[0.7017, 0.9055]} & 0.9688 {\tiny[0.9534, 0.9855]} \\
 & 2 & 4 & 0.3 & 0.5341 {\tiny[0.5084, 0.5646]} & 0.5249 {\tiny[0.4965, 0.5582]} & 0.7270 {\tiny[0.6548, 0.8064]} & 0.7270 {\tiny[0.6420, 0.8194]} & 0.9827 {\tiny[0.9664, 0.9992]} \\
 & 2 & 8 & 0.3 & 0.4970 {\tiny[0.4797, 0.5155]} & 0.4553 {\tiny[0.4363, 0.4752]} & 0.7524 {\tiny[0.7059, 0.8012]} & 0.6512 {\tiny[0.6036, 0.7000]} & 0.9161 {\tiny[0.8936, 0.9393]} \\
 & 2 & 12 & 0.3 & 0.5213 {\tiny[0.4999, 0.5432]} & 0.4861 {\tiny[0.4654, 0.5073]} & 0.6966 {\tiny[0.6469, 0.7462]} & 0.6728 {\tiny[0.6271, 0.7195]} & 0.9326 {\tiny[0.9068, 0.9591]} \\
 & 2 & 16 & 0.3 & 0.5069 {\tiny[0.4889, 0.5265]} & 0.4711 {\tiny[0.4552, 0.4881]} & 0.7121 {\tiny[0.6775, 0.7485]} & 0.6569 {\tiny[0.6252, 0.6897]} & 0.9293 {\tiny[0.9114, 0.9478]} \\
 & 3 & 1 & 0.3 & 0.7931 {\tiny[0.7748, 0.8133]} & 0.7876 {\tiny[0.7681, 0.8100]} & 0.8393 {\tiny[0.6667, 1.0217]} & 0.9464 {\tiny[0.7797, 1.1395]} & 0.9930 {\tiny[0.9755, 1.0115]} \\
 & 3 & 2 & 0.3 & 0.6268 {\tiny[0.5728, 0.6792]} & 0.6169 {\tiny[0.5585, 0.6732]} & 0.8082 {\tiny[0.7041, 0.9242]} & 0.7740 {\tiny[0.6536, 0.9160]} & 0.9841 {\tiny[0.9686, 1.0003]} \\
 & 3 & 4 & 0.3 & 0.5332 {\tiny[0.5054, 0.5644]} & 0.5088 {\tiny[0.4788, 0.5434]} & 0.7660 {\tiny[0.6930, 0.8488]} & 0.7400 {\tiny[0.6737, 0.8140]} & 0.9543 {\tiny[0.9388, 0.9710]} \\
 & 3 & 8 & 0.3 & 0.5214 {\tiny[0.5049, 0.5393]} & 0.4944 {\tiny[0.4764, 0.5136]} & 0.7707 {\tiny[0.7240, 0.8195]} & 0.7062 {\tiny[0.6664, 0.7478]} & 0.9483 {\tiny[0.9297, 0.9670]} \\
 & 3 & 12 & 0.3 & 0.5216 {\tiny[0.5059, 0.5371]} & 0.4726 {\tiny[0.4568, 0.4887]} & 0.7647 {\tiny[0.7225, 0.8089]} & 0.7136 {\tiny[0.6732, 0.7547]} & 0.9060 {\tiny[0.8883, 0.9253]} \\
 & 3 & 16 & 0.3 & 0.5228 {\tiny[0.5034, 0.5433]} & 0.4801 {\tiny[0.4642, 0.4976]} & 0.7176 {\tiny[0.6853, 0.7521]} & 0.6610 {\tiny[0.6322, 0.6923]} & 0.9183 {\tiny[0.8957, 0.9420]} \\
 & 4 & 1 & 0.3 & 0.7666 {\tiny[0.7495, 0.7854]} & 0.7731 {\tiny[0.7530, 0.7937]} & 0.6753 {\tiny[0.5361, 0.8272]} & 0.7403 {\tiny[0.6111, 0.9000]} & 1.0084 {\tiny[0.9888, 1.0273]} \\
 & 4 & 2 & 0.3 & 0.5857 {\tiny[0.5500, 0.6241]} & 0.5791 {\tiny[0.5391, 0.6226]} & 0.6594 {\tiny[0.5848, 0.7436]} & 0.6725 {\tiny[0.5927, 0.7562]} & 0.9888 {\tiny[0.9750, 1.0034]} \\
 & 4 & 4 & 0.3 & 0.5209 {\tiny[0.4960, 0.5498]} & 0.5016 {\tiny[0.4742, 0.5328]} & 0.6895 {\tiny[0.6419, 0.7386]} & 0.6758 {\tiny[0.6277, 0.7293]} & 0.9629 {\tiny[0.9443, 0.9814]} \\
 & 4 & 8 & 0.3 & 0.5002 {\tiny[0.4837, 0.5180]} & 0.4724 {\tiny[0.4559, 0.4901]} & 0.6570 {\tiny[0.6157, 0.7020]} & 0.6246 {\tiny[0.5880, 0.6646]} & 0.9444 {\tiny[0.9268, 0.9621]} \\
 & 4 & 12 & 0.3 & 0.5008 {\tiny[0.4843, 0.5182]} & 0.4650 {\tiny[0.4485, 0.4818]} & 0.6968 {\tiny[0.6594, 0.7363]} & 0.6489 {\tiny[0.6131, 0.6869]} & 0.9286 {\tiny[0.9136, 0.9438]} \\
 & 4 & 16 & 0.3 & 0.5034 {\tiny[0.4891, 0.5184]} & 0.4597 {\tiny[0.4465, 0.4736]} & 0.7130 {\tiny[0.6782, 0.7503]} & 0.6415 {\tiny[0.6154, 0.6692]} & 0.9132 {\tiny[0.8906, 0.9352]} \\
 & 2 & 8 & 0.1 & 0.3577 {\tiny[0.3211, 0.3985]} & 0.3433 {\tiny[0.3068, 0.3865]} & 0.6471 {\tiny[0.5413, 0.7647]} & 0.6176 {\tiny[0.5299, 0.7200]} & 0.9598 {\tiny[0.9208, 0.9983]} \\
 & 2 & 8 & 0.2 & 0.4432 {\tiny[0.4222, 0.4645]} & 0.4113 {\tiny[0.3877, 0.4368]} & 0.6761 {\tiny[0.6091, 0.7461]} & 0.6375 {\tiny[0.5802, 0.7014]} & 0.9280 {\tiny[0.9029, 0.9546]} \\
 & 3 & 8 & 0.1 & 0.3730 {\tiny[0.3398, 0.4099]} & 0.3553 {\tiny[0.3211, 0.3935]} & 0.6667 {\tiny[0.5630, 0.7810]} & 0.6515 {\tiny[0.5692, 0.7422]} & 0.9526 {\tiny[0.9235, 0.9824]} \\
 & 3 & 8 & 0.2 & 0.4574 {\tiny[0.4375, 0.4786]} & 0.4370 {\tiny[0.4150, 0.4610]} & 0.7174 {\tiny[0.6534, 0.7857]} & 0.6696 {\tiny[0.6037, 0.7408]} & 0.9553 {\tiny[0.9380, 0.9736]} \\
 & 4 & 8 & 0.1 & 0.4011 {\tiny[0.3748, 0.4315]} & 0.3700 {\tiny[0.3416, 0.4022]} & 0.7328 {\tiny[0.5714, 0.9077]} & 0.6565 {\tiny[0.5246, 0.8145]} & 0.9225 {\tiny[0.8894, 0.9546]} \\
 & 4 & 8 & 0.2 & 0.4695 {\tiny[0.4522, 0.4886]} & 0.4397 {\tiny[0.4218, 0.4600]} & 0.6831 {\tiny[0.6065, 0.7634]} & 0.6496 {\tiny[0.5857, 0.7168]} & 0.9366 {\tiny[0.9204, 0.9537]} \\
\midrule
5 & 2 & 1 & 0.3 & 0.5672 {\tiny[0.5253, 0.6161]} & 0.4652 {\tiny[0.4278, 0.5073]} & 0.5000 {\tiny[0.1429, 1.0000]} & 1.0000 {\tiny[0.5556, 2.0000]} & 0.8202 {\tiny[0.7644, 0.8780]} \\
 & 2 & 2 & 0.3 & 0.3245 {\tiny[0.2963, 0.3594]} & 0.2473 {\tiny[0.2237, 0.2801]} & 0.5000 {\tiny[0.3158, 0.7188]} & 0.4474 {\tiny[0.2667, 0.6452]} & 0.7620 {\tiny[0.7072, 0.8240]} \\
 & 2 & 4 & 0.3 & 0.2675 {\tiny[0.2404, 0.2955]} & 0.1913 {\tiny[0.1706, 0.2134]} & 0.5352 {\tiny[0.3929, 0.6949]} & 0.3239 {\tiny[0.2143, 0.4531]} & 0.7150 {\tiny[0.6647, 0.7705]} \\
 & 2 & 8 & 0.3 & 0.2194 {\tiny[0.1993, 0.2404]} & 0.1321 {\tiny[0.1202, 0.1444]} & 0.4725 {\tiny[0.3783, 0.5767]} & 0.2477 {\tiny[0.1852, 0.3128]} & 0.6022 {\tiny[0.5558, 0.6538]} \\
 & 2 & 12 & 0.3 & 0.2325 {\tiny[0.2099, 0.2550]} & 0.1500 {\tiny[0.1347, 0.1662]} & 0.4740 {\tiny[0.4005, 0.5564]} & 0.3045 {\tiny[0.2525, 0.3559]} & 0.6454 {\tiny[0.5932, 0.7024]} \\
 & 2 & 16 & 0.3 & 0.2434 {\tiny[0.2240, 0.2640]} & 0.1455 {\tiny[0.1328, 0.1581]} & 0.5165 {\tiny[0.4493, 0.5935]} & 0.3017 {\tiny[0.2597, 0.3466]} & 0.5977 {\tiny[0.5472, 0.6520]} \\
 & 3 & 1 & 0.3 & 0.5618 {\tiny[0.5203, 0.6047]} & 0.5219 {\tiny[0.4817, 0.5633]} & 0.3750 {\tiny[0.0588, 1.0000]} & 0.6250 {\tiny[0.1818, 1.5000]} & 0.9291 {\tiny[0.8697, 0.9887]} \\
 & 3 & 2 & 0.3 & 0.3660 {\tiny[0.3341, 0.4047]} & 0.2827 {\tiny[0.2506, 0.3228]} & 0.4444 {\tiny[0.2593, 0.6818]} & 0.3333 {\tiny[0.1429, 0.5556]} & 0.7724 {\tiny[0.7239, 0.8258]} \\
 & 3 & 4 & 0.3 & 0.2728 {\tiny[0.2459, 0.3006]} & 0.1870 {\tiny[0.1685, 0.2074]} & 0.5833 {\tiny[0.4526, 0.7246]} & 0.3958 {\tiny[0.2947, 0.5135]} & 0.6854 {\tiny[0.6357, 0.7385]} \\
 & 3 & 8 & 0.3 & 0.2515 {\tiny[0.2315, 0.2713]} & 0.1573 {\tiny[0.1428, 0.1729]} & 0.5415 {\tiny[0.4562, 0.6385]} & 0.3636 {\tiny[0.2925, 0.4424]} & 0.6257 {\tiny[0.5823, 0.6724]} \\
 & 3 & 12 & 0.3 & 0.2672 {\tiny[0.2497, 0.2855]} & 0.1526 {\tiny[0.1371, 0.1688]} & 0.5457 {\tiny[0.4787, 0.6162]} & 0.3368 {\tiny[0.2861, 0.3932]} & 0.5710 {\tiny[0.5233, 0.6193]} \\
 & 3 & 16 & 0.3 & 0.2451 {\tiny[0.2249, 0.2659]} & 0.1562 {\tiny[0.1390, 0.1740]} & 0.5268 {\tiny[0.4675, 0.5909]} & 0.3212 {\tiny[0.2689, 0.3768]} & 0.6371 {\tiny[0.5902, 0.6886]} \\
 & 4 & 1 & 0.3 & 0.5380 {\tiny[0.4973, 0.5799]} & 0.4860 {\tiny[0.4490, 0.5254]} & 1.1667 {\tiny[0.5000, 4.0000]} & 0.6667 {\tiny[0.2500, 1.6667]} & 0.9033 {\tiny[0.8585, 0.9527]} \\
 & 4 & 2 & 0.3 & 0.3256 {\tiny[0.2947, 0.3603]} & 0.2659 {\tiny[0.2391, 0.2972]} & 0.6410 {\tiny[0.4594, 0.8372]} & 0.4872 {\tiny[0.3000, 0.6944]} & 0.8165 {\tiny[0.7696, 0.8659]} \\
 & 4 & 4 & 0.3 & 0.2258 {\tiny[0.2044, 0.2515]} & 0.1629 {\tiny[0.1460, 0.1833]} & 0.4074 {\tiny[0.2881, 0.5421]} & 0.3333 {\tiny[0.2273, 0.4505]} & 0.7214 {\tiny[0.6770, 0.7663]} \\
 & 4 & 8 & 0.3 & 0.2175 {\tiny[0.1972, 0.2397]} & 0.1405 {\tiny[0.1278, 0.1547]} & 0.4645 {\tiny[0.3830, 0.5599]} & 0.3258 {\tiny[0.2669, 0.3931]} & 0.6462 {\tiny[0.5985, 0.6966]} \\
 & 4 & 12 & 0.3 & 0.2058 {\tiny[0.1872, 0.2267]} & 0.1318 {\tiny[0.1192, 0.1458]} & 0.4304 {\tiny[0.3694, 0.5000]} & 0.3035 {\tiny[0.2568, 0.3561]} & 0.6406 {\tiny[0.5813, 0.7012]} \\
 & 4 & 16 & 0.3 & 0.2261 {\tiny[0.2058, 0.2478]} & 0.1308 {\tiny[0.1189, 0.1441]} & 0.4882 {\tiny[0.4291, 0.5505]} & 0.2759 {\tiny[0.2398, 0.3145]} & 0.5784 {\tiny[0.5379, 0.6235]} \\
 & 2 & 8 & 0.1 & 0.1816 {\tiny[0.1421, 0.2282]} & 0.1173 {\tiny[0.0861, 0.1529]} & 0.1667 {\tiny[0.0345, 0.5000]} & 0.1667 {\tiny[0.0345, 0.5000]} & 0.6462 {\tiny[0.4844, 0.8254]} \\
 & 2 & 8 & 0.2 & 0.2006 {\tiny[0.1797, 0.2239]} & 0.1166 {\tiny[0.1020, 0.1328]} & 0.5469 {\tiny[0.4000, 0.7368]} & 0.2812 {\tiny[0.1695, 0.4068]} & 0.5813 {\tiny[0.5197, 0.6464]} \\
 & 3 & 8 & 0.1 & 0.1704 {\tiny[0.1359, 0.2080]} & 0.0914 {\tiny[0.0649, 0.1224]} & 0.5556 {\tiny[0.1429, 0.8889]} & 0.1111 {\tiny[0.0370, 0.4000]} & 0.5362 {\tiny[0.3968, 0.7031]} \\
 & 3 & 8 & 0.2 & 0.2310 {\tiny[0.2144, 0.2491]} & 0.1478 {\tiny[0.1321, 0.1649]} & 0.3506 {\tiny[0.2297, 0.4795]} & 0.2727 {\tiny[0.1571, 0.4091]} & 0.6399 {\tiny[0.5784, 0.7019]} \\
 & 4 & 8 & 0.1 & 0.1811 {\tiny[0.1392, 0.2265]} & 0.1032 {\tiny[0.0732, 0.1362]} & 0.4545 {\tiny[0.1250, 1.0000]} & 0.2727 {\tiny[0.0588, 0.5714]} & 0.5698 {\tiny[0.4396, 0.7222]} \\
 & 4 & 8 & 0.2 & 0.1906 {\tiny[0.1721, 0.2110]} & 0.1283 {\tiny[0.1145, 0.1438]} & 0.4713 {\tiny[0.3506, 0.5977]} & 0.2989 {\tiny[0.1915, 0.4157]} & 0.6733 {\tiny[0.6246, 0.7231]} \\
\bottomrule
\end{tabular}

%% file: tables/tab_app_pt_ratios_b.tex
\begin{tabular}{@{}cccclllll@{}}
\toprule
& & & & \multicolumn{2}{c}{against LoMatching} & \multicolumn{2}{c}{against \bplom} & composition \\
\cmidrule(lr){5-6}\cmidrule(lr){7-8}
$d$ & $K$ & $L$ & $p$ (\%) & network & composition & network & composition & / network \\
\midrule
7 & 2 & 1 & 0.3 & 0.3718 {\tiny[0.2162, 0.5676]} & 0.3590 {\tiny[0.2235, 0.5288]} & 1.0000$^{\mathrm{d}}$ {\tiny[1.0000, 1.0000]} & 1.0000$^{\mathrm{d}}$ {\tiny[1.0000, 1.0000]} & 0.9655 {\tiny[0.6785, 1.4375]} \\
 & 2 & 2 & 0.3 & 0.1923 {\tiny[0.1564, 0.2388]} & 0.0936 {\tiny[0.0656, 0.1277]} & 0.3333 {\tiny[0.0769, 1.0000]} & 0.3333 {\tiny[0.0769, 1.0000]} & 0.4870 {\tiny[0.3519, 0.6306]} \\
 & 2 & 4 & 0.3 & 0.1698 {\tiny[0.1339, 0.2132]} & 0.0742 {\tiny[0.0552, 0.0957]} & 0.4444 {\tiny[0.1000, 1.1250]} & 0.2222 {\tiny[0.0435, 0.6667]} & 0.4370 {\tiny[0.3333, 0.5600]} \\
 & 2 & 8 & 0.3 & 0.1439 {\tiny[0.1172, 0.1744]} & 0.0396 {\tiny[0.0276, 0.0532]} & 0.4400 {\tiny[0.2558, 0.6889]} & 0.1400 {\tiny[0.0492, 0.2558]} & 0.2753 {\tiny[0.1991, 0.3671]} \\
 & 2 & 12 & 0.3 & 0.1481 {\tiny[0.1212, 0.1782]} & 0.0477 {\tiny[0.0324, 0.0655]} & 0.3457 {\tiny[0.2267, 0.4804]} & 0.1728 {\tiny[0.0976, 0.2653]} & 0.3219 {\tiny[0.2340, 0.4161]} \\
 & 2 & 16 & 0.3 & 0.1186 {\tiny[0.0974, 0.1433]} & 0.0412 {\tiny[0.0277, 0.0577]} & 0.3814 {\tiny[0.2500, 0.5758]} & 0.1443 {\tiny[0.0702, 0.2472]} & 0.3478 {\tiny[0.2441, 0.4655]} \\
 & 3 & 1 & 0.3 & 0.4167 {\tiny[0.2772, 0.5811]} & 0.2262 {\tiny[0.1319, 0.3380]} & 0.3333$^{\mathrm{z}}$ {\tiny[0.1429, 1.0000]} & 0.3333$^{\mathrm{z}}$ {\tiny[0.1429, 1.0000]} & 0.5429 {\tiny[0.3235, 0.8667]} \\
 & 3 & 2 & 0.3 & 0.2601 {\tiny[0.2114, 0.3108]} & 0.1278 {\tiny[0.0977, 0.1637]} & 0.6250 {\tiny[0.1667, 1.2500]} & 0.2500 {\tiny[0.0667, 0.5000]} & 0.4914 {\tiny[0.3621, 0.6774]} \\
 & 3 & 4 & 0.3 & 0.1941 {\tiny[0.1572, 0.2360]} & 0.0728 {\tiny[0.0535, 0.0956]} & 0.6923 {\tiny[0.3158, 1.5000]} & 0.2308 {\tiny[0.0435, 0.6667]} & 0.3750 {\tiny[0.3015, 0.4586]} \\
 & 3 & 8 & 0.3 & 0.1577 {\tiny[0.1332, 0.1855]} & 0.0462 {\tiny[0.0341, 0.0606]} & 0.6400 {\tiny[0.4082, 0.9362]} & 0.1800 {\tiny[0.0784, 0.3077]} & 0.2927 {\tiny[0.2287, 0.3672]} \\
 & 3 & 12 & 0.3 & 0.1552 {\tiny[0.1259, 0.1876]} & 0.0423 {\tiny[0.0289, 0.0583]} & 0.4105 {\tiny[0.2674, 0.6053]} & 0.0737 {\tiny[0.0215, 0.1481]} & 0.2727 {\tiny[0.1978, 0.3553]} \\
 & 3 & 16 & 0.3 & 0.1422 {\tiny[0.1202, 0.1668]} & 0.0428 {\tiny[0.0317, 0.0556]} & 0.3706 {\tiny[0.2641, 0.5074]} & 0.1412 {\tiny[0.0856, 0.2143]} & 0.3011 {\tiny[0.2303, 0.3799]} \\
 & 4 & 1 & 0.3 & 0.4316 {\tiny[0.3067, 0.5698]} & 0.3053 {\tiny[0.1895, 0.4572]} & 1.0000$^{\mathrm{d}}$ {\tiny[1.0000, 1.0000]} & 1.0000$^{\mathrm{d}}$ {\tiny[1.0000, 1.0000]} & 0.7073 {\tiny[0.5000, 1.0000]} \\
 & 4 & 2 & 0.3 & 0.2351 {\tiny[0.1935, 0.2839]} & 0.1261 {\tiny[0.1017, 0.1558]} & 0.5000 {\tiny[0.0909, 1.0000]} & 0.5000 {\tiny[0.0909, 1.0000]} & 0.5361 {\tiny[0.4514, 0.6338]} \\
 & 4 & 4 & 0.3 & 0.1244 {\tiny[0.1024, 0.1530]} & 0.0450 {\tiny[0.0346, 0.0576]} & 0.6667 {\tiny[0.3478, 1.2000]} & 0.0667 {\tiny[0.0244, 0.2222]} & 0.3621 {\tiny[0.2826, 0.4495]} \\
 & 4 & 8 & 0.3 & 0.1191 {\tiny[0.0979, 0.1428]} & 0.0485 {\tiny[0.0379, 0.0602]} & 0.3768 {\tiny[0.2432, 0.5211]} & 0.1739 {\tiny[0.0778, 0.2771]} & 0.4070 {\tiny[0.3333, 0.4917]} \\
 & 4 & 12 & 0.3 & 0.1080 {\tiny[0.0851, 0.1365]} & 0.0371 {\tiny[0.0276, 0.0485]} & 0.3153 {\tiny[0.1892, 0.5000]} & 0.0991 {\tiny[0.0397, 0.1818]} & 0.3441 {\tiny[0.2566, 0.4453]} \\
 & 4 & 16 & 0.3 & 0.1169 {\tiny[0.1007, 0.1337]} & 0.0330 {\tiny[0.0254, 0.0420]} & 0.3529 {\tiny[0.2844, 0.4303]} & 0.0840 {\tiny[0.0480, 0.1327]} & 0.2826 {\tiny[0.2189, 0.3532]} \\
 & 2 & 8 & 0.1 & 0.1020 {\tiny[0.0227, 0.2857]} & 0.0101$^{\mathrm{z}}$ {\tiny[0.0054, 0.0222]} & 0.3333$^{\mathrm{z}}$ {\tiny[0.1429, 1.0000]} & 0.3333$^{\mathrm{z}}$ {\tiny[0.1429, 1.0000]} & 0.0909$^{\mathrm{z}}$ {\tiny[0.0476, 0.3333]} \\
 & 2 & 8 & 0.2 & 0.1204 {\tiny[0.0907, 0.1542]} & 0.0336 {\tiny[0.0178, 0.0530]} & 0.6923 {\tiny[0.2353, 1.7143]} & 0.0769 {\tiny[0.0270, 0.2727]} & 0.2794 {\tiny[0.1646, 0.4000]} \\
 & 3 & 8 & 0.1 & 0.1628 {\tiny[0.0488, 0.3235]} & 0.0233 {\tiny[0.0092, 0.0833]} & 1.0000$^{\mathrm{d}}$ {\tiny[1.0000, 1.0000]} & 1.0000$^{\mathrm{d}}$ {\tiny[1.0000, 1.0000]} & 0.1429 {\tiny[0.0476, 0.5000]} \\
 & 3 & 8 & 0.2 & 0.1181 {\tiny[0.0899, 0.1505]} & 0.0382 {\tiny[0.0241, 0.0535]} & 0.0909 {\tiny[0.0286, 0.5000]} & 0.1818 {\tiny[0.0400, 0.5000]} & 0.3235 {\tiny[0.2133, 0.4426]} \\
 & 4 & 8 & 0.1 & 0.0638 {\tiny[0.0115, 0.1452]} & 0.0213 {\tiny[0.0092, 0.0638]} & 1.0000 {\tiny[0.1429, 7.0000]} & 0.3333$^{\mathrm{z}}$ {\tiny[0.1429, 1.0000]} & 0.3333 {\tiny[0.1111, 1.0000]} \\
 & 4 & 8 & 0.2 & 0.1168 {\tiny[0.0880, 0.1516]} & 0.0453 {\tiny[0.0321, 0.0608]} & 0.5833 {\tiny[0.2500, 1.0000]} & 0.0833 {\tiny[0.0303, 0.2857]} & 0.3875 {\tiny[0.2805, 0.5072]} \\
\midrule
9 & 2 & 1 & 0.3 & 0.8000 {\tiny[0.2500, 2.0000]} & 0.0909$^{\mathrm{z}}$ {\tiny[0.0476, 0.3333]} & 1.0000$^{\mathrm{d}}$ {\tiny[1.0000, 1.0000]} & 1.0000$^{\mathrm{d}}$ {\tiny[1.0000, 1.0000]} & 0.1111$^{\mathrm{z}}$ {\tiny[0.0588, 0.3333]} \\
 & 2 & 2 & 0.3 & 0.2264 {\tiny[0.1403, 0.3421]} & 0.0849 {\tiny[0.0360, 0.1444]} & 0.3333$^{\mathrm{z}}$ {\tiny[0.1429, 1.0000]} & 0.3333$^{\mathrm{z}}$ {\tiny[0.1429, 1.0000]} & 0.3750 {\tiny[0.1429, 0.7391]} \\
 & 2 & 4 & 0.3 & 0.1099 {\tiny[0.0628, 0.1702]} & 0.0105 {\tiny[0.0024, 0.0265]} & 0.6667 {\tiny[0.2000, 1.0000]} & 0.1429$^{\mathrm{z}}$ {\tiny[0.0667, 1.0000]} & 0.0952 {\tiny[0.0213, 0.2353]} \\
 & 2 & 8 & 0.3 & 0.0923 {\tiny[0.0543, 0.1497]} & 0.0135 {\tiny[0.0040, 0.0289]} & 0.1739 {\tiny[0.0345, 0.5833]} & 0.0213$^{\mathrm{z}}$ {\tiny[0.0112, 0.0588]} & 0.1463 {\tiny[0.0500, 0.2564]} \\
 & 2 & 12 & 0.3 & 0.0932 {\tiny[0.0559, 0.1470]} & 0.0151 {\tiny[0.0034, 0.0326]} & 0.2609 {\tiny[0.0937, 1.0000]} & 0.0213$^{\mathrm{z}}$ {\tiny[0.0097, 0.0769]} & 0.1622 {\tiny[0.0385, 0.2973]} \\
 & 2 & 16 & 0.3 & 0.0995 {\tiny[0.0587, 0.1687]} & 0.0023 {\tiny[0.0008, 0.0096]} & 0.3273 {\tiny[0.1235, 1.0626]} & 0.0090$^{\mathrm{z}}$ {\tiny[0.0041, 0.0286]} & 0.0233 {\tiny[0.0090, 0.0833]} \\
 & 3 & 1 & 0.3 & 0.7500 {\tiny[0.2000, 1.0000]} & 0.1111$^{\mathrm{z}}$ {\tiny[0.0526, 1.0000]} & 1.0000$^{\mathrm{d}}$ {\tiny[1.0000, 1.0000]} & 1.0000$^{\mathrm{d}}$ {\tiny[1.0000, 1.0000]} & 0.1429$^{\mathrm{z}}$ {\tiny[0.0588, 1.0000]} \\
 & 3 & 2 & 0.3 & 0.1928 {\tiny[0.1067, 0.2927]} & 0.0361 {\tiny[0.0060, 0.0923]} & 1.0000$^{\mathrm{d}}$ {\tiny[1.0000, 1.0000]} & 1.0000$^{\mathrm{d}}$ {\tiny[1.0000, 1.0000]} & 0.1875 {\tiny[0.0370, 0.5000]} \\
 & 3 & 4 & 0.3 & 0.1863 {\tiny[0.1100, 0.2894]} & 0.0190 {\tiny[0.0035, 0.0406]} & 0.3333$^{\mathrm{z}}$ {\tiny[0.1429, 1.0000]} & 0.3333$^{\mathrm{z}}$ {\tiny[0.1429, 1.0000]} & 0.1020 {\tiny[0.0208, 0.1930]} \\
 & 3 & 8 & 0.3 & 0.1748 {\tiny[0.1322, 0.2219]} & 0.0340 {\tiny[0.0158, 0.0575]} & 0.8571 {\tiny[0.4000, 1.6364]} & 0.2143 {\tiny[0.0435, 0.4286]} & 0.1944 {\tiny[0.0909, 0.3284]} \\
 & 3 & 12 & 0.3 & 0.1195 {\tiny[0.0692, 0.1969]} & 0.0239 {\tiny[0.0097, 0.0471]} & 0.3226 {\tiny[0.1000, 1.0833]} & 0.0645 {\tiny[0.0108, 0.3000]} & 0.2000 {\tiny[0.1064, 0.3077]} \\
 & 3 & 16 & 0.3 & 0.1120 {\tiny[0.0726, 0.1719]} & 0.0114 {\tiny[0.0034, 0.0247]} & 0.2319 {\tiny[0.0875, 0.7943]} & 0.0145 {\tiny[0.0043, 0.0811]} & 0.1017 {\tiny[0.0328, 0.1957]} \\
 & 4 & 1 & 0.3 & 0.6667 {\tiny[0.1538, 2.0000]} & 0.2222 {\tiny[0.0435, 0.7146]} & 1.0000$^{\mathrm{d}}$ {\tiny[1.0000, 1.0000]} & 1.0000$^{\mathrm{d}}$ {\tiny[1.0000, 1.0000]} & 0.3333 {\tiny[0.0667, 1.5000]} \\
 & 4 & 2 & 0.3 & 0.2215 {\tiny[0.1385, 0.3185]} & 0.0537 {\tiny[0.0219, 0.0927]} & 1.0000$^{\mathrm{d}}$ {\tiny[1.0000, 1.0000]} & 1.0000$^{\mathrm{d}}$ {\tiny[1.0000, 1.0000]} & 0.2424 {\tiny[0.1034, 0.4444]} \\
 & 4 & 4 & 0.3 & 0.1404 {\tiny[0.1013, 0.1887]} & 0.0197 {\tiny[0.0076, 0.0355]} & 1.0000 {\tiny[0.1667, 7.0000]} & 0.1429$^{\mathrm{z}}$ {\tiny[0.0588, 1.0000]} & 0.1404 {\tiny[0.0580, 0.2400]} \\
 & 4 & 8 & 0.3 & 0.0818 {\tiny[0.0551, 0.1133]} & 0.0123 {\tiny[0.0042, 0.0214]} & 0.7143 {\tiny[0.1429, 3.0000]} & 0.1429 {\tiny[0.0435, 1.0000]} & 0.1500 {\tiny[0.0541, 0.2667]} \\
 & 4 & 12 & 0.3 & 0.0874 {\tiny[0.0483, 0.1495]} & 0.0132 {\tiny[0.0052, 0.0273]} & 0.2881 {\tiny[0.0968, 1.0000]} & 0.0508 {\tiny[0.0081, 0.2381]} & 0.1515 {\tiny[0.0690, 0.2500]} \\
 & 4 & 16 & 0.3 & 0.0832 {\tiny[0.0534, 0.1326]} & 0.0120 {\tiny[0.0056, 0.0221]} & 0.1918 {\tiny[0.1121, 0.5366]} & 0.0274 {\tiny[0.0050, 0.1136]} & 0.1446 {\tiny[0.0750, 0.2297]} \\
 & 2 & 8 & 0.1 & 0.0286$^{\mathrm{z}}$ {\tiny[0.0110, 0.3333]} & 0.0286$^{\mathrm{z}}$ {\tiny[0.0110, 0.3333]} & 1.0000$^{\mathrm{d}}$ {\tiny[1.0000, 1.0000]} & 1.0000$^{\mathrm{d}}$ {\tiny[1.0000, 1.0000]} & 1.0000$^{\mathrm{d}}$ {\tiny[1.0000, 1.0000]} \\
 & 2 & 8 & 0.2 & 0.0691 {\tiny[0.0318, 0.1622]} & 0.0023$^{\mathrm{z}}$ {\tiny[0.0012, 0.0047]} & 0.3333 {\tiny[0.1429, 5.0000]} & 0.0769$^{\mathrm{z}}$ {\tiny[0.0286, 1.0000]} & 0.0323$^{\mathrm{z}}$ {\tiny[0.0222, 0.0588]} \\
 & 3 & 8 & 0.1 & 1.0000$^{\mathrm{d}}$ {\tiny[1.0000, 1.0000]} & 1.0000$^{\mathrm{d}}$ {\tiny[1.0000, 1.0000]} & 1.0000$^{\mathrm{d}}$ {\tiny[1.0000, 1.0000]} & 1.0000$^{\mathrm{d}}$ {\tiny[1.0000, 1.0000]} & 1.0000$^{\mathrm{d}}$ {\tiny[1.0000, 1.0000]} \\
 & 3 & 8 & 0.2 & 0.0902 {\tiny[0.0379, 0.1553]} & 0.0246 {\tiny[0.0037, 0.0660]} & 1.0000 {\tiny[0.1429, 7.0000]} & 1.0000 {\tiny[0.1429, 7.0000]} & 0.2727 {\tiny[0.0435, 0.6250]} \\
 & 4 & 8 & 0.1 & 0.1429$^{\mathrm{z}}$ {\tiny[0.0667, 1.0000]} & 0.1429$^{\mathrm{z}}$ {\tiny[0.0667, 1.0000]} & 1.0000$^{\mathrm{d}}$ {\tiny[1.0000, 1.0000]} & 1.0000$^{\mathrm{d}}$ {\tiny[1.0000, 1.0000]} & 1.0000$^{\mathrm{d}}$ {\tiny[1.0000, 1.0000]} \\
 & 4 & 8 & 0.2 & 0.1318 {\tiny[0.0619, 0.2298]} & 0.0388 {\tiny[0.0079, 0.0815]} & 0.5000 {\tiny[0.1429, 1.0000]} & 0.5000 {\tiny[0.1429, 1.0000]} & 0.2941 {\tiny[0.0909, 0.6000]} \\
\bottomrule
\end{tabular}

%% file: tables/tab_app_value.tex
\begin{tabular}{@{}cclcrrrrrrr@{}}
\toprule
& & & & & \multicolumn{2}{c}{development failures} & \multicolumn{2}{c}{fresh test} & \multicolumn{2}{c}{fresh test, $f=20\,\%$} \\
\cmidrule(lr){6-7}\cmidrule(lr){8-9}\cmidrule(lr){10-11}
benchmark & $d$ & platform & $f$ & cap per layer & this row & best row & failures & $\mu$s per layer & failures & $\mu$s per layer \\
\midrule
I & 3 & Grace & 1\,\% & 0.001\,s & 30\,357 & 30\,057 (10\,\%) & 118\,764 & 1.2 & 117\,766 & 1.8 \\
I & 5 & Grace & 2\,\% & 0.05\,s & 2893 & 2874 (10\,\%) & 11\,662 & 6.8 & 11\,585 & 24.3 \\
I & 7 & Grace & 2\,\% & none & 264 & 262 (5\,\%) & 1087 & 27.1 & 1080 & 96.4 \\
I & 9 & Grace & 2\,\% & none & 34 & 34 (2\,\%) & 104 & 66.4 & 102 & 270.0 \\
I & 3 & Spark & 2\,\% & none & 30\,188 & 30\,057 (10\,\%) & 118\,101 & 4.4 & 117\,770 & 6.6 \\
I & 5 & Spark & 2\,\% & 0.05\,s & 2893 & 2874 (10\,\%) & 11\,662 & 24.8 & 11\,585 & 86.0 \\
I & 7 & Spark & 2\,\% & none & 264 & 262 (5\,\%) & 1087 & 93.7 & 1237 & 155.1 \\
I & 9 & Spark & 2\,\% & none & 34 & 34 (2\,\%) & 104 & 236.1 & 310 & 287.6 \\
\midrule
II & 3 & Grace & 0.5\,\% & 0.002\,s & 6349 & 6300 (10\,\%) & 2818 & 1.2 & 2763 & 3.3 \\
II & 5 & Grace & 5\,\% & 0.02\,s & 4490 & 4484 (10\,\%) & 1520 & 20.8 & 1519 & 49.3 \\
II & 7 & Grace & 10\,\% & none & 358 & 357 (20\,\%) & 1186 & 152.7 & 1186 & 198.0 \\
II & 3 & Spark & 0.5\,\% & 0.002\,s & 6349 & 6300 (10\,\%) & 2818 & 5.1 & 2763 & 9.9 \\
II & 5 & Spark & 5\,\% & 0.02\,s & 4491 & 4485 (10\,\%) & 1520 & 72.6 & 1550 & 107.8 \\
II & 7 & Spark & 2\,\% & none & 372 & 371 (5\,\%) & 1209 & 193.8 & 3605 & 159.9 \\
\bottomrule
\end{tabular}

%% file: tables/tab_rt_mr1.tex
\begin{tabular}{@{}cccrcccccrc@{}}
\toprule
& & \multicolumn{5}{c}{Grace platform} & & \multicolumn{3}{c}{Spark platform} \\
\cmidrule(lr){3-7}\cmidrule(lr){9-11}
$d$ & $f$ & cap & failures & vs network & vs \lom & $\mu$s & & cap & failures & $\mu$s \\
\midrule
3 & 0\,\% & --- & 122\,872 & --- & 0.5634 [0.5578, 0.5696] & 1.2 &  & --- & 122\,872 & 4.3 \\
 & 0.5\,\% & none & 119\,708 & 0.9803 [0.9771, 0.9836] & 0.5523 [0.5463, 0.5589] & 1.2 &  & none & 119\,708 & 4.3 \\
 & 1\,\% & 0.001\,s$^{\circ}$ & 118\,764 & 0.9711 [0.9675, 0.9748] & 0.5471 [0.5412, 0.5536] & 1.2 &  & none & 118\,768 & 4.3 \\
 & 2\,\% & 0.001\,s & 118\,097 & 0.9636 [0.9596, 0.9677] & 0.5429 [0.5369, 0.5495] & 1.2 &  & none$^{\circ}$ & 118\,101 & 4.4 \\
 & 5\,\% & 0.001\,s & 117\,738 & 0.9593 [0.9550, 0.9638] & 0.5405 [0.5346, 0.5472] & 1.3 &  & none & 117\,742 & 4.8 \\
 & 10\,\% & 0.001\,s$^{\ast}$ & 117\,759 & 0.9595 [0.9551, 0.9640] & 0.5406 [0.5346, 0.5472] & 1.5 &  & none$^{\ast}$ & 117\,763 & 5.5 \\
 & 20\,\% & 0.001\,s & 117\,766 & 0.9596 [0.9551, 0.9641] & 0.5406 [0.5347, 0.5473] & 1.8 &  & none & 117\,770 & 6.6 \\
\midrule
5 & 0\,\% & --- & 16\,772 & --- & 0.2889 [0.2826, 0.2954] & 2.2 &  & --- & 16\,772 & 10.5 \\
 & 0.5\,\% & 0.05\,s & 12\,379 & 0.7573 [0.7452, 0.7694] & 0.2188 [0.2135, 0.2243] & 3.3 &  & 0.05\,s & 12\,379 & 13.4 \\
 & 1\,\% & 0.05\,s & 11\,884 & 0.7235 [0.7108, 0.7360] & 0.2090 [0.2040, 0.2143] & 4.6 &  & 0.05\,s & 11\,884 & 17.5 \\
 & 2\,\% & 0.05\,s$^{\circ}$ & 11\,662 & 0.7081 [0.6953, 0.7207] & 0.2045 [0.1996, 0.2099] & 6.8 &  & 0.05\,s$^{\circ}$ & 11\,662 & 24.8 \\
 & 5\,\% & 0.05\,s & 11\,590 & 0.7028 [0.6898, 0.7157] & 0.2030 [0.1980, 0.2083] & 11.6 &  & 0.05\,s & 11\,590 & 41.3 \\
 & 10\,\% & 0.05\,s$^{\ast}$ & 11\,580 & 0.7021 [0.6891, 0.7149] & 0.2028 [0.1978, 0.2081] & 17.1 &  & 0.05\,s$^{\ast}$ & 11\,580 & 60.4 \\
 & 20\,\% & 0.05\,s & 11\,585 & 0.7025 [0.6895, 0.7154] & 0.2030 [0.1979, 0.2082] & 24.3 &  & 0.05\,s & 11\,585 & 86.0 \\
\midrule
7 & 0\,\% & --- & 2722 & --- & 0.1772 [0.1677, 0.1868] & 3.3 &  & --- & 2722 & 19.3 \\
 & 0.5\,\% & none & 1155 & 0.4576 [0.4280, 0.4846] & 0.0811 [0.0750, 0.0864] & 12.1 &  & none & 1155 & 41.4 \\
 & 1\,\% & none & 1103 & 0.4380 [0.4085, 0.4644] & 0.0776 [0.0717, 0.0828] & 18.3 &  & none & 1103 & 62.3 \\
 & 2\,\% & none$^{\circ}$ & 1087 & 0.4317 [0.4025, 0.4576] & 0.0765 [0.0706, 0.0816] & 27.1 &  & none$^{\circ}$ & 1087 & 93.7 \\
 & 5\,\% & none$^{\ast}$ & 1084 & 0.4301 [0.4008, 0.4562] & 0.0762 [0.0704, 0.0814] & 45.7 &  & none$^{\ast}$ & 1084 & 161.0 \\
 & 10\,\% & none & 1080 & 0.4284 [0.3989, 0.4544] & 0.0759 [0.0701, 0.0811] & 67.1 &  & 0.05\,s & 1086 & 176.6 \\
 & 20\,\% & none & 1080 & 0.4284 [0.3989, 0.4544] & 0.0759 [0.0701, 0.0811] & 96.4 &  & 0.005\,s & 1237 & 155.1 \\
\midrule
9 & 0\,\% & --- & 697 & --- & 0.1462 [0.1215, 0.1740] & 5.1 &  & --- & 697 & 32.4 \\
 & 0.5\,\% & none & 115 & 0.1710 [0.1269, 0.2035] & 0.0250 [0.0182, 0.0306] & 26.8 &  & none & 115 & 96.0 \\
 & 1\,\% & none & 110 & 0.1633 [0.1213, 0.1946] & 0.0239 [0.0174, 0.0293] & 42.5 &  & none & 110 & 151.5 \\
 & 2\,\% & none$^{\circ\ast}$ & 104 & 0.1541 [0.1123, 0.1836] & 0.0225 [0.0161, 0.0276] & 66.4 &  & none$^{\circ\ast}$ & 104 & 236.1 \\
 & 5\,\% & none & 103 & 0.1534 [0.1118, 0.1828] & 0.0224 [0.0160, 0.0275] & 119.1 &  & 0.1\,s & 106 & 299.1 \\
 & 10\,\% & none & 103 & 0.1534 [0.1118, 0.1828] & 0.0224 [0.0160, 0.0275] & 180.2 &  & 0.02\,s & 143 & 287.1 \\
 & 20\,\% & none & 102 & 0.1525 [0.1109, 0.1819] & 0.0223 [0.0159, 0.0274] & 270.0 &  & 0.005\,s & 310 & 287.6 \\
\bottomrule
\end{tabular}

%% file: tables/tab_rt_mr2.tex
\begin{tabular}{@{}cccrccccccrc@{}}
\toprule
& & \multicolumn{6}{c}{Grace platform} & & \multicolumn{3}{c}{Spark platform} \\
\cmidrule(lr){3-8}\cmidrule(lr){10-12}
$d$ & $f$ & cap & failures & measure (a) & vs network & vs \lom & $\mu$s & & cap & failures & $\mu$s \\
\midrule
3 & 0\,\% & --- & 2846 & 0.875 [0.804, 0.952] & --- & 0.560 [0.541, 0.581] & 1.2 &  & --- & 2846 & 5.0 \\
 & 0.5\,\% & 0.002\,s$^{\circ}$ & 2818 & 0.866 [0.796, 0.943] & 0.992 [0.981, 1.003] & 0.556 [0.536, 0.576] & 1.2 &  & 0.002\,s$^{\circ}$ & 2818 & 5.1 \\
 & 1\,\% & 0.002\,s & 2797 & 0.860 [0.790, 0.936] & 0.987 [0.972, 1.001] & 0.553 [0.533, 0.573] & 1.3 &  & 0.002\,s & 2797 & 5.1 \\
 & 2\,\% & none & 2771 & 0.852 [0.783, 0.926] & 0.976 [0.957, 0.995] & 0.547 [0.528, 0.567] & 1.7 &  & none & 2771 & 5.3 \\
 & 5\,\% & none & 2749 & 0.845 [0.778, 0.918] & 0.965 [0.941, 0.990] & 0.541 [0.522, 0.561] & 2.2 &  & none & 2749 & 5.9 \\
 & 10\,\% & none$^{\ast}$ & 2760 & 0.848 [0.780, 0.922] & 0.969 [0.943, 0.997] & 0.543 [0.524, 0.563] & 2.6 &  & none$^{\ast}$ & 2760 & 7.1 \\
 & 20\,\% & none & 2763 & 0.849 [0.781, 0.923] & 0.971 [0.944, 0.999] & 0.544 [0.525, 0.564] & 3.3 &  & none & 2763 & 9.9 \\
\midrule
5 & 0\,\% & --- & 2060 & 0.956 [0.838, 1.091] & --- & 0.259 [0.246, 0.272] & 2.4 &  & --- & 2060 & 13.0 \\
 & 0.5\,\% & 0.02\,s & 1654 & 0.767 [0.672, 0.876] & 0.821 [0.795, 0.848] & 0.212 [0.201, 0.225] & 3.7 &  & 0.02\,s & 1654 & 15.9 \\
 & 1\,\% & 0.02\,s & 1590 & 0.738 [0.645, 0.844] & 0.770 [0.739, 0.802] & 0.199 [0.188, 0.211] & 5.8 &  & 0.02\,s & 1590 & 22.1 \\
 & 2\,\% & 0.02\,s & 1543 & 0.716 [0.626, 0.819] & 0.731 [0.698, 0.766] & 0.189 [0.178, 0.201] & 10.3 &  & 0.02\,s & 1543 & 36.4 \\
 & 5\,\% & 0.02\,s$^{\circ}$ & 1520 & 0.705 [0.617, 0.807] & 0.714 [0.679, 0.751] & 0.185 [0.174, 0.196] & 20.8 &  & 0.02\,s$^{\circ}$ & 1520 & 72.6 \\
 & 10\,\% & 0.02\,s$^{\ast}$ & 1519 & 0.705 [0.616, 0.806] & 0.710 [0.675, 0.747] & 0.184 [0.173, 0.195] & 33.0 &  & 0.02\,s$^{\ast}$ & 1519 & 115.8 \\
 & 20\,\% & 0.02\,s & 1519 & 0.705 [0.616, 0.806] & 0.710 [0.675, 0.748] & 0.184 [0.173, 0.195] & 49.3 &  & 0.002\,s & 1550 & 107.8 \\
\midrule
7 & 0\,\% & --- & 2946 & 0.743 [0.482, 1.146] & --- & 0.170 [0.162, 0.178] & 4.1 &  & --- & 2946 & 24.4 \\
 & 0.5\,\% & none & 1323 & 0.334 [0.216, 0.517] & 0.449 [0.423, 0.476] & 0.076 [0.071, 0.081] & 19.8 &  & none & 1323 & 68.0 \\
 & 1\,\% & none & 1231 & 0.311 [0.201, 0.481] & 0.397 [0.372, 0.423] & 0.067 [0.063, 0.072] & 33.6 &  & none & 1231 & 116.0 \\
 & 2\,\% & none & 1209 & 0.305 [0.197, 0.473] & 0.385 [0.360, 0.412] & 0.065 [0.061, 0.070] & 55.4 &  & none$^{\circ\ast}$ & 1209 & 193.8 \\
 & 5\,\% & none & 1195 & 0.302 [0.194, 0.468] & 0.377 [0.352, 0.404] & 0.064 [0.059, 0.069] & 101.1 &  & 0.02\,s & 1224 & 194.1 \\
 & 10\,\% & none$^{\circ}$ & 1186 & 0.299 [0.193, 0.464] & 0.372 [0.347, 0.399] & 0.063 [0.059, 0.068] & 152.7 &  & 0.005\,s & 1356 & 196.6 \\
 & 20\,\% & 0.1\,s$^{\ast}$ & 1186 & 0.299 [0.193, 0.464] & 0.371 [0.346, 0.398] & 0.063 [0.058, 0.068] & 198.0 &  & 0.001\,s$^{\dagger}$ & 3605 & 159.9 \\
\bottomrule
\end{tabular}

%% file: tables/tab_app_live.tex
\begin{tabular}{@{}cclcrcrcrcr@{}}
\toprule
& & & \multicolumn{4}{c}{RTX PRO 6000, 32 EPYC 9555 cores, $N=32$} & \multicolumn{4}{c}{DGX Spark, 20 cores, $N=20$} \\
\cmidrule(lr){4-7}\cmidrule(lr){8-11}
benchmark & $d$ & point & $f$, cap & $\mu$s & live / sim & failures & $f$, cap & $\mu$s & live / sim & failures \\
\midrule
I & 3 & main & 1\,\%, 0.001\,s & 2.4 & 0.985 & 352 / 352 & 2\,\%, none & 7.1 & 0.900 & 340 / 340 \\
 &  & first test & 10\,\%, 0.001\,s & 7.7 & 0.976 & 334 / 334 & 10\,\%, none & 12.2 & 0.941 & 334 / 334 \\
 & 5 & main & 2\,\%, 0.05\,s & 13.4 & 0.944 & 36 / 36 & 2\,\%, 0.05\,s & 25.3 & 0.910 & 36 / 36 \\
 &  & first test & 10\,\%, 0.05\,s & 26.1 & 0.942 & 35 / 35 & 10\,\%, 0.05\,s & 49.9 & 1.003 & 35 / 35 \\
 & 7 & main & 2\,\%, none & 52.8 & 0.950 & 7 / 7 & 2\,\%, none & 103.8 & 1.096 & 7 / 7 \\
 &  & first test & 5\,\%, none & 76.5 & 0.962 & 7 / 7 & 5\,\%, none & 162.4 & 1.140 & 7 / 7 \\
 & 9 & main & 2\,\%, none & 121.7 & 0.938 & 0 / 0 & 2\,\%, none & 257.4 & 1.068 & 0 / 0 \\
 &  & first test & 2\,\%, none & 121.7 & 0.938 & 0 / 0 & 2\,\%, none & 257.4 & 1.068 & 0 / 0 \\
\midrule
II & 3 & main & 0.5\,\%, 0.002\,s & 1.7 & 0.991 & 309 / 309 & 0.5\,\%, 0.002\,s & 5.3 & 0.997 & 309 / 309 \\
 &  & first test & 10\,\%, none & 6.8 & 0.875 & 302 / 302 & 10\,\%, none & 13.3 & 1.097 & 302 / 302 \\
 & 5 & main & 5\,\%, 0.02\,s & 23.6 & 0.928 & 45 / 45 & 5\,\%, 0.02\,s & 49.9 & 0.953 & 45 / 45 \\
 &  & first test & 10\,\%, 0.02\,s & 40.2 & 0.929 & 45 / 45 & 10\,\%, 0.02\,s & 84.3 & 0.988 & 45 / 45 \\
 & 7 & main & 10\,\%, none & 218.2 & 0.938 & 4 / 4 & 2\,\%, none & 199.7 & 1.086 & 5 / 4 \\
 &  & first test & 20\,\%, 0.1\,s & 267.6 & 0.962 & 4 / 4 & 2\,\%, none & 199.7 & 1.086 & 5 / 4 \\
\bottomrule
\end{tabular}

%% file: tables/tab_app_choices.tex
\begin{tabular}{@{}cccccrrrrrrc@{}}
\toprule
benchmark & $d$ & platform & $f$ & cap per layer & composition & network & shots & routed & $\mu$s per layer & p99 (ms) & below budget \\
\midrule
I & 3 & Grace & 10\,\% & 0.001\,s & 30\,057 & 31\,315 & 10\,958\,400 & 1\,095\,928 & 1.6 & 38 & 56/56 \\
I & 3 & Spark & 10\,\% & none & 30\,057 & 31\,315 & 10\,958\,400 & 1\,095\,928 & 5.9 & 181 & 56/56 \\
I & 5 & Grace & 10\,\% & 0.05\,s & 2874 & 4218 & 10\,958\,400 & 1\,095\,846 & 15.4 & 1125 & 56/56 \\
I & 5 & Spark & 10\,\% & 0.05\,s & 2874 & 4218 & 10\,958\,400 & 1\,095\,846 & 54.3 & 2343 & 56/56 \\
I & 7 & Grace & 5\,\% & none & 262 & 695 & 10\,958\,400 & 547\,920 & 44.9 & 4535 & 56/56 \\
I & 7 & Spark & 5\,\% & none & 262 & 695 & 10\,958\,400 & 547\,920 & 157.6 & 8612 & 49/56 \\
I & 9 & Grace & 2\,\% & none & 34 & 204 & 10\,958\,400 & 219\,168 & 65.2 & 7648 & 56/56 \\
I & 9 & Spark & 2\,\% & none & 34 & 204 & 10\,958\,400 & 219\,168 & 231.2 & 15\,639 & 47/56 \\
II & 3 & Grace & 10\,\% & none & 6300 & 6393 & 688\,000 & 68\,801 & 2.6 & 160 & 56/56 \\
II & 3 & Spark & 10\,\% & none & 6300 & 6393 & 688\,000 & 68\,801 & 7.7 & 220 & 56/56 \\
II & 5 & Grace & 10\,\% & 0.02\,s & 4484 & 6213 & 5\,712\,000 & 571\,200 & 26.6 & 859 & 56/56 \\
II & 5 & Spark & 10\,\% & 0.02\,s & 4485 & 6213 & 5\,712\,000 & 571\,200 & 93.9 & 2099 & 47/56 \\
II & 7 & Grace & 20\,\% & 0.1\,s & 357 & 876 & 3\,720\,000 & 744\,000 & 183.5 & 6100 & 55/56 \\
II & 7 & Spark & 2\,\% & none & 372 & 876 & 3\,720\,000 & 74\,400 & 176.5 & 7147 & 41/56 \\
\bottomrule
\end{tabular}